%% file: ms.tex
\documentclass{ptephy_v1}

\usepackage{url} 

\begin{document}
\title{Gravitational Wave Physics and Astronomy in the nascent era}
\input{author}
\date{\today}
\begin{abstract}%




The detections of gravitational waves (GW) by LIGO/Virgo collaborations provide  various possibilities to physics and astronomy.  
We are quite sure that GW observations will develop a lot both 
in precision and in number owing to the continuous works for the improvement of detectors, including the expectation to the newly joined detector, KAGRA, and the planned detector, LIGO-India.  In this occasion, we review the fundamental outcomes and prospects of gravitational wave physics and astronomy. We survey the development focusing on 
representative sources of gravitational waves: 
binary black holes, binary neutron stars, and supernovae. 
We also summarize the role of gravitational wave observations 
as a probe of new physics.  

\end{abstract}

\subjectindex{xxxx, xxx}

\maketitle

\tableofcontents

\newpage
\section{Introduction}\label{ptepGW_sec1}
\input{ptepGW_section1.tex}

\newpage
\section{Binary Black Holes}\label{ptepGW_sec2}
\label{ptepGW_secA01A02}
\input{ptepGW_sectionA01A02.tex}

\subsection{Theoretical study on binary black hole formation}
\label{ptepGW_secA03}
\input{ptepGW_sectionA03.tex}
\section{Binary Neutron Stars }\label{ptepGW_sec3}
\subsection{Exploring NS physics}
\label{ptepGW_secB01}
\input{ptepGW_sectionB01.tex}

\subsection{Association of a short-duration gamma-ray burst}
\label{ptepGW_secB02}
\input{ptepGW_sectionB02.tex}
\subsection{Optical/Infrared Counterparts and Heavy Element Nucleosynthesis}
\label{ptepGW_secB03}
\input{ptepGW_sectionB03.tex}
\section{Supernovae}\label{ptepGW_sec4}
\subsection{GW signatures from core-collapse supernovae}
\label{ptepGW_secC01}
\input{ptepGW_sectionC01.tex}
\subsection{Understanding Supernovae via Neutrino Emissions}
\label{ptepGW_secC02}
\input{ptepGW_sectionC02.tex}

\section{Probes of new physics}
\subsection{Test of gravity theories using gravitational wave data}
\input{ptepGW_section5.tex}

\subsection{Dark matter}
\label{ptepGW_secA02}
\input{ptepGW_sectionA02.tex}

\section{Conclusion}\label{ptepGW_sec5}
\input{ptepGW_conclusion.tex}


\section*{Acknowledgment}
This work is supported by the project 
``GW physics and astronomy: Genesis", which was selected as one of the ``Grant-in-Aid for Scientific Research on Innovative Areas" (JP15H06357-JP15H06365, FY 2017--2021) by Japan Society for the Promotion of Science (JSPS). 

\input{reference_PTEPGW.tex}

\end{document}

%% file: author.tex
\author[1]{Makoto Arimoto}
\author[2]{Hideki Asada}
\author[3]{Michael L. Cherry}
\author[4]{Michiko S. Fujii}
\author[5]{Yasushi Fukazawa}
\author[6]{Akira Harada}
\author[7]{Kazuhiro Hayama}
\author[8]{Takashi Hosokawa}
\author[9]{Kunihito Ioka}
\author[10]{Yoichi Itoh}
\author[11]{Nobuyuki Kanda}
\author[12]{Koji S. Kawabata}
\author[6]{Kyohei Kawaguchi}
\author[13]{Nobuyuki Kawai}
\author[14]{Tsutomu Kobayashi}
\author[15]{Kazunori Kohri}
\author[16,17]{Yusuke Koshio}
\author[7]{Kei Kotake}
\author[4]{Jun Kumamoto}
\author[18]{Masahiro N. Machida}
\author[19]{Hideo Matsufuru}
\author[20]{Tatehiro Mihara}
\author[12]{Masaki Mori}
\author[21]{Tomoki Morokuma}
\author[9,17]{Shinji Mukohyama}
\author[22]{Hiroyuki Nakano}
\author[6]{Tatsuya Narikawa}
\author[23]{Hitoshi Negoro}
\author[24]{Atsushi Nishizawa}
\author[25]{Takayuki Ohgami}
\author[26]{Kazuyuki Omukai}
\author[27]{Takanori Sakamoto}
\author[20]{Shigeyuki Sako}
\author[12]{Mahito Sasada}
\author[28]{Yuichiro Sekiguchi}
\author[26]{Motoko Serino}
\author[29]{Jiro Soda}
\author[26]{Satoshi Sugita}
\author[30]{Kohsuke Sumiyoshi}
\author[24]{Hajime Susa}
\author[12]{Teruaki Suyama}
\author[31]{Hirotaka Takahashi}
\author[9]{Kazuya Takahashi}
\author[32]{Tomoya Takiwaki}
\author[8,9]{Takahiro Tanaka}
\author[25]{Masaomi Tanaka}
\author[33]{Ataru Tanikawa}
\author[24,16]{Nozomu Tominaga}
\author[6]{Nami Uchikata}
\author[34]{Yousuke Utsumi}
\author[17]{Mark R. Vagins}
\author[8]{Kei Yamada}
\author[35]{Michitoshi Yoshida}

\affil[1]{Faculty of Mathematics and Physics, Institute of Science and Engineering, Kanazawa University, Kakuma, Kanazawa, Ishikawa 920-1192, Japan}

\affil[2]{Graduate School of Science and Technology, Hirosaki University, Aomori 036-8561, Japan}

\affil[3]{Department of Physics and Astronomy, Louisiana State University, 202 Nicholson Hall, Baton Rouge, Louisiana 70803, USA}

\affil[4]{Department of Astronomy, Graduate School of Science, The University of Tokyo, 7-3-1 Hongo, Bunkyo-ku, Tokyo 113-0033, Japan}

\affil[5]{Department of Physics, Hiroshima University, 1-3-1 Kagamiyama, Higashi-Hiroshima, Hiroshima 739-8526, Japan}

\affil[6]{Institute for Cosmic Ray Research, The University of Tokyo, 5-1-5 Kashiwanoha, Kashiwa, Chiba 277-8582, Japan}

\affil[7]{Department of Applied Physics, Research Institute of Explosive Stellar Phenomena, Fukuoka University, 8-19-1, Jonan, Nanakuma, Fukuoka 814-0180, Japan}

\affil[8]{Graduate School of Science, Kyoto University, Kyoto 606-8502, Japan}

\affil[9]{Center for Gravitational Physics, Yukawa Institute for Theoretical Physics, Kyoto University, 606-8502, Kyoto, Japan}

\affil[10]{Nishi-harima Astronomical Observatory, Center for Astronomy, University of Hyogo, 407-2 Nishi-gaichi, Sayo, Hyogo 679-5313, Japan}

\affil[11]{Department of Physics, Osaka City University, Osaka, Osaka, 558-8585, Japan, Nambu Yoichiro Institute of Theoretical and Experimental Physics, Osaka City University, Osaka, Osaka, 558-8585, Japan}

\affil[12]{Hiroshima Astrophysical Science Center, Hiroshima University, 1-3-1 Kagamiyama, Higashi-Hiroshima, Hiroshima 739-8526, Japan}

\affil[13]{Department of Physics, Tokyo Institute of Technology, 2-12-1 Ookayama, Meguro-ku, Tokyo 152-8551, Japan}

\affil[14]{Department of Physics, Rikkyo University, Toshima, Tokyo 171-8501, Japan}

\affil[15]{KEK and Sokendai, 1-1 Oho, Tsukuba 305-0801, Japan}

\affil[16]{Department of Physics, Okayama University, Okayama, Okayama 700-8530, Japan}

\affil[17]{Kavli Institute for the Physics and Mathematics of the Universe (WPI), The University of Tokyo Institutes for Advanced Study, University of Tokyo, Kashiwa, Chiba 277-8583, Japan}

\affil[18]{Department of Earth and Planetary Sciences, Faculty of Sciences, Kyushu University, Nishi, Fukuoka 819-0395, Japan}

\affil[19]{High Energy Accelerator Research Organization (KEK), Oho 1-1, Tsukuba 305-0801, Japan}

\affil[20]{MAXI team, Institute of Physical and Chemical Research (RIKEN), 2-1, Hirosawa, Wako, Saitama 351-0198, Japan}

\affil[21]{Institute of Astronomy, Graduate School of Science, The University of Tokyo, 2-21-1, Osawa, Mitaka, Tokyo 181-0015, Japan}

\affil[22]{Faculty of Law, Ryukoku University, 67 Fukakusa Tsukamoto-cho, Fushimi-ku, Kyoto 612-8577, Japan}

\affil[23]{Department of Physics, Nihon University, 1-8 Kanda-Surugadai, Chiyoda-ku, Tokyo, 101-8308, Japan}

\affil[24]{Research Center for the Early Universe (RESCEU), School of Science, The University of Tokyo, Tokyo 113-0033, Japan}

\affil[25]{Department of Physics, Faculty of Science and Engineering, Konan University, 8-9-1 Okamoto, Kobe, Hyogo 658-8501, Japan}

\affil[26]{Astronomical Institute, Graduate School of Science, Tohoku University, Aoba, Sendai 980-8578, Japan}

\affil[27]{College of Science and Engineering, Department of Physics and Mathematics, Aoyama Gakuin University, 5-10-1 Fuchinobe, Chuo-ku, Sagamihara, Kanagawa 252-5258, Japan}


\affil[28]{Department of Physics, Faculty of Science, Toho University, 2-2-1 Miyama, Funabashi, Chiba 274-8510, Japan}

\affil[29]{Department of Physics, Kobe University, Kobe 657-8501, Japan}

\affil[30]{National Institute of Technology, Numazu College, Ooka 3600, Numazu, Shizuoka 410-8501, Japan}

\affil[31]{Research Center for Space Science, Advanced Research Laboratories, Tokyo City University,
8-15-1, Todoroki, Setagaya, Tokyo 158-0082, Japan, and
Department of Information and Management Systems Engineering, Nagaoka University of Technology, Niigata 940-2188, Japan}

\affil[32]{Division of Science, National Astronomical Observatory of Japan, 2-21-1, Osawa, Mitaka, Tokyo 181-8588, Japan}

\affil[33]{Department of Earth Science and Astronomy, College of Arts and Sciences, The University of Tokyo, 3-8-1 Komaba, Meguro-ku, Tokyo 153-8902, Japan}

\affil[34]{Kavli Institute for Particle Astrophysics and Cosmology, SLAC National Accelerator Laboratory, Stanford University, 2575 Sand Hill Road, Menlo Park, CA 94025, USA}

\affil[35]{Subaru Telescope, National Astronomical Observatory of Japan, 650 North A'ohoku Place, Hilo, Hawaii 96720, USA}

%% file: ptepGW_section1.tex
\newcommand{\blue}[1]{#1}

\begin{table}[bh]
    \centering
    \caption{Observing terms of LIGO/Virgo detectors. Detectors column: H1 at Hanford, Washington, and L1 at Livingston, Louisiana. V1 at Cascina, Italy. }
    \label{table:LIGOVirgo1}
    \begin{tabular}{ll|ll}
    \hline
   & Observation Period & detectors & events reported   \\
    \hline
    O1& Sep 2015 to Jan 2016 &  H1, L1 & 3 BBHs \\
    O2& Nov 2016 to Sep 2017 &  H1, L1 & 3 BBHs \\
    O2& Aug 2017   &  H1, L1, V1 & 4 BBHs, 1 BNS \\
    O3a& Apr 2019 to Sep 2019 &  H1, L1, V1 & many BBHs, 1 BNS\\
    O3b& Nov 2019 to Apr 2020 &  H1, L1, V1 & N/A \\
    \hline
    \end{tabular}
\end{table}

\subsection{Gravitational wave from binary black holes}
The first gravitational wave (GW) event from the binary black holes (BBHs), GW150914, which was discovered by LIGO/Virgo collaboration brought us many surprises~\cite{GW150914}. 
Regarding binary neutron star (BNS) coalescences, it was believed that there is not much uncertainty in estimating their event rate compared with the case of BBHs. 
Even \blue{for BNS coalescences}, the uncertainty in the estimated event rate was expected to be about one order of magnitude. 
On the other hand, although it is conceivable that BBHs should exist in our universe, 
the estimation of the GW event rate had no observational support, differently from the case of BNSs. 
Hence, BBHs were not considered to be a promised GW source. 
It was somewhat surprising that such a system was the first directly detected GW source. 
However, it was even more surprising that those black holes (BHs) were heavier than many people expected, i.e., each component BH has a mass about 30$M_\odot$. 
The mass of the BH candidates found so far by electromagnetic observations is estimated to be at most about 20$M_\odot$~\cite{Farr:2010tu}.
Although 
there were some candidate objects that did not contradict with even heavier mass within the observational \blue{error,} there is no object that 
can be said to have about 30$M_\odot$. 
The observation of GWs has changed our understanding about the mass distribution of BHs in the universe~\cite{rosswog17,hotokezaka18}. 
Since then, the number of BBH observations officially announced by the LIGO/Virgo collaborations has been increasing~\cite{LIGOScientific:2018mvr}. 
Not all of them are massive BHs, but the succeeding 
observations indicate that the first GW event, GW150914, was not so special.

As a result of this discovery, how those BBHs detected by GWs 
were formed in the history of the universe has become one of the great mysteries of astrophysics. 
A variety of BBH formation scenarios have been proposed~~\cite{abbott16}. 
The most promising candidate for the formation channel would be the one through the binary interaction between 
heavy star binaries formed from gases with a low metallicity. 
Even for this leading scenario, there are various uncertainties: the binary formation rate from the gas cloud, 
stellar evolution including mass loss processes, binary interaction to form close binaries, formation process of BHs, and so on. 
The binary evolution can be so complex, changing its orbital radii and eccentricity. 
Therefore, it is a theoretical challenge to predict the distributions of BH masses, orbital separation and eccentricity at the formation of a BH binary, 
to make comparison with observations. As alternative scenarios, we can think of the formation of BH binaries in star clusters~\cite{2015PhRvL.115e1101R} and their formation in the primordial universe~\cite{Sasaki:2018dmp}. Many variants of these scenarios can be thought of. 
In the future, as the details of the theoretical model become better understood, and as GW observations accumulate the information 
about the event rate, the mass and distance distributions, the distribution of BH spins, and so on,  
we would be able to identify the contributions from respective formation channels of BBHs. 

\subsection{Gravitational wave from binary neutron stars}
The first GW event from a BNS merger GW170817 observed on August 17, 2017, marked a splendid opening of GW astronomy~\cite{GW170817}. 
BNS coalescence was counted as a guaranteed GW source. 
The discovery was followed by simultaneous observation of gamma-ray bursts, identification of the electromagnetic counterpart 
by optical and infrared telescopes~\cite{abbott17MMA}, long-term follow-up observations by radio waves and X-rays, detection of apparent superluminal motion~\cite{Mooley:2018qfh,Ghirlanda:2018uyx}. 
How BNS coalescence proceed was revealed to be something close to what was theoretically predicted. 
In the study of BNS coalescence, weak gravity approximation is not appropriate any more. 
Hence, numerical relativity is an indispensable research tool with the knowledge of equation of state for the 
high density nuclear matter. At the same time various radiation fields, photons and neutrinos, originating from high density matter 
must be solved and the magnetic fields also can play an important role in the simulations. 
Therefore combining various techniques for numerical simulation is required.
In the case of GW170817, from its optical and infrared light curves, 
it is thought that the ejecta mass was about $0.05M_\odot$ and a part of them possessed high neutron excess,
resulting in production of significant amount of $r$-process elements. 
The amount of the produced $r$-process elements may be sufficient to explain almost all of them in the universe. 
In addition, numerical simulations show that the merged binary becomes a massive neutron star (NS) supported by rotation, 
and the material stripped around it forms a disk. 
The simultaneous (1.7 second after the coalescence time) observation of a short gamma-ray burst (SGRB) indicates the connection between SGRB and BNS coalescence.  
Although the observed SGRB was extraordinary weak, this can be understood because the SGRB jet was not pointing at the line of 
sight toward us~\cite{Ioka:2017nzl,Ioka:2019jljb}. This interpretation gained a strong support because apparent superluminal motion was observed by radio waves~\cite{Mooley:2018qfh,Ghirlanda:2018uyx}. 
The detection of apparent superluminal motion is also a strong evidence for that a relativistic jet has been launched.
In addition, the GW observations at this event also provided restrictions on the equation of state of dense nuclear matter, 
mostly from the constraint on the tidal deformability of NSs just before the coalescence~\cite{GW170817}. 
Here again, numerical relativity technique played an important role. 
It is also an important theoretical challenge to derive the equation of state that can be compared with phenomena from the first principle of QCD. 
High-energy gamma-rays, neutrinos, and cosmic rays were not detected in GW170817
~\cite{2017ApJ...850L..22A,2017ApJ...850L..35A,2018ApJ...861...85A,2018ApJ...857L...4A},
although BNS mergers and SGRBs are considered to be sources of these particles
~\cite{2018ApJ...854...60M,2018PhRvD..98d3020K,2018MNRAS.476.1191B,2016ApJ...827...83K}.
We do not further discuss these multi-messenger signals in this review.

\subsection{Gravitational wave from Supernovae}
Although GWs from supernovae have not been detected yet, they are important targets 
of gravitational wave observations. 
After more than a half-century of continuing efforts ever since the seminal work by Colgate and White (1966) ~\cite{colgate}, 
theory and neutrino radiation-hydrodynamics simulations of core-collapse supernovae (CCSNe)
are now converging to a point that multi-dimensional (multi-D)
 hydrodynamics instabilities play a key role in the 
neutrino mechanism~\cite{bethe}, the most favoured scenario 
to trigger explosions.
In fact, self-consistent 
simulations in three spatial dimensions (3D)
 now report shock revival of the stalled bounce shock into explosion
 by the multi-D neutrino mechanism
(see~\cite{bernhard16,janka16,burrows13,Kotake12} for reviews).

From observational point of view, mounting evidence from multi-wavelength electromagnetic-wave (EM) observations (such as blast/ejecta morphologies, line profiles, polarizations) are pointing toward CCSNe being generally aspherical (e.g.,~\cite{casA,tanaka17} and references therein). Nevertheless, these EM signals are secondary as a probe into the supernova inner-workings, providing information only after a shock break-out from the massive stars. On the other hand, the GWs from CCSNe are direct observables, which imprint a live information of the central engine. Furthermore coincident detection of CCSN neutrinos with GWs is important not only
   for significantly enhancing the
   detectability of the GWs (e.g.,~\cite{naka016}), but also for unraveling the hydrodynamics evolution from the onset of core-collapse toward the forming compact objects
   (neutron stars and black holes, see~\cite{mirizzi} for a review).
Current estimates of CCSN 
rates in our Milky way predict one event every $\sim 40 \pm 10$ 
year~\cite{horiuchi_kneller}.
 While rare, they provide us unique opportunity to study 
CCSNe using a full set of multi-messenger observables
including GWs, neutrinos, and nuclear gamma-rays~\cite{gehrels}.

\subsection{Test of general relativity}
The direct detection of GWs has opened a new window for tests of general relativity (GR). 
In order to detect GW signals, we need theoretical prediction for the GW waveform in advance. 
In more complex theory of gravitation that extends GR, 
theoretically predicting the waveform is more difficult. 
Our current knowledge about the waveform in modified gravity is limited, but some 
possible deviations in the waveform from GR have been predicted. 
GW observations so far have not suggested any deviation from GR~\cite{LIGOScientific:2019fpa}. 
However, the lack of deviation itself can have important physical meaning. 
Besides constraints on the deviation in the waveform, 
an important constraint was also derived from the speed of propagation of GWs, 
i.e., from the detection of short gamma-ray burst 
associated with GW170817 event at a distance of about 40 Mpc.  
The fractional deviation of the propagation speed of GWs from that of light is less than $\approx 10^{-15}$~\cite{Monitor:2017mdv}, which excludes various possible 
extensions of GR~\cite{Creminelli:2017sry,Sakstein:2017xjx,Baker:2017hug,Langlois:2017dyl,Crisostomi:2017lbg,Dima:2017pwp}. 

\blue{GWs from binaries can be used as the standard siren~\cite{Schutz:1986gp}, providing an alternative method 
to measure the cosmic expansion history. GW170817 already gives us an estimate of the Hubble constant~\cite{Abbott:2017xzu,Hotokezaka:2018dfi}, but we do not further discuss the aspects of testing cosmological models using GWs in this review.}

\subsection{Organization of this article}
In this review paper we overview the vast area of newly developing gravitational wave physics and astronomy mentioned above.
This paper is organized as follows in Sec. 2, 3 and 4, we focus on 
three major sources of gravitational waves, i.e., BBHs, BNSs, and SNe. 
We discuss various topics emanating from the detection of gravitational waves, 
including the future prospects. In Sec. 5 we focus more on the new physics 
examined by using GWs as a probe. We discuss the test of gravity and 
implication to the dark matter physics. Sec. 6 is devoted to a 
short conclusion.

%% file: ptepGW_sectionA01A02.tex
\subsection{Gravitational waves from binary black holes}
LIGO/Virgo has reported GWs from BBH mergers as a 
result of O1 and O2 observing run~\cite{LIGOScientific:2018mvr} that is called
``Gravitational-Wave Transient Catalog'' (GWTC)-1. 
The best fit parameters for these binaries are listed in Table~\ref{table:LIGOVirgo}. 
We have large errors in the parameter estimation,
especially for the individual masses and the effective spin parameter,
because the signal-to-noise ratios (SNRs) are not so high.
There are results reported by alternative searches. One of them reports other candidates of GW events~\cite{Venumadhav:2019lyq}.

\begin{table}[th]
    \centering
    \caption{BBH merger events observed during LIGO/Virgo O1 and O2. We show only the median values
    for the individual masses $m_1$ and $m_2$, chirp mass ${\cal M}$,  effective spin $\chi_{\rm eff}$, final BH mass $M_{{\rm fi}}$, and its angular momentum parameter $a_{{\rm fi}}$.  These numbers are of source frame.  The estimated distance in Mpc and red-shift $z$ are also shown. The 90\% credible intervals have been reported in Table III of Ref.~\cite{LIGOScientific:2018mvr}.}
    \label{table:LIGOVirgo}
    \begin{tabular}{c|cccc|cc|rc}
    \hline
    Event & $m_1/M_\odot$ & $m_2/M_\odot$ & ${\cal M}/M_\odot$ & $\chi_{\rm eff}$ & $M_{{\rm fi}}/M_\odot$ & $a_{{\rm fi}}$ & \multicolumn{2}{c}{distance (Mpc,$z$)}\\
    \hline
    GW150914 & 35.6 & 30.6 & 28.6 & -0.01 & 63.1 & 0.69 & 430~Mpc&0.09 \\
    GW151012 & 23.2 & 13.6 & 15.2 & 0.05 &35.7 & 0.67 & 1060~Mpc& 0.20 \\
    GW151226 & 13.7 & 7.7 & 8.9 & 0.18 & 20.5 & 0.74 & 440~Mpc& 0.09\\
    GW170104 & 30.8 & 20.0 & 21.4 & -0.04 & 49.1 & 0.66 & 960~Mpc&0.19 \\
    GW170608 & 11.0 & 7.6 & 7.9 & 0.03 & 17.8 & 0.69& 320~Mpc&0.07 \\
    GW170729 & 50.2 & 34.0 & 35.4 & 0.37 & 80.3 & 0.81& 2750~Mpc&0.48 \\
    GW170809 & 35.0 & 23.8 & 24.9 & 0.08 & 56.4 & 0.70& 990~Mpc&0.20\\
    GW170814 & 30.6 & 25.2 & 24.1 & 0.07 & 53.4 & 0.72 & 580~Mpc&0.12\\
    GW170818 & 35.4 & 26.7 & 26.5 & -0.09 & 59.8 & 0.67 & 1020~Mpc& 0.20 \\
    GW170823 & 39.5 & 29.0 & 29.2 & 0.09 & 65.6 & 0.71 & 1850~Mpc& 0.34\\
    \hline
    \end{tabular}
\end{table}

\begin{figure}[th]
    \centering
   \includegraphics[width=0.99\textwidth]{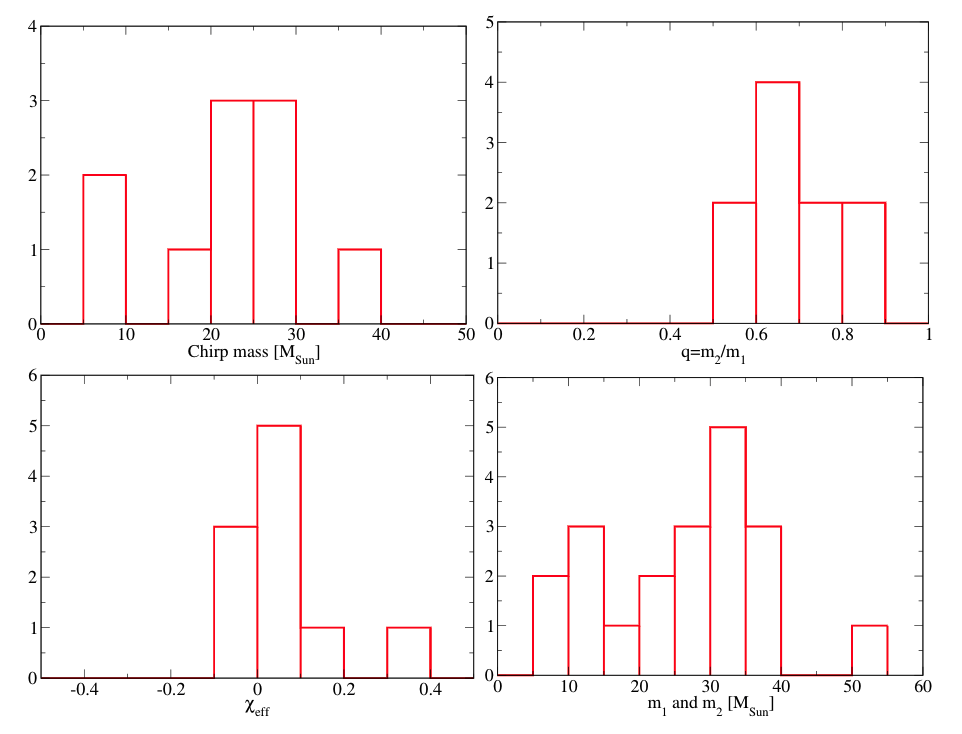}
    \caption{Distribution of the median values of the chirp mass ${\cal M}$, mass ratio $q=m_2/m_1$, effective spin $\chi_{\rm eff}$, and individual masses (which are redundant though) of BBHs observed during LIGO/Virgo O1 and O2.
    It should be noted that there are large errors in the current observations. Even for the chirp mass which is the most sensitive parameter in the GW waveform of BBH mergers, the 90\% credible interval gives $\lesssim$ 20\% difference in the estimation.}
    \label{fig:A01_1}
\end{figure}

The obtained mass and spin distributions are summarized in Fig.~\ref{fig:A01_1}.
The chirp mass is defined by 
\begin{equation}
     {\cal M}=\mu^{3/5} M^{2/5}\,, 
\end{equation}
with $M=m_1+m_2$ and $\mu=m_1 m_2/M$,
where $m_1$ and $m_2$ are the component masses. 
The effective spin parameter is defined by
\begin{equation}
   \chi_{\rm eff}=\frac{1}{M}
   \left(\frac{\mbox{\boldmath{$S$}}_1}{m_1}+\frac{\mbox{\boldmath{$S$}}_2}{m_2}\right)
   \cdot \hat{\mbox{\boldmath $L$}} \,, \label{definition-chi-eff}
\end{equation}
where $\mbox{\boldmath{$S$}}_1$ and $\mbox{\boldmath{$S$}}_2$ are 
spin angular momentum vectors of the respective bodies and 
$\hat{\mbox{\boldmath $L$}}$ 
is the 
unit vector pointing at the direction of the orbital angular momentum of the 
binary. 

The reported event rate by the LIGO/Virgo Collaborations is 
ranging from $9.7$ to $101$ $\mbox{Gpc}^{-3}\mbox{yr}^{-1}$ for 
BBH mergers. 
The masses of the observed BHs are relatively large compared with the masses of 
stellar mass BH candidates found by X-ray observations (see, e.g., Ref.~\cite{Farr:2010tu} and references therein). 
The origin of BBHs 
is an important new issue of debate raised by the GW observations. A promising 
formation path is the one through the stellar evolution of low metallicity 
stars, which is 
the main topic of Section~\ref{ptepGW_secA03}. There are various formation channels 
and the main focus is whether or not the stellar dynamics in the star cluster plays 
a crucial role. An alternative is the formation of BH binaries in the 
early universe, which is the so-called primordial BH scenario. The formation 
process of primordial BH binaries is discussed in Section~\ref{ptepGW_secA02}.

%% file: ptepGW_sectionA03.tex
Mergers of BBHs are the dominant population of gravitational wave sources. Currently, two evolutionary pathways are envisaged as origins of BBHs.
First is the so-called in-situ BBH formation scenario, where massive star binaries are formed and each member star becomes a black hole (BH) after stellar evolution.
Second is the so-called ex-situ scenario, where single stars formed in a dense star cluster become BHs after stellar evolution and those BHs become binaries as a result of dynamical interaction.  
Here, we describe our current understanding of those scenarios in this order. 

\subsubsection{Formation of isolated stellar binaries}

\paragraph A 

Binary star formation at low metallicities

\begin{figure}
  \begin{center}
    \includegraphics[scale=0.75]{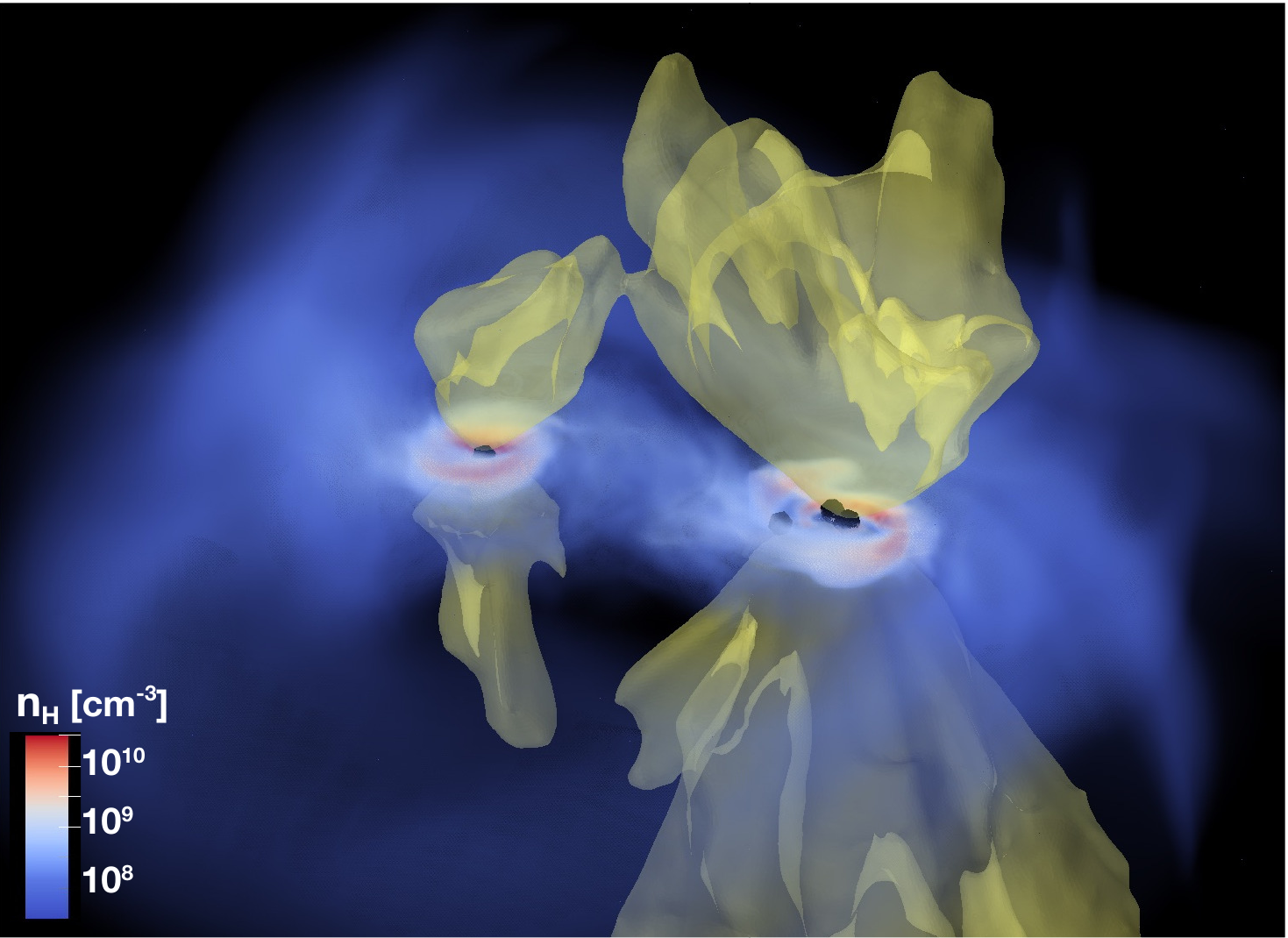}
\caption{
  \label{fig:hosokawa1}
An example of the latest 3D radiation-hydrodynamics simulations of the primordial star formation, in which the metallicity is exactly zero (from Sugimura et al. submitted \citep{sgmr20}), showing a snapshot at the epoch when massive protostars in a binary system accrete the gas under the radiative feedback.
Presented are the volume-rendered 3D distribution of the mass density, together with ionization fronts (yellow surfaces) and protostars (black dots).}
\end{center}
\end{figure}

Whereas massive stellar binaries are possible astrophysical progenitors of BH-BH binaries that finally merge, their formation process is still uncertain. 
In particular, major focus is put on formation of such systems at low metallicities $Z \lesssim 0.1~Z_\odot$, which are required for yielding massive ($\simeq$ a few $\times 10~M_\odot$) BHs avoiding significant mass loss during the stellar evolution. 
Observations suggest that a large fraction of massive stars are born in binaries in the solar neighborhood. Some of them are in very tight systems with separations of only $\lesssim 0.1$~AU, and are possible candidates that finally evolve into BH binaries that merge within the Hubble time.
On the other hand, since the delay time until the BBH merger is potentially as long as the Hubble timescale, depending on the initial separation, a part of the progenitor stars may have been born in the early universe. 
Therefore, our goal is to elucidate the formation process of massive and tight binary stars in a variety of low-metallicity environments, including the local and early universe.


A straightforward method which ultimately leads us to the goal is directly following the evolution with numerical simulations. One of the key processes is the gas gravitational fragmentation that potentially leads to bearing multiple stars, some of which are to be in binaries. Magnetic fields, if any, should be important because it transfers the angular momentum of the gas, affecting the orbital evolution of newly forming binaries. The stellar radiative feedback is also critical to determine how massive binaries finally appear, halting the mass accretion onto the stars \citep{MT08,H11,H16}. We consequently need to solve the interplay between gas dynamics, magnetic fields, and radiative fields in full 3D spatially resolving a very wide dynamic range. Ultimately integrating the evolution over $\sim$~Myr is required to cover the whole evolution of the star formation process. Although this is still challenging, recent studies are advancing to incorporate the above processes as much as possible with limited computational resources.
For instance, Fig.~\ref{fig:hosokawa1} shows a snapshot in our recent 3D radiation hydrodynamic simulations using the adaptive-mesh-refinement code SFUMATO \citep{sgmr20}. In this particular case, we follow the long-term ($\sim 0.1$~Myr) evolution in the protostellar accretion stage of the primordial star formation. We see a multiple stellar system including a massive binary system that has appeared after the disk fragmentation in the earlier stage. The double bipolar H$\,$II regions are growing from the two massive stars that are still accreting the surrounding gas under the radiative feedback. 
Although this provides a glimpse of the binary formation in the early universe, the newborn binary has a large separation: the two massive stars exciting the bipolar H$\,$II regions are separated by more than $10^3$~AU. The formation process of the massive and tight binaries, some of which should evolve to the gravitational wave sources, is yet to be revealed. Obviously further studies and simulations are necessary for explaining how such systems are formed. We describe our recent studies and future prospects in what follows.

\paragraph B

Role of magnetic fields in close binary formation

Observations of nearby star-forming regions indicate that a large fraction of stars are members of binary systems \citep{duchene13} and  
the binary frequency is even higher for high-mass stars than for low-mass ones \citep{sana12}. 
Theoretically, numerical simulations have shown that a binary stellar system forms as a result of fragmentation of a molecular cloud core during the gravitational collapse \citep{matsumoto03}.
In rapidly rotating and gravitationally unstable cores, 
fragmentation occurs more easily  \citep{miyama84,tsuribe99} and the accretion rate is higher after the protostars are formed.  
The rotation velocity increases and the cloud becomes disk-like during the collapse owing to the angular momentum conservation.
This promotes fragmentation and subsequent binary formation.

According to numerical simulations, the binary separation continues to increase with time as a parcel of gas with larger angular momentum falls onto the central region or proto-binary system at later times \citep{matsumoto03,machida08} and the outcome is  
a wide binary system with a separation typically greater than $\gg 100$\,AU. 
In contrast, observationally high-mass stars tend to be in close-binary systems with the separation 1-10\,AU \citep{duchene13}. 
This discrepancy between theory and observations can be circumvented if  we take non-ideal magnetohydrodynamics (MHD) effect into account. 

\begin{figure}
  \begin{center}
    \includegraphics[scale=0.75]{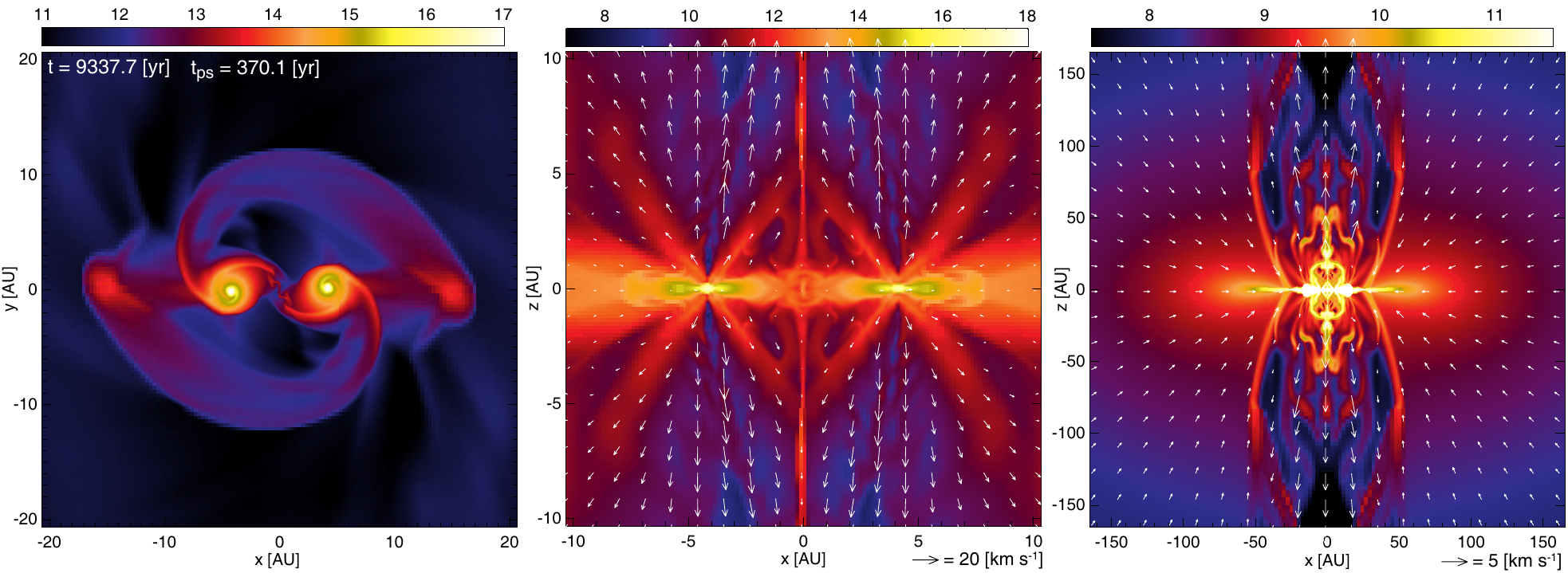}
\caption{
  \label{fig:machida1}
Proto-binary formation in the MHD simulation. 
Shown are density (color) and velocity (arrows; middle and right panels) distributions on the $z=0$ (left) and $y=0$ (middle and right) plane at three different 
spatial scales. 
The elapsed time after the onset of prestellar cloud $t$ and that after protostar formation $t_{\rm ps}$ are described in the left panel. 
}
\end{center}
\end{figure}

Recent non-ideal MHD simulations for present-day star formation showed that the excess angular momentum is transported from the proto-binary system by magnetic braking and by outflow launching and, as a consequence, the close binary system can be formed in a highly unstable magnetized cloud.  
Figures~\ref{fig:machida1} is a snapshot of a proto-binary system in an example of such simulations \citep{saiki20}. 
In the figure, we can see that powerful jet is driven by each of the protostars.
The separation of the binary system is about 10\,AU at this epoch. 
On the equatorial plane, both the circumstellar and circumbinary disks can be confirmed (the left panel of Fig.~\ref{fig:machida1}). 
Two cavity-like structures created by the jets are seen in the middle panel of Fig.~\ref{fig:machida1}.
The velocity of the jets driven by the protostars exceeds  $>100$\,km\,s$^{-1}$. 
Those high-velocity components are enclosed by the low-velocity outflow of  $\sim10$\,km\,s$^{-1}$
driven by the circumbinary disk.
The angular momentum of the binary system is extracted from the system both by the large-scale wide-angle outflows and small-scale collimated jets. 
The binary jet/outflow structure is also reported from recent ALMA observations \citep{tobin19}.

\paragraph C

Stellar system formed from metal-free gas

Next we see the nature of first stellar systems formed from the primordial gas and discuss possible formation of first star binaries. 
The collapse of a slightly gravitationally unstable gas cloud ($n_{\rm H} =1.4\times 10^4{\rm cm^{-3}}$ and  $T=200$K) is simulated by the SPH method. 
Armed with elaborate sink particle creation scheme and usage of the barotropic equation of state pre-calculated by a one-zone model, 
we succeeded in extending the time integration up to 25000 yr in 
the accretion phase, $\sim 17$ times longer than the previous calculation \citep{susa2019}.

\begin{figure}
  \begin{center}
    \includegraphics[scale=0.4]{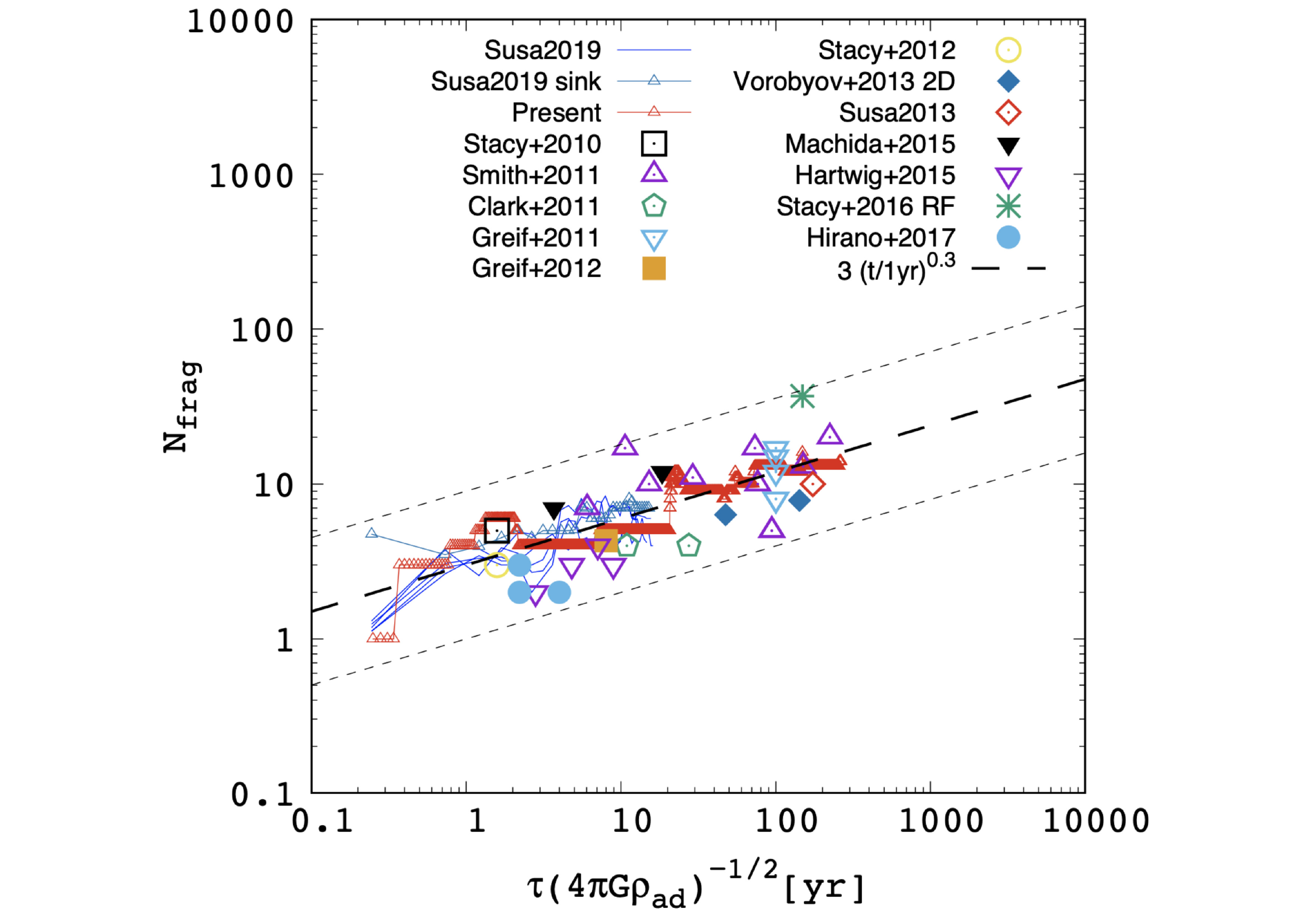}
\caption{
  \label{fig:susa1}
Number of fragments in a first-star-forming cloud as function of the scaled time. We assume $\rho_{\rm ad}$ is $2\times 10^{-5} {\rm (g/cm^{-3})}$, above which the gas behaves adiabatic.
Red line with small triangles denotes the present result, and the 
thin lines correspond to our previous result\citep{susa2019}. Results in the literature\citep{stacy2012,smith2011,clark2011a,greif2011,greif2012,vorobyov2013,susa2013,machida2015,hartwig2015a,stacy2016,hirano2017} are also shown.}
\end{center}
\end{figure}

 Figure~\ref{fig:susa1} shows the number of fragments, i.e., sinks, as functions of the scaled time normalized by the free-fall time at 
 the density $n_{\rm th}=10^{15}{\rm cm}^{-3}$, above which the temperature rises adiabatically. 
 The results can be scaled using the scaled time according to the scale-free nature of gravity \citep{susa2019}. 
 Note that the integration time of the present simulation is also the longest so far in terms of the scaled time. 
 We can see a simple trend that the number of fragments is proportional to the elapsed time after the central density exceeds $n_{\rm th}$. 
 This trend has already been known previously \citep{susa2019}, but now we confirm that the relation holds even after 10 times longer integration time, although 
 the simulation is done for one particular realization.  This result is also generally consistent with results of other simulations in the literature.

\begin{figure}
  \begin{center}
    \includegraphics[scale=0.4]{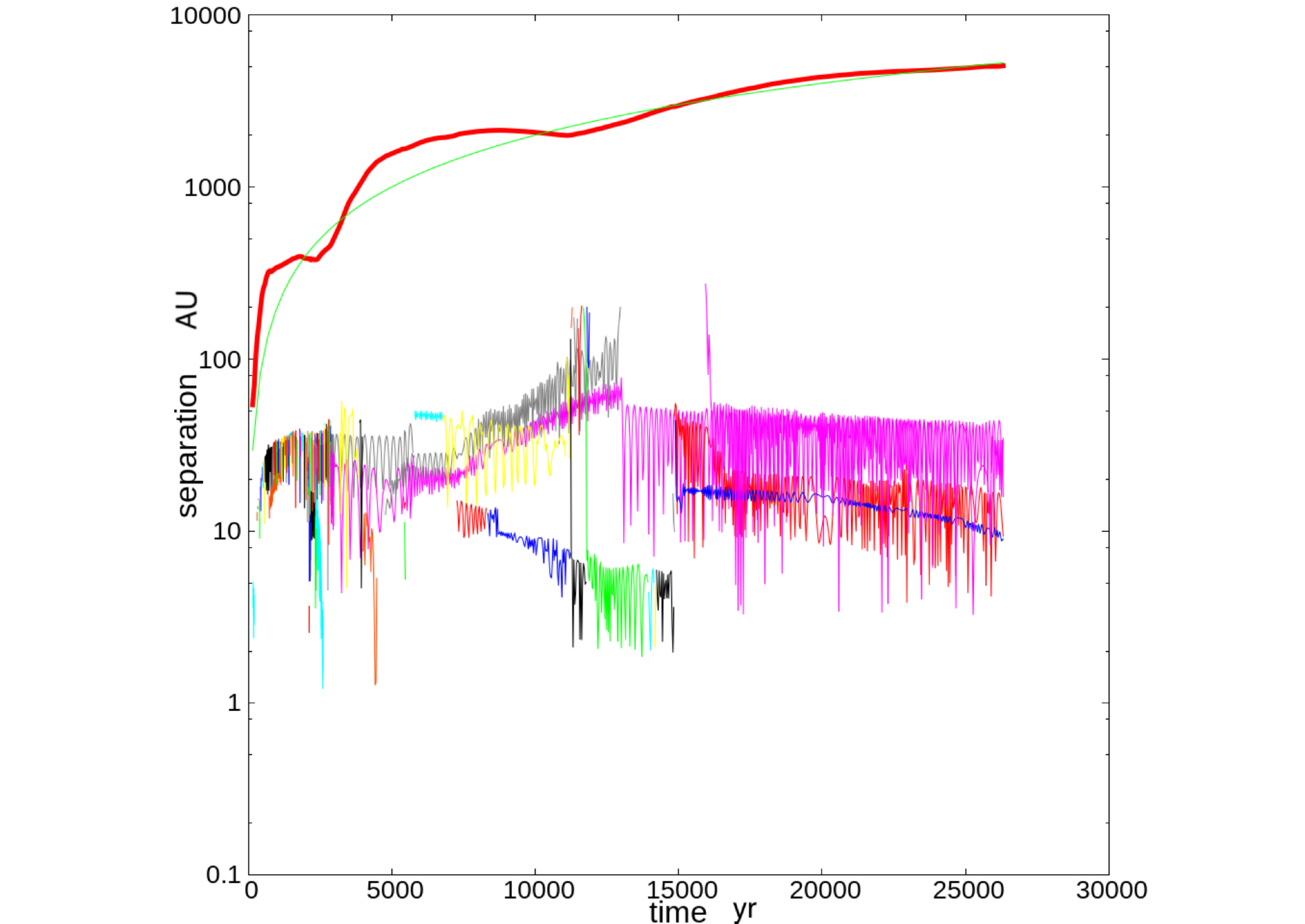}
\caption{
  \label{fig:susa2}
The separations of binaries identified in the present simulation as functions of time. Different colors denote different binaries. Thick red solid curve shows the mean square average of the distance between the stars and the center of mass of the stellar system whereas the green thin curve shows the distance in proportion to the elapsed time since the first sink formation.
}
\end{center}
\end{figure}

Several systems of binaries are identified in this cluster. 
Figure~\ref{fig:susa2} shows the time evolution of their separations.
Different colors correspond to different pair of binaries. 
The members of the binaries easily change as a result of the close encounter/merger with other stars. 
The thick red curve at the top of the panel shows the typical size of this cluster, which grows in proportion to the elapsed time after the first protostar formation, as is predicted by the conservation of angular momentum and the gas distribution at the end of the collapse phase. 
On the other hand, the binary separations also grow when the binary systems acquire the angular momentum as a result of the gas accretion, which is, however, counterbalanced by the effect of encounters with other stars that makes the binary tighter. Hence, the separations are kept rather constant during the computation. 
The separations observed in this particular run are $\sim$ a few 10 AU, which is too wide to end up with BBHs that coalesce within a Hubble time. In the present simulation, however, we adopt a lower threshold density than the true adiabatic density for computational reasons and the actual binary separation should be scaled from our nominal value. After the scaling, the time should be 100 times shorter and the separation $\sim 70$ times less. Thus, our binary separation corresponds to a few times $0.1$ AU after scaling, although the integrated time is only 250 yr. If this separation is kept constant for another $\sim 5000$ yr when the surrounding gas is swept away by the radiative feedback, these binaries could be progenitors of merging BBHs observed in gravitational waves.

To draw firm conclusions about the nature of formed stellar systems, we need to extend the numerical simulation in time until 
the radiative feedback from the protostar terminates the accretion to the system at several thousand years after the formation of the first protostar. 
We should also consider the effects of turbulence and magnetic fields. 
In fact, the presence of a coherent magnetic field in primordial environment \blue{induces} a strong magnetic breaking effect \citep{machida_doi2013}, leading to less fragmentation in the accretion disk. 
Also recent simulations predict more fragmentation 
in case with strong turbulence. 
Those are the issues to be addressed in the future.

\subsubsection{Binary black hole formation by dynamical interactions in dense star clusters}

\paragraph A

Binary black holes formed in star clusters

Dense star clusters are one of promising formation sites of BBHs. Especially, globular clusters (GCs) have been expected as a source of BBHs merging within the Hubble time. 
In these days, most studies on BBH formation in GCs are being done with the Monte-Carlo method \citep{2015PhRvL.115e1101R,2017MNRAS.464L..36A}. Those works suggested a local merger rate density of $\sim 5$\,Gpc$^{-3}$\,yr$^{-1}$ \citep{2017MNRAS.464L..36A, Rodriguez16}. They also predicted BBHs formed in GCs have several distinct features from those formed under isolated environments. The GC-originated BBHs have more isotropic spin-orbit misalignment angles \citep{2016ApJ...832L...2R}. Some of those BBHs may contain massive ($\sim 100M_\odot$) BHs locating in the so-called ``the pair-instability mass gap'' \citep{Rodriguez19}, 
or may have finite eccentricities at a GW frequency of $\sim 10$ Hz \citep{Samsing18,Rodriguez18}. 
Such Monte-Carlo simulations, however, need careful calibrations of some parameters from comparison with the $N$-body results \citep{2013MNRAS.431.2184G,2016MNRAS.463.2109R}. 
On the other hand, direct $N$-body simulations with a realistic number of particles (i.e., $N\sim 10^6$) are still numerically too expensive with cost of $N^2$ and with very small binary orbital period compared with the GC lifetime. 
As a result, currently, there is only a single example 
of million-body runs performed using a GPU cluster \citep{2016MNRAS.458.1450W}, although the situation may change in near future by adoption of tree-direct hybrid methods \citep{2011PASJ...63..881O,2015ComAC...2....6I} and novel schemes for binary integration \citep{2020MNRAS.493.3398W}.

On the other hand, open clusters (OCs) have not been expected to produce such tight BBHs because they are less massive and less dense compared with GCs.
From the mass function of star clusters, however, there are hundred times more OCs than GCs.
Therefore, we thought it might be interesting to examine the formation rate of BBHs in OCs seriously.

\begin{figure}
  \begin{center}
    \includegraphics[scale=0.2]{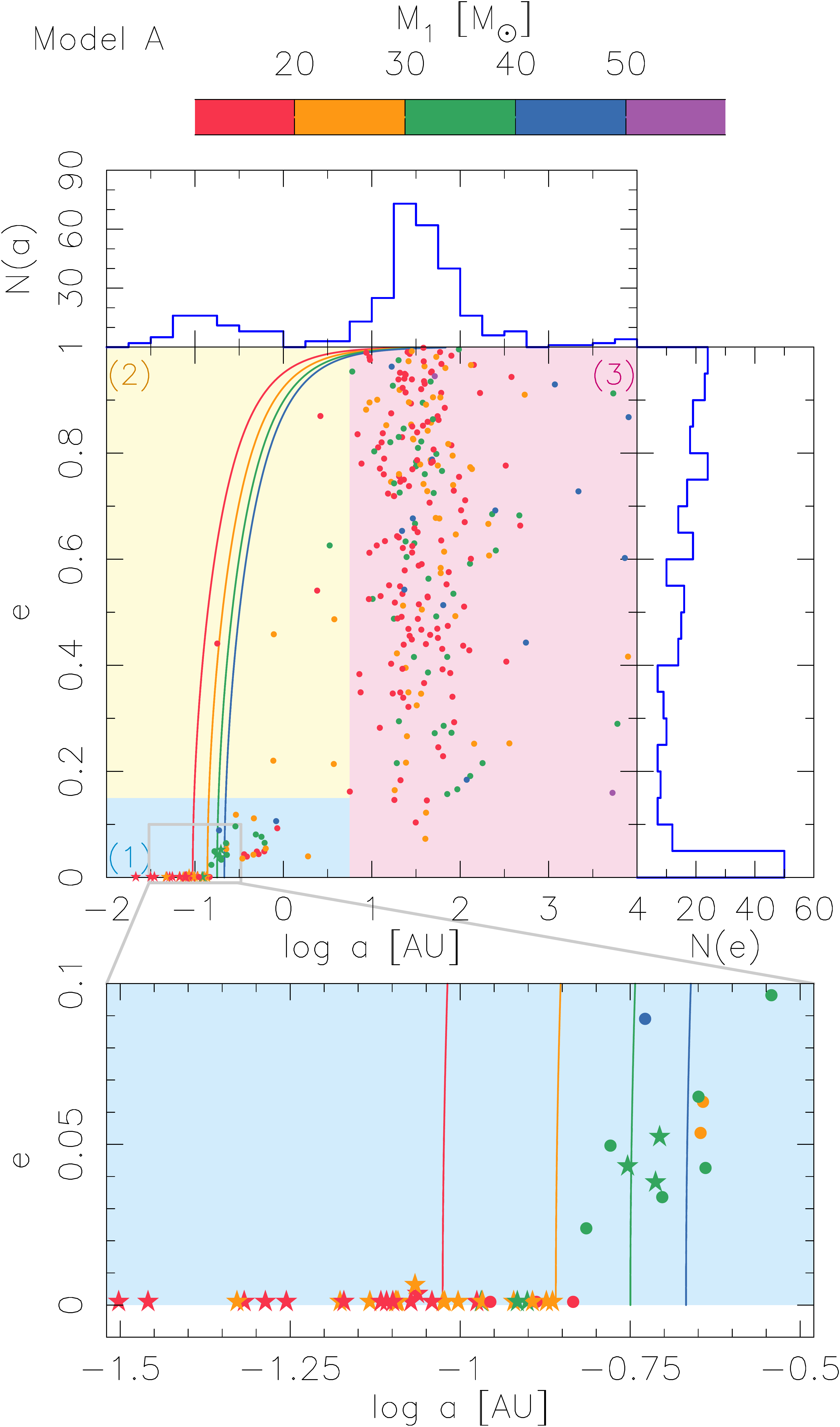}
\caption{
  \label{fig:BBHdist}
Semi-major axis ($a$) and eccentricity ($e$) distribution of BBHs obtained from our simulations (from Fig.~1 of \citep{Kumamoto19}). Star symbols show BBHs with a merger time less than 10\,Gyr. Colors of the symbols indicate the primary mass ($M_1$) of the BBHs. Colored background indicates classification depending on their formation mechanisms illustrated in \ref{fig:BinEvo}. Colored curves show the relation between $a$ and $e$ to give $t_{\rm GW}=10$\,Gyr for BBHs with masses of 15, 25, 35, and 45$M_{\odot}$ (red, yellow, green, and blue).
  }
\end{center}
\end{figure}

We performed a series of $N$-body simulations of OCs using a direct $N$-body code, {\tt NBODY6++GPU}, which includes both stellar and binary evolution models \cite{2015MNRAS.450.4070W}. As an initial condition, we adopted a Plummer model \cite{1911MNRAS..71..460P} with a total mass of $2.5\times10^3M_{\odot}$ with 0.1 solar metallicity. 
We calculated 360 realizations per a given set of the parameters and obtained $\sim 1$ BBH per cluster on average. 
The merger time of the BBHs due to gravitational wave emission is given as a function of the semi-major axis ($a$) and eccentricity ($e$) of BBHs \cite{1963PhRv..131..435P}. In Fig. \ref{fig:BBHdist}, 
the distributions of $a$ and $e$ of BBHs formed in the simulations are shown. 
We can see that most of BBHs distribute in red-colored region (Region 3), which correspond to those dynamically formed and the characteristic value of their semi-major axis is set by the condition that the potential depth of binaries is of the order of that of the cluster.
Owing to shallow potential depth of OCs, their separation is too large for the BBHs to merge within the Hubble time. 
On the other hand, we found some BBHs that can merge within the Hubble time in blue-colored region (Region 1), in which BBHs have very low eccentricity and small semi-major axis. Those are formed via binary stellar evolution such as stable and unstable mass transfer. Some of them experience a dynamical encounter after the binary stellar evolution and have large eccentricity (yellow-colored, Region 2). 
Three formation channels above are summarized in Fig. \ref{fig:BBHdist}.

In a GC, more BBHs are dynamically formed and, owing to the deeper potential well, they tend to have smaller value of semi-major axis and some of them merge within the Hubble time. 
On the other hand, in open clusters, the major channel for BBH mergers is different: a binary consisting of two massive main-sequence (MS) stars are first formed and becomes tight as a result of the binary stellar evolution.
If this separation is kept until BBHs are formed after the stellar lifetime, they can merge within the Hubble time. 

From our results, 
we can estimate a merger rate of BBHs formed in OCs at 10\,Gyr ago to be 0.3\,yr$^{-1}$\,Gpc$^{-3}$ \cite{Kumamoto19}. This value corresponds to 20--50\,\% of those formed in GCs~cite{Rodriguez16}. 
This means BBHs formed in OCs contribute to GW events much more than previously expected.

\begin{figure}
  \begin{center}
    \includegraphics[scale=0.8]{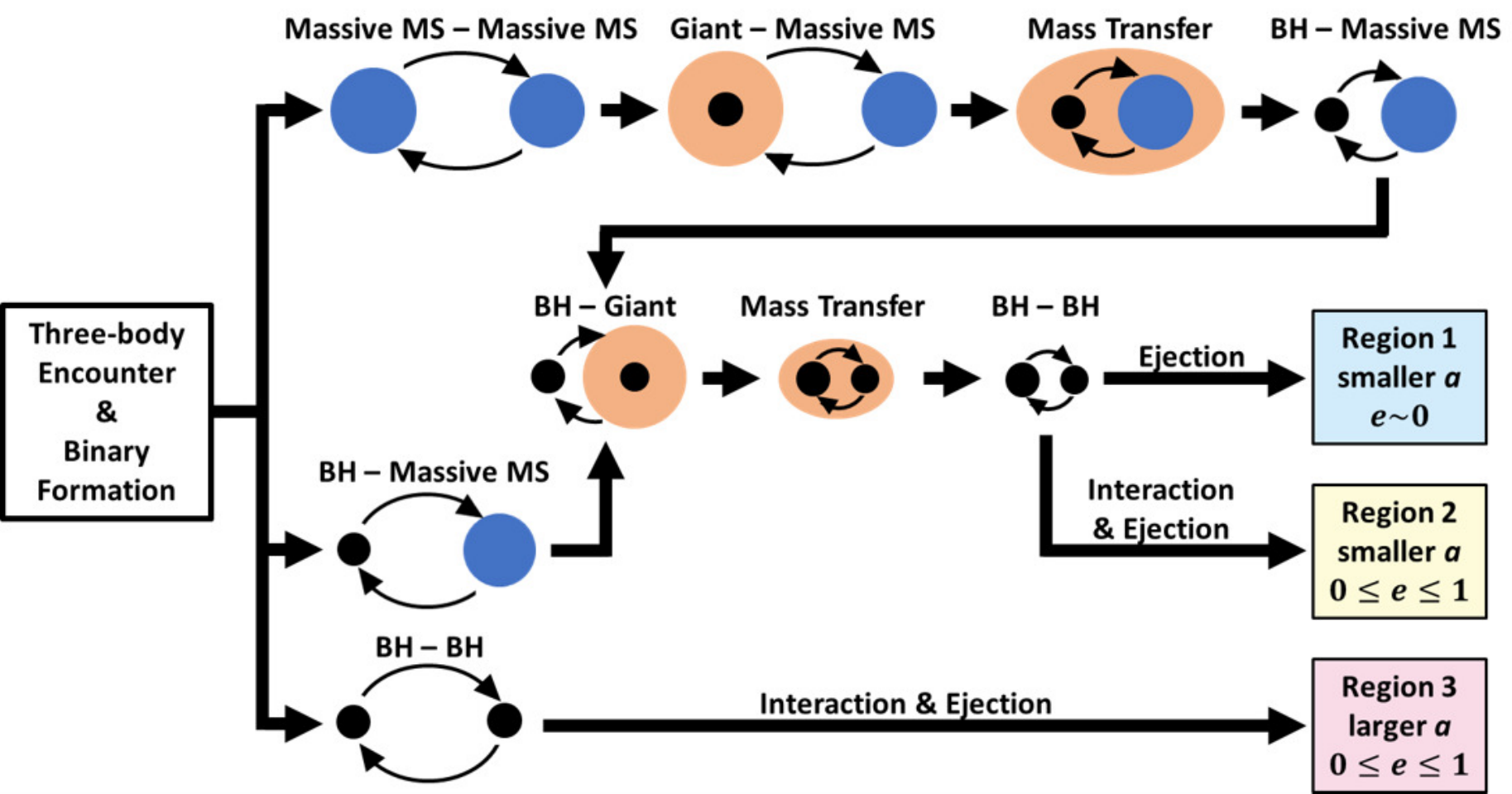}
\caption{
  \label{fig:BinEvo}
Schematic figure of the formation processes of three different types of BBHs. (From Fig. 3. of \citep{Kumamoto19})
  }
\end{center}
\end{figure}

We also repeat similar simulations but with different metallicities ranging from zero to one solar metallicity. 
Using the BBH formation rate and distribution of their orbital parameters,
and assuming cosmic metallicity evolution with some dispersion \cite{2019MNRAS.488.5300C} and a cosmic star formation rate density \citep{2017ApJ...840...39M}, we can estimate the merger rate of BBHs formed in OCs at the present universe by summing up the contributions from all OCs. The estimated event rate is $\sim 70~{\rm yr}^{-1} {\rm Gpc}^{-3}$ \cite{Kumamoto20}, comparable to that estimated from LIGO/Virgo O2 and O3, $64.0_{-33.0}^{+73.5}~{\rm yr}^{-1}{\rm Gpc}^{-3}$ \cite{2019ApJ...882L..24A}. 

Our simulations also give a prediction on BH-MS binary distribution observable with Gaia \citep{2016A&A...595A...1G}. In the next data release of Gaia (Gaia DR3), binaries will be included. Using the results of our simulations, we estimated that $\sim 10$ BH-MS binaries would be detectable with Gaia \cite{Shikauchi20}, even if we assume relatively stringent observational constraints \cite{2018ApJ...861...21Y}. In contrast to BH-MS binary formed in galactic fields, those formed in OCs would have smaller mass of the MS companion ($<5M_{\odot}$), longer orbital period ($> 1$\,yr), and higher eccentricity ($>0.1$). 

\paragraph B

Binary population synthesis model of extremely low-metallicity star

Binary population synthesis (BPS) is a powerful tool to predict merger rates of BBHs formed in galactic 
fields (e.g.,~\cite{Belczynski16}). Coupled with star cluster simulations, BPS can also predict the merger rate of BBHs formed in dense stellar clusters (e.g.,~\cite{Rodriguez16}).
So far, most studies have focused only on BBHs formed under Pop~I and II environments ($0.01 \lesssim Z/Z_\odot \lesssim 1$).
On the other hand, it has been suggested that Pop~III-originated BBHs have distinct features from Pop~I/II-originated BBHs
\citep{Kinugawa14}. 
Thus, it is important to bridge the metallicity gap (i.e. $0 < Z/Z_\odot \lesssim 0.01$) in order to elucidate origins of merging BBHs. 
Hereafter, we call stars in this metallicity gap the extreme metal-poor (EMP) stars. 
In order to bridge the metallicity gap, we have made a table for evolutionary tracks of stars with $-8 \le \log(Z/Z_\odot) \le -2$. 
Note that there has been no evolution track computation for EMP stars before. We have compiled those evolutionary tracks as a set of fitting formulae, and incorporated them into {\tt BSE} and {\tt NBODY6} (\cite{Hurley02,Aarseth03}, respectively). 
The {\tt BSE} code is based on several BPS calculations (e.g.,~\cite{Belczynski16,Kinugawa14}), and the {\tt NBODY6} code and its derivatives are widely used for obtaining BBH merger rates in dense star clusters (e.g.,~\cite{Banerjee10,Tanikawa13,Bae14,Fujii17,Park17,Hong18,Kumamoto19}). 
Here, we describe these fitting formulae briefly (see \cite{Tanikawa19}, in detail).

We first make stellar evolution models as a reference for fitting
formulae by means of 1 dimensional (1D) simulation. For the 1D
simulation, we use the {\tt HOSHI} code
\citep{Takahashi16,Takahashi18,Takahashi19,Yoshida19}. We follow the
time evolution of stars with $8 \le M/M_\odot \le 160$ for
$-8 \le \log(Z/Z_\odot) \le -2$. We start the calculation from the
zero-age main-sequence (ZAMS) and stop the calculation at the
carbon-ignition time. We do not include stellar wind mass loss. We
take into account the post carbon-ignition evolution (e.g. supernova)
and stellar wind mass loss as post processing. This is the same as in the {\tt BSE} and {\tt NBODY6} codes incorporating the evolution of Pop~I/II stars.
The evolution of EMP stars differs from that of Pop~I/II in the  following respects: neither Hertzsprung-gap (HG) nor blue-loop phase exists and the envelope of massive ($M \sim 20M_\odot$) stars remains radiative until the end of their lives. This has significant impact on binary evolution, such as the mass accretion rate in the Roche-lobe overflow phase, and occurrence of post-MS mergers and common envelope evolution.

\begin{figure}
  \begin{center}
    \includegraphics[scale=1.0]{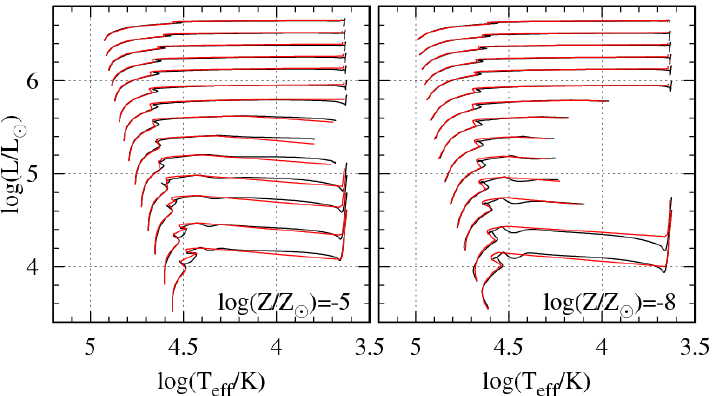}
    \caption{HR diagram for comparison between stellar evolution
      models (black) and fitting formulae (red). The stellar evolution
      in this figure ends at the carbon-ignition time, and stellar
      wind mass loss is not taken into
      account. \label{fig:tanikawa}}
  \end{center}
\end{figure}

We use the above stellar evolution models as reference, and make
fitting formulae of EMP stars.  Figure~\ref{fig:tanikawa} shows
stellar evolution on HR diagram in the cases of $\log(Z/Z_\odot)=-5$ and $-8$ for comparison with 
the results obtained from 1D calculations and the fitting formulae. 
Note that the stellar evolution in this figure is terminated at the carbon-ignition time, and stellar wind mass loss is not taken into account.  The tracks by fitting formulae show
good agreement with those by 1D calculations, with the maximum deviation less than $20$\%. This error is satisfactory for BPS and star cluster
simulations coupled with BPS, considering larger
uncertainties coming from 
uncertainties in tidal interaction, common envelope evolution,
stellar wind mass loss, natal kicks of neutron star and black holes, supernova mass loss including pair instability, pulsational
pair instability supernovae, and so on.

\subsubsection{\blue{Outlook}}

\blue{To distinguish the possible formation channels described above, we need further statistical constraints with increasing the number of BBHs identified by GW detections. For instance, binary population synthesis models predict that the Pop III progenitor stars provide characteristic features in resulting chirp-mass distributions. A most prominent feature is a peak around $30M_\odot$ \citep{Kinugawa14}, which may be still consistent with the latest LIGO/Virgo O3a results. Future GW observations will impose more constraints on other properties of BBHs, such as mass ratio, separation, spin, and eccentricity, etc. Maturing both in-situ and ex-situ formation channels is indispensable for understanding the origins of BBHs, making full use of the observational constraints.}

%% file: ptepGW_sectionB01.tex

\subsubsection{BNS waveform modeling based on numerical-relativity simulations}

GWs from binary neutron stars (BNSs) contain rich information of the NSs. In particular, the tidal deformability of NSs has been proposed as one of the most promising quantities related to the equation of state that can be extracted from the GW observation~\cite{Lai:1993pa,Mora:2003wt,Flanagan:2007ix,Read:2009yp,Damour:2009wj,Hinderer:2009ca,Vines:2011ud,Damour:2012yf,Bini:2012gu,Favata:2013rwa,Yagi:2013baa,Read:2013zra,Bini:2014zxa,Bernuzzi:2014owa,Wade:2014vqa,Lackey:2014fwa}. 
\blue{The tidal deformability $\lambda$ is defined through $Q_{ij} = -\lambda E_{ij}$, where $Q_{ij}$ and $E_{ij}$ are the induced quadrupole moment and the external tidal field, respectively.
The tidal deformability can also be written in terms of the radius of the neutron star, $R$, and its Love number, $k_2$, as $\lambda = 2 k_2 R^5$.
Therefore, the behavior of $\lambda$ depends on the inner structure of a NS, its EOS. For example, for the soft (stiff) EOSs, the radius of a NS is small (large), therefore, the NS is less (largely) deformed, i.e. $\lambda$ takes small (large) value.}  
The simultaneous measurement of the mass and the tidal deformability of the NSs provides a substantial constraint on the equation of state of nuclear matter which is yet poorly understood~\cite{Lattimer:2012nd,Lattimer:2015nhk}.
 
To extract the tidal deformability of NSs from the observed signals, an accurate theoretical waveform template is crucial. For this purpose, many efforts have been made to model the gravitational waveforms of a NS merger in the past few decades. The waveforms including the linear-order tidal effects are derived by post-Newtonian (PN) calculation for the early inspiral stage~\cite{Flanagan:2007ix,Vines:2011ud}. Furthermore, the waveforms employing the {\it effective-one-body} (EOB) formalism are derived to incorporate higher-order PN effects~\cite{Damour:2009wj,Bini:2012gu,Damour:2012yf,Bini:2014zxa,Bernuzzi:2014owa,Hinderer:2016eia,Steinhoff:2016rfi,Dietrich:2017feu,Nagar:2018zoe,Messina:2019uby}. However, these waveform model could not be accurate enough for the estimation of the tidal deformability due to a significant systematic error of the unknown higher-order PN terms~\cite{Favata:2013rwa,Yagi:2013baa,Wade:2014vqa,Lackey:2014fwa}. Furthermore, the tidal correction to the waveforms is derived based on the PN calculation which is only valid in the regime where the orbital separation is sufficiently large. Since the tidal effect is most significant just before the merger, it is necessary to examine the accuracy of these waveform models.
 
Numerical-relativity (NR) simulation is the unique method to predict the tidal effects in a regime where the non-linear effect of hydrodynamics should be taken into account in the framework of general relativity~\cite{Thierfelder:2011yi,Baiotti:2011am,Bernuzzi:2012ci,Radice:2013hxh,Hotokezaka:2015xka,Haas:2016cop,Hotokezaka:2016bzh,Dietrich:2017feu,Dietrich:2017aum,Kiuchi:2017pte,Dietrich:2019kaq}. The recent progress in numerical techniques and computational resources enabled us to perform the high-precision NR simulations for BNS mergers and to generate the waveforms with systematic errors sufficiently small for the GW data analysis. 

In our recent work~\cite{Kiuchi:2017pte,Kawaguchi:2018gvj,Kiuchi:2019kzt}, we perform NR simulations for NS-NS mergers systematically varying the mass and the tidal deformability of NSs: 6 and 4 models that cover the symmetric mass ratio, $\eta$, in the range of $0.247$--$0.25$, the binary with the total mass of $\approx2.7$ and $2.5\,M_\odot$, and 5 different EOSs which cover the binary tidal deformability in the range of $\approx300$--$1900$. Here, the binary tidal deformability, ${\tilde \Lambda}$, is defined by
\begin{align}
{\tilde \Lambda}&=\frac{8}{13}\left[\left(1+7\eta-31\eta^2\right)\left(\Lambda_1+\Lambda_2\right)\right.\nonumber
\\&-\left.\sqrt{1-4\eta}\left(1+9\eta-11\eta^2\right)\left(\Lambda_1-\Lambda_2\right)\right],
\label{eq:symlambda}
\end{align}
where \blue{$\Lambda_i=\lambda_i/m_i^5\,~(i=1,2)$} are the tidal deformability of each individual star (Note that $m_1 < m_2$). By this work the waveforms from NS-NS mergers for more than $15$ inspiral orbits are obtained for a wide range of binary parameters.

We carefully estimate the error budgets of the waveforms obtained by the NR simulations. In our NR simulation code, the field equations are solved under discretization. We found that the error due to the finite grid resolution is the largest error budget among the error sources of the numerical simulation. To estimate the error due to the finite grid resolution, we perform the simulations with various gird resolutions and checked the convergence property of the obtained waveform data. As the results, we find that the phase error of the waveforms is a sub-radian order~\cite{Kiuchi:2017pte,Kawaguchi:2018gvj,Kiuchi:2019kzt}.

\begin{figure}
\centering
 \includegraphics[width=0.9\linewidth]{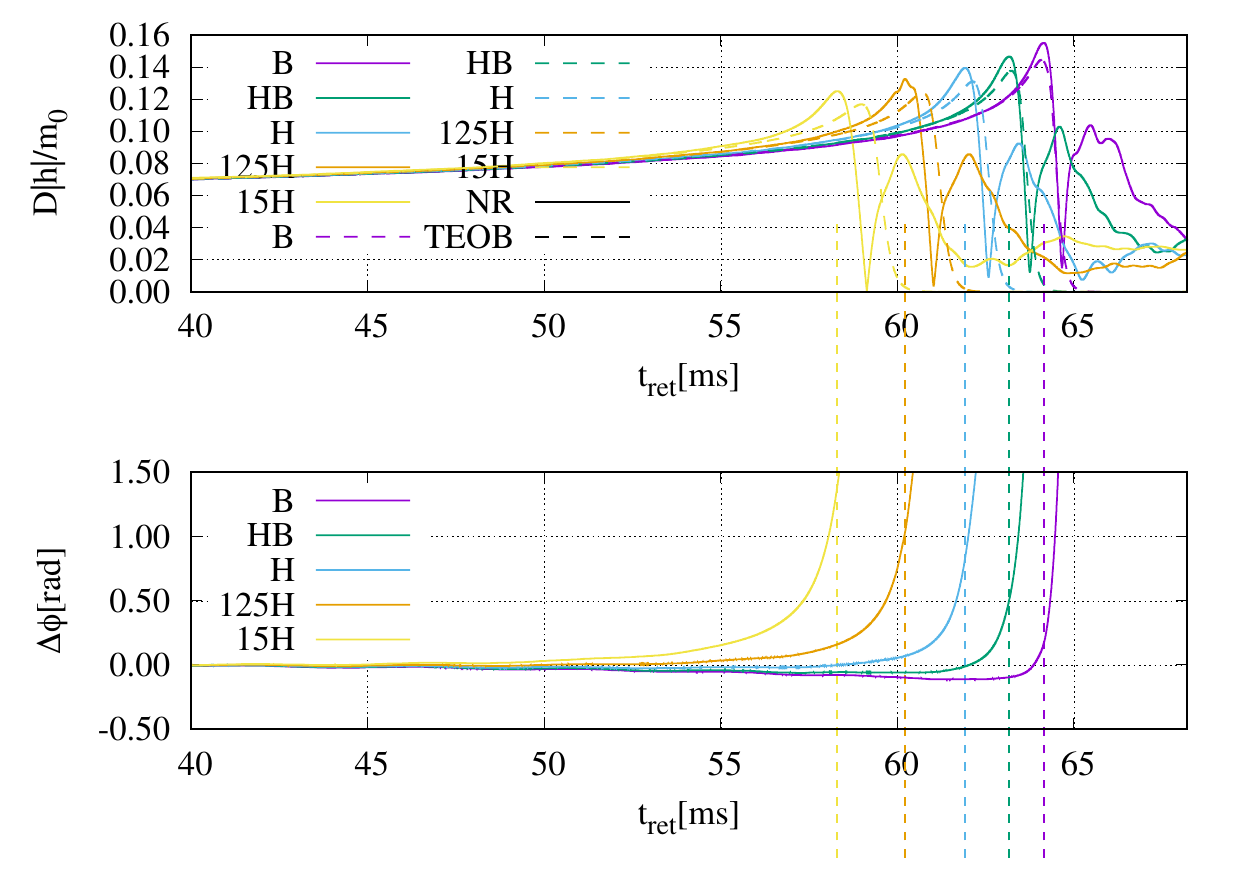} 
 \caption{Comparison between NR and SEOBNRv2T waveforms for the models with the total mass of $\approx 2.5\,M_\odot$ and the symmetric mass ratio of $0.247$. The top and bottom panels show the amplitude of NR and SEOBNRv2T waveforms and the phase difference between them, respectively. Here, $D$ and $m_0$ denote the distance to the source and the total mass of the binary. The vertical dashed lines are the peak amplitude time when the GW amplitude reaches a peak in each model.}
\label{fig:eobcomp}
\end{figure}

Employing the obtained high-precision NR waveforms, we examined the validity of the analytical waveform models for NS binary. We show that the latest tidal-EOB (SEOBNRv2T~\cite{Hinderer:2016eia,Steinhoff:2016rfi}) waveforms can be accurate even up to $\approx3\,{\rm ms}$ before the onset of merger~\cite{Kiuchi:2017pte} (see  Fig.~\ref{fig:eobcomp} for an example). However, the phase difference between the tidal-EOB waveforms and the NR results is still larger than $\approx 1\,{\rm rad}$ after two NSs come into contact for the case that the NS radii are larger than $\approx13\,{\rm km}$. This finding indicated that further improvement of the waveform model would be needed to suppress the systematic error in the measurement of the tidal deformability.

For this purpose, we develop a \blue{phenomenological} model for GWs from inspiraling NSs based on our high-precision NR simulations. \blue{Our phenomenological} waveform model is derived in the frequency domain as in the Phenom-series for binary black holes~\cite{Khan:2015jqa} for convenience in data analysis. We decompose a frequency-domain gravitational waveform, ${\tilde h}\left(f\right)$, into the frequency-domain amplitude, $A\left(f\right)$, and phase, $\Psi\left(f\right)$, (with an ambiguity in the origin of the phase) by 
\begin{align}
	{\tilde h}\left(f\right)=A\left(f\right) {\rm e}^{-i\Psi\left(f\right)},
\end{align}
and we define the corrections of the NS tidal deformation to GW amplitude and phase by
\begin{align}
	A_{\rm tidal}\left(f\right)=A\left(f\right)-A_{\rm BBH}\left(f\right)\label{eq:defAtidal}
\end{align}
and 
\begin{align}
	\Psi_{\rm tidal}\left(f\right)=\Psi\left(f\right)-\Psi_{\rm BBH}\left(f\right),\label{eq:defphitidal}
\end{align}
respectively. Here, $A_{\rm BBH}\left(f\right)$ and $\Psi_{\rm BBH}\left(f\right)$ are the GW amplitude and phase of a binary black hole with the same mass as the BNS, respectively (hereafter referred to as the point-particle parts). In this work, we employ the SEOBNRv2 waveforms~\cite{Taracchini:2013rva} as the fiducial point-particle part of GWs.

For the tidal part phase and amplitude of \blue{our phenomenological waveform model}, we employ the following function forms motivated by the 2.5 PN order post-Newtonian formula Refs.~\cite{Damour:2012yf} and~\cite{Hotokezaka:2016bzh} for the tidal-part amplitude and phase correction:
\begin{align}
&\Psi^{\rm tidal}=\frac{3}{128\eta}\left[-\frac{39}{2}{\tilde \Lambda}\left(1+a\,{\tilde \Lambda}^{2/3} x^p \right)\right]x^{5/2}\nonumber
\\&\times\left(1+\frac{3115}{1248}x-\pi x^{3/2}+\frac{28024205}{3302208}x^2 -\frac{4283}{1092}\pi x^{5/2}\right)\label{eq:phimodel}
\end{align}
for the phase correction and
\begin{align}
	A^{\rm tidal}&=\sqrt{\frac{5\pi\eta}{24}}\frac{m_0^2}{D_{\rm eff}} {\tilde \Lambda} x^{-7/4}\nonumber\\
	&\times \left(-\frac{27}{16}x^{5}-\frac{449}{64}x^{6}-b\,x^q\right)\label{eq:Amodel}
\end{align}
for the amplitude correction. 
Here, $a$, $p$, $b$, and $q$ are the free parameters of the models.

Since our NR waveforms only contain the waveforms of which frequency is larger than $\approx 400\,{\rm Hz}$, they are too short to calibrate the \blue{phenomenological} waveform model only by themselves. Thus, \blue{for this purpose}, \blue{we construct hybrid waveforms employing our NR waveforms  and the effective-one-body waveforms of Refs.~\cite{Hinderer:2016eia,Steinhoff:2016rfi} (SEOBNRv2T) as the high and low-frequency parts, respectively, and use them to calibrate the phenomenological waveform model.} The hybridization of the waveforms is performed in the time-domain by the procedure described in Ref.~\cite{Hotokezaka:2016bzh,Kawaguchi:2018gvj}.

\begin{figure}
\centering
 \includegraphics[width=.45\linewidth]{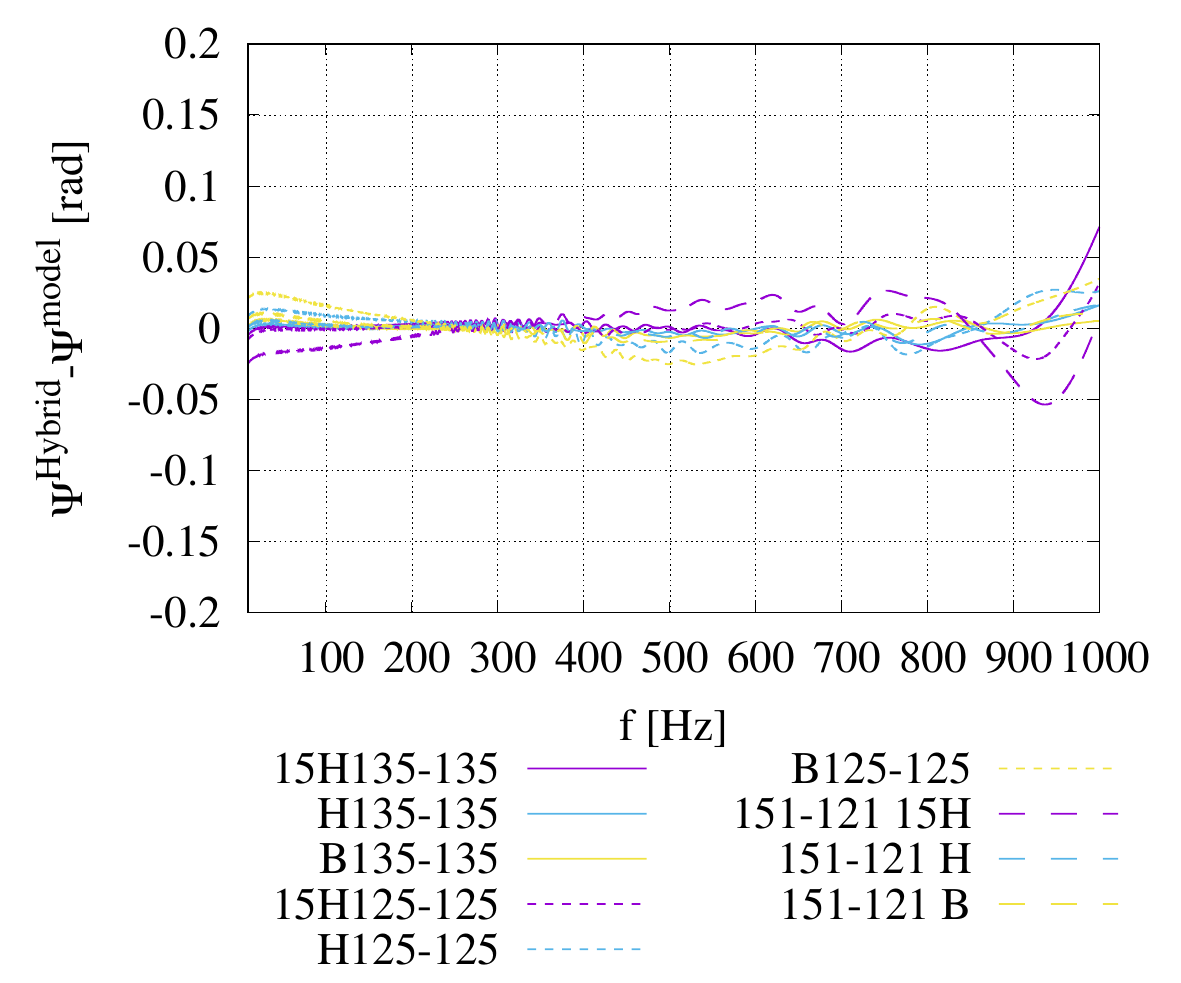} 
 \includegraphics[width=.45\linewidth]{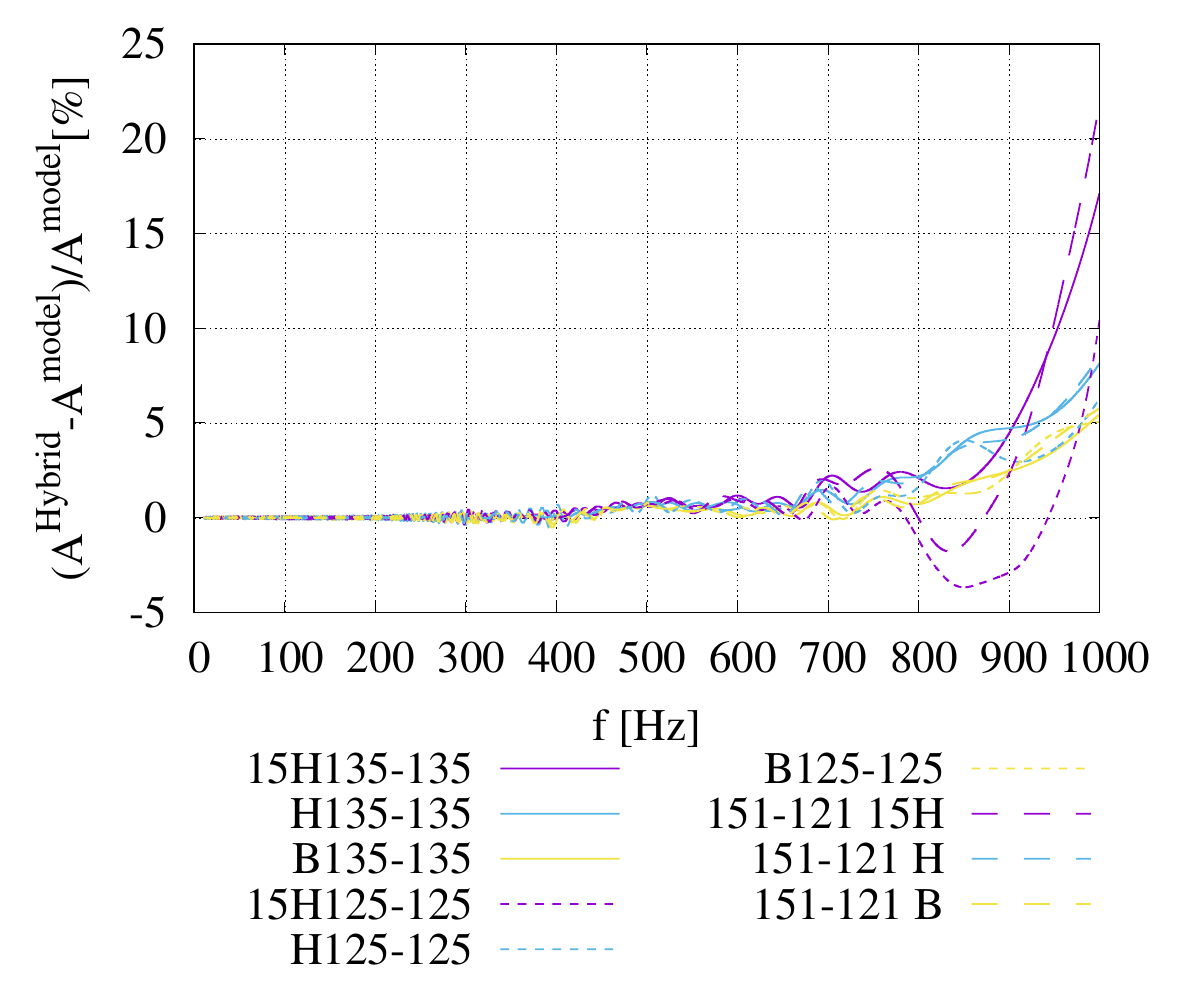} 
 \caption{(Left) Difference in the tidal-part phase between the hybrid waveforms and our analytical model for the various binary \blue{configurations} (see~\cite{Kiuchi:2017pte} and~\cite{Kawaguchi:2018gvj} for the details of the models). Phase differences are plotted after the alignment in the frequency range of 10–1000 Hz. (Right) Relative difference of amplitude between the hybrid waveforms and our analytical model.}
\label{fig:modelcomp}
\end{figure}

To focus on the inspiral-phase waveform and to avoid the contamination from the post-merger waveforms, which would have large uncertainties, we restrict the GW frequency range in $10$--$1000\,{\rm Hz}$. The fitting parameters are determined by employing the hybrid waveforms with the largest value of binary tidal deformability in the models studied in this work. By performing the least square fit with respect to the phase difference and relative difference of the amplitude, we obtained $a=12.55$, $p=4.240$, $b=4251$, and $q=7.890$. Employing these parameters, we find that the phase and amplitude \blue{corrections} to the frequency domain waveforms are reproduced within $0.1\,{\rm rad}$ and $25\%$ for all the binary \blue{configurations} studied in this work (see Fig.~\ref{fig:modelcomp}).

To validate our \blue{phenomenological} waveform model more quantitatively from the view point of the data analysis, we calculate the mismatch between the our waveform models and the hybrid waveforms, ${\bar F}$, defined by
\begin{align}
	{\bar F}=1-\max_{\phi_0,t_0}\frac{\left({\tilde h}_1\middle|{\tilde h}_2{\rm e}^{2\pi i f  t_0 +i \phi_0}\right)}{||{\tilde h}_1||\,||{\tilde h}_2||},\label{eq:mismatch}
\end{align}
where $(\cdot|\cdot)$ and $||\cdot||$ are defined by
\begin{align}
	\left({\tilde h}_1\middle|{\tilde h}_2\right)=4{\rm Re}\left[\int_{f_{\rm min}}^{f_{\rm max}}
	\frac{{\tilde h}_1\left(f\right){\tilde h}^*_2\left(f\right)}{S_{\rm n}\left(f\right)}df\right],\label{eq:inp}
\end{align}
and
\begin{align}
	\rho=||{\tilde h}||=\sqrt{\left({\tilde h}\middle|{\tilde h}\right)}.
\end{align} 
Here $S_{\rm n}$ denotes the one-sided noise spectrum density of the detector, and we employ the noise spectrum density of the {\tt ZERO\_DETUNED\_HIGH\_POWER} configuration of advanced LIGO~\cite{aLIGOnoise} for it. We found that the mismatch between our \blue{phenomenological} waveform model and the hybrid waveforms is always within $10^{-5}$~\cite{Kawaguchi:2018gvj,Kiuchi:2019kzt}.

\begin{figure}
\centering
 \includegraphics[width=.5\linewidth]{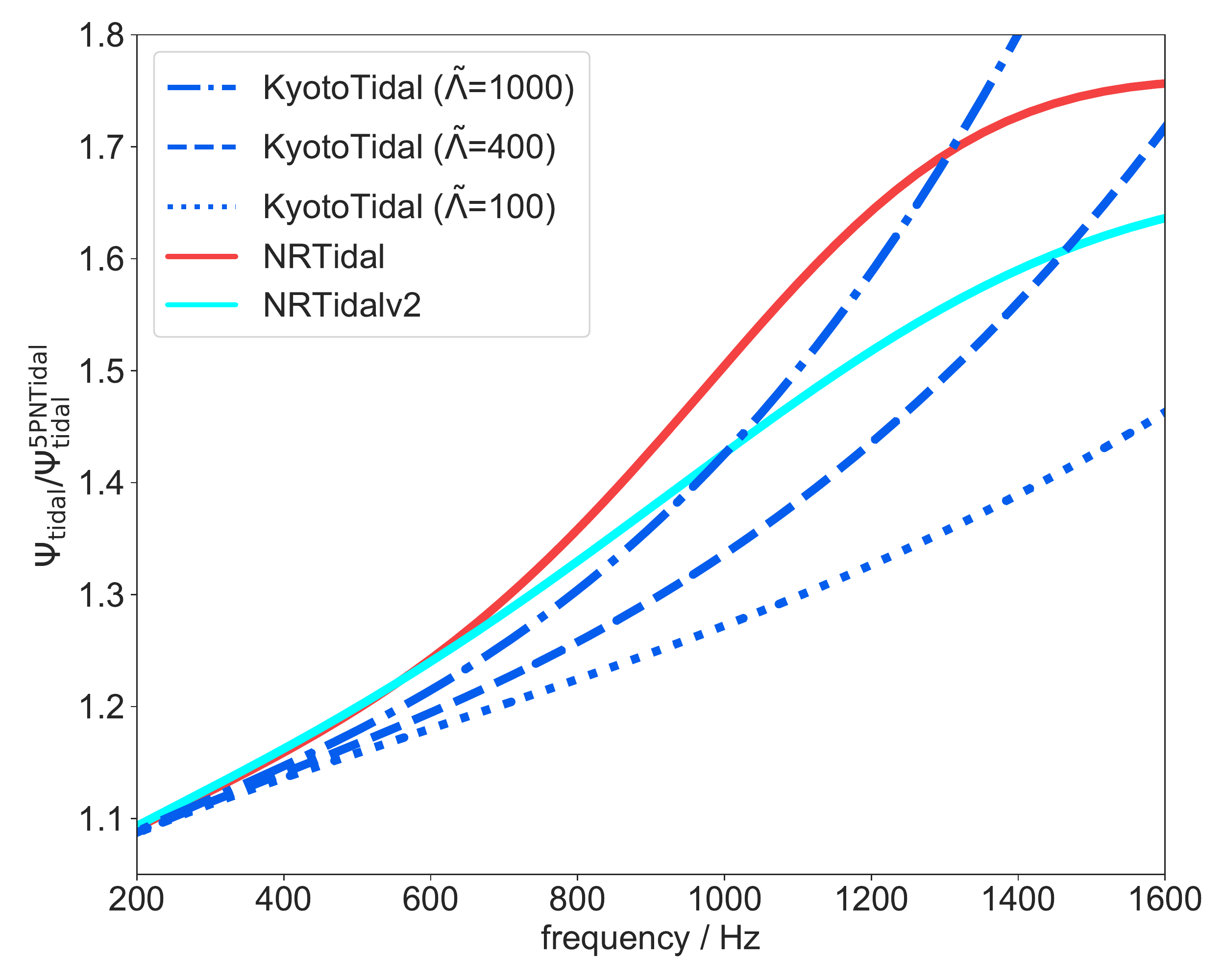} 
 \caption{Tidal phase in the frequency domain normalized by the leading, Newtonian (relative 5PN-order) tidal phase formula.  Our analytical model (referred to as Kyoto tidal) assuming ${\tilde \Lambda}=$ 1000 (dot-dashed, blue), 400 (dashed, blue), and 100 (dotted, blue), the NRTidal model (solid, red ~\cite{Dietrich:2017aum}) \blue{and} the NRTidalv2 model (solid, cyan ~\cite{Dietrich:2019kaq}) are shown in the plot. Note that latter two models are independent of ${\tilde \Lambda}$ when normalized by the leading tidal phase.}
\label{fig:modelcomp2}
\end{figure}

Gravitational waveform models for BNSs based on NR waveforms are also derived in~\cite{Dietrich:2017aum} and~\cite{Dietrich:2019kaq} in a similar manner. The main difference between our and their work is the difference of the NR waveforms and the tidal-EOB waveforms used for the model calibration. Figure~\ref{fig:modelcomp2} compares our tidal phase correction model and those of~\cite{Dietrich:2017aum} and~\cite{Dietrich:2019kaq} normalized by the leading-order PN term (see also ~\cite{Narikawa:2019xng} and~\cite{Dietrich:2019kaq}). Figure~\ref{fig:modelcomp2} shows that those models agree with each other within $10$--$20\%$ for 10--1000 Hz. Indeed, those models give consistent results for the current detector sensitivity as shown in the next subsection (see also~\cite{Dietrich:2019kaq}).

\subsubsection{Measuring tidal deformability from GW}

As mentioned above, NSs provide unique laboratories to study the properties of ultra-dense matter.
To constrain the equation of state of NSs effectively, we need accurate waveform models for the data analysis.
We assume that the waveform of the GWs can be decomposed into the point-particle, spin and tidal parts.
The point-particle and the spin parts are the same as those of the binary black hole case, on the other hand, tidal part is  that of the BNS case because of the matter effects of NSs.
The standard waveform, \blue{so-called} \texttt{TaylorF2} (hereafter \texttt{TF2}), is derived by the post-Newtonian (PN) approximation \blue{\cite{Sathyaprakash:1991mt}}.
The point-particle \cite{Buonanno:2009zt,Blanchet:2013haa}, 
and spin parts \cite{Bohe:2013cla, Arun:2008kb, Mikoczi:2005dn} are calculated up to the 3.5PN-order, 
while the effect of the tidal part (let us call it \texttt{PNTidal}) enters from the 5PN-order, calculated up to the 7.5 PN order \cite{Damour:2012yf}.
Since the tidal effects become important at the late inspiral stage where the PN approximation becomes invalid, we need better waveform models to estimate tidal deformability accurately.
%
%

Dietrich {\it et al.} \cite{Dietrich:2017aum, Dietrich:2019kaq} (\texttt{NRTidal}) and Kawaguchi {\it et al.} \cite{Kawaguchi:2018gvj} (\texttt{KyotoTidal}) independently construct new waveform models in which the tidal part is improved by the calibration using NR waveforms.
In \texttt{KyotoTidal}, nonlinear effects of $\tilde{\Lambda}$ is taken into account.
The lack of the point-particle and spin parts beyond the 4PN-order also bias the measurement of binary quantities.
Kawaguchi {\it et al.} have also constructed higher PN-order  
correction terms of the point-particle part
by fitting the \texttt{SEOBNRv2} waveform model \cite{Purrer:2015tud,Taracchini:2013rva}, \blue{so-called} \texttt{TF2+}.
Here, \texttt{TF2+\_KyotoTidal} is calibrated 
in the frequency range of 10-1000 Hz, 
because they focus on the inspiral phase to avoid uncertainties in the post-merger phase.
%
Several studies have also derived constraints on the NS EOS via measuring tidal deformability from GW170817 \cite{Margalit:2017dij, Bauswein:2017vtn, Ruiz:2017due, Annala:2017llu, Zhou:2017pha, Fattoyev:2017jql, Paschalidis:2017qmb, Nandi:2017rhy, Most:2018hfd, Raithel:2018ncd, Landry:2018prl}.

\begin{figure}
\centering
 \includegraphics[width=.5\linewidth]{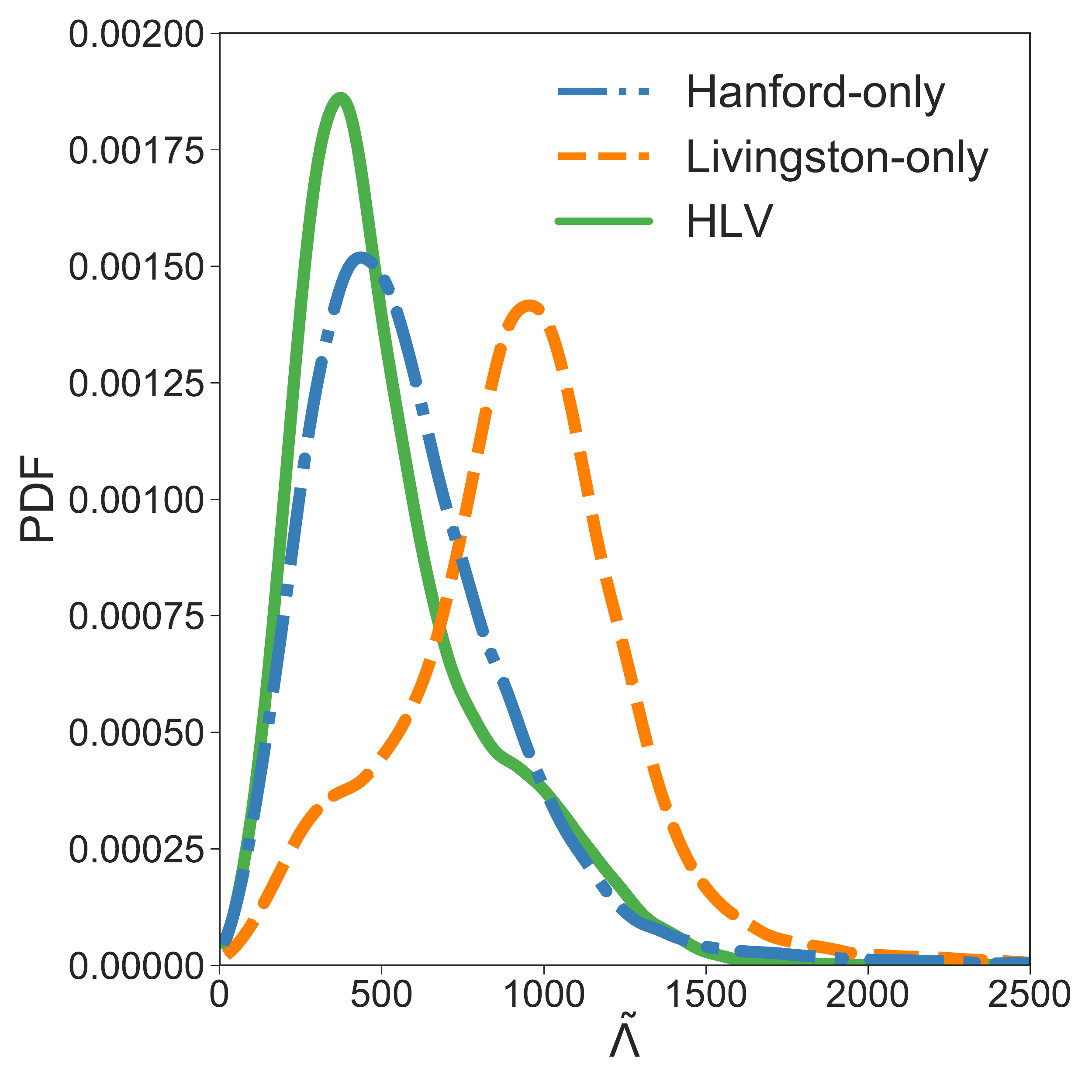} 
 \caption{Marginalized posterior probability distribution of binary tidal deformability,
 $\tilde{\Lambda}$ for GW170817, derived by data of different detectors with
 $f_{\rm max}$ = 2048 Hz. (Reprinted with permission from \cite{Narikawa:2018yzt}. \copyright  (2017) by the American Physical Society.)
}
\label{fig:twin}
\end{figure}

\begin{figure}
\centering
\includegraphics[width=.5\linewidth]{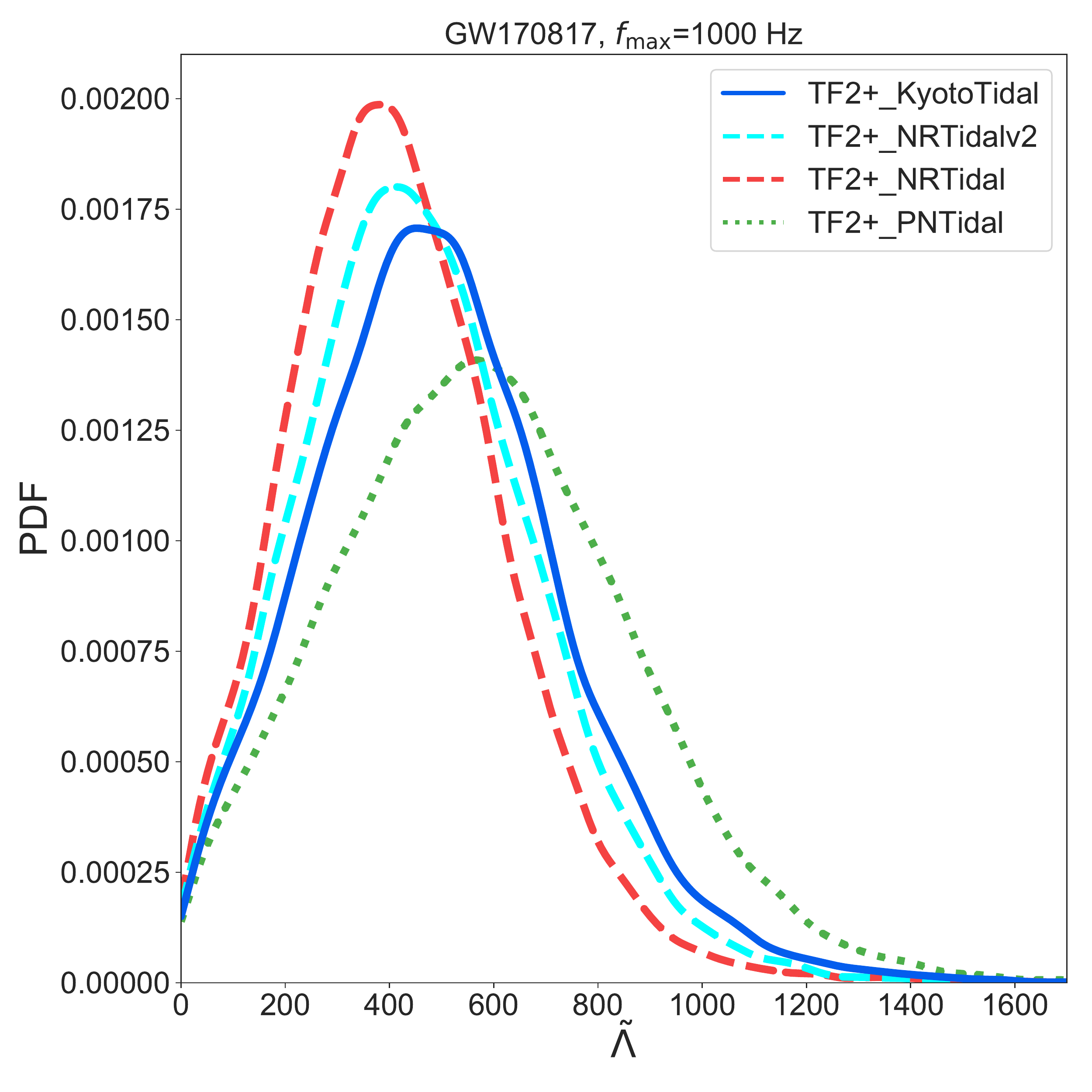} \caption{Marginalized posterior PDFs of binary tidal deformability, $\tilde{\Lambda}$ for GW170817, estimated by different waveform models for $f_{\rm max}=1000~{\rm Hz}$.
}
\label{fig:compare_Lambdatilde}
\end{figure}

Narikawa {\it et al.} separately analyze the GW data of a binary-neutron-star merger, GW170817, 
by each of the Advanced LIGO twin detectors \cite{Narikawa:2018yzt}.
They found that the posterior probability distributions of the binary tidal deformability for 
the Hanford and Livingston detectors \blue{are distinctively different}
 as shown in Fig.~\ref{fig:twin}.
\blue{They have suggested that the difference between the detectors cannot be understood by statistical error only, since only the posterior with the Livingston detector does not change smoothly as the upper cutoff frequency increases.
Their findings suggest that further research of the noise properties
in the high-frequency region of the Livingston data
is needed to extract information on the tidal deformability of GW170817.}
Narikawa {\it et al.} also found that there is a difference in the estimates of $\tilde{\Lambda}$ 
for GW170817 between NR calibrated waveform models (the \texttt{KyotoTidal} and \texttt{NRTidalv2} models) \blue{as shown in Fig.~\ref{fig:compare_Lambdatilde}}
\cite{Narikawa:2019xng}.
Here, \texttt{NRTidalv2} model is an upgrade of the \texttt{NRTidal} model \cite{Dietrich:2019kaq}.
\blue{The order of peak values of $\tilde{\Lambda}$ for the different waveform models in Fig.~\ref{fig:compare_Lambdatilde} can be understood by the difference in the magnitude of the tidal phase 
at around $\tilde \Lambda\approx 600-900$ shown in Fig.~\ref{fig:modelcomp2}. However, the difference is smaller than the
statistical error.}

The second BNS merger event GW190425 was observed during O3a \cite{Abbott:2020uma}.
The total mass of the system was about 3.4$M_{\odot}$, which is larger than the total mass of any other known BNS system.
Since the LIGO Hanford detector was offline at the time of the event, parameter estimation was performed by using the data from the LIGO Livingston and Virgo detectors.
The EOS with large tidal deformability $\tilde{\Lambda} > 1100$ was disfavored, which is consistent with the several analyses done for GW170817.
For the low-spin prior (component spin is enforced as $\chi_{1,2}\sim U[-0.05,~0.05]$), the 90\% upper limit of GW190425 becomes much smaller than that of GW170817, $\tilde{\Lambda} \le 600$.
Since the total mass of the system was large, the possibility of BH-NS system is discussed in \cite{Abbott:2020uma}.

%% file: ptepGW_sectionB02.tex
\subsubsection{High-energy Observations}

The observations of GW170817 by high-energy instruments, {\it Fermi}, {\it MAXI},  {\it Swift}, {\it CALET} and 
{\it Chandra}, are reviewed in \blue{this subsection}. 

\begin{figure}
  \begin{center}
    \includegraphics[width=90mm]{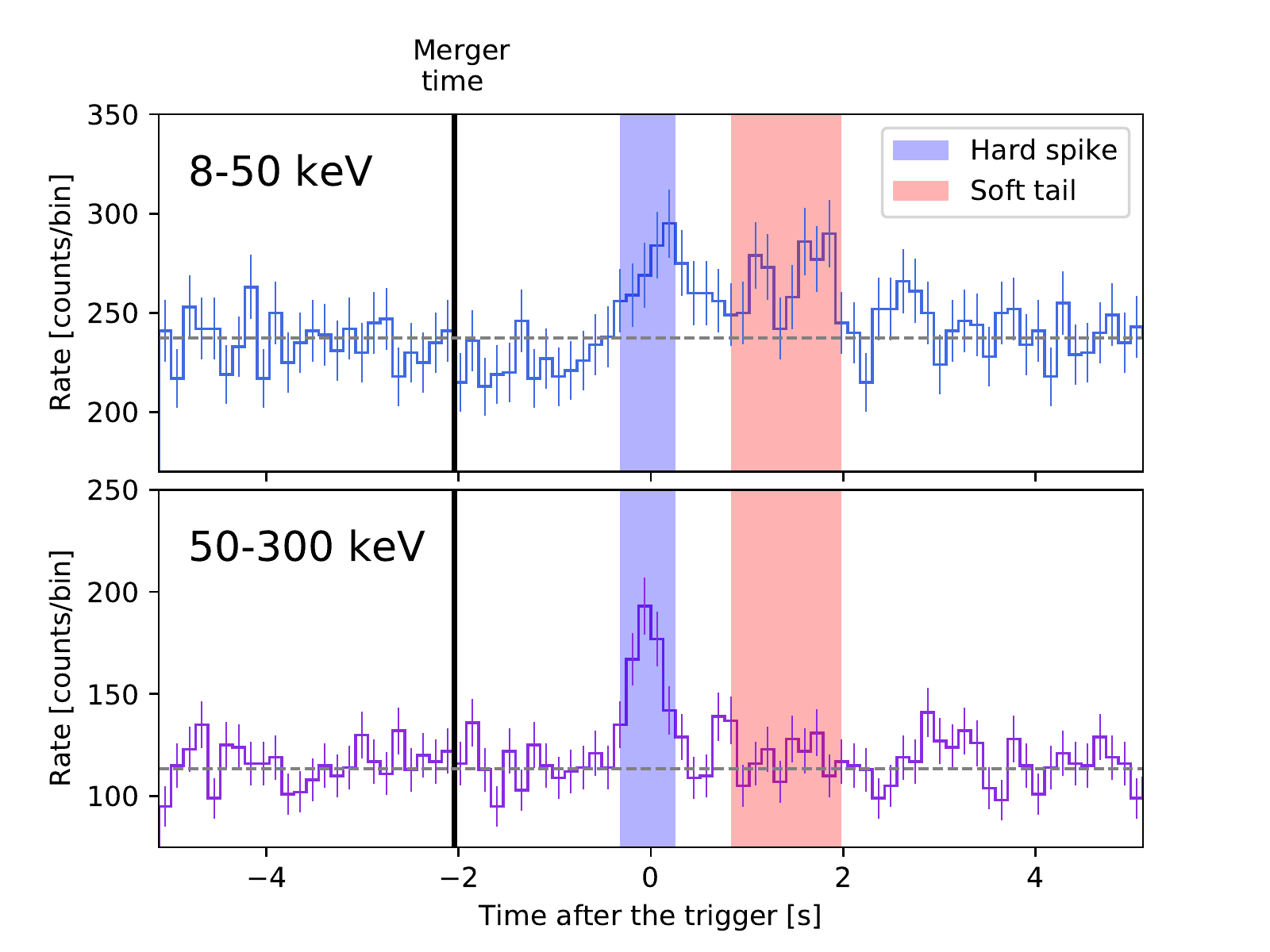}
\caption{
  \label{fig:LC_GRB170817A}
Light curve of GRB 170817A detected by GBM in the 8--50keV and 50--300keV bands. Emission consists of a hard pulse and a weak soft tail. The vertical solid line represents the time of the binary neutron star merger. The horizontal dashed line  represents the averaged background level.
  }
\end{center}
\end{figure}

\paragraph{Fermi}
On  August 17, 2017, at 12:41:06.5 (UTC), hard X-ray emission from GRB 170817A triggered {\it Fermi} Gamma-ray Burst Monitor  (GBM with a coverage of 8keV--40MeV; \cite{2009ApJ...702..791M}) near the detection limit  \citep{Goldstein:2017mmi}. The emission  with a short hard pulse and a soft tail lasted for $\sim$2 s, as shown in Fig.~\ref{fig:LC_GRB170817A}. From the obtained time duration, this GRB is classified as a short GRB. 
Also the SPectrometer on board {\it INTEGRAL} Anti-Coincidence Shield (SPI-ACS) detected a hard pulse completely coincident with the {\it Fermi}/GBM one \cite{Savchenko:2017ffs}.

Amazingly, the binary coalescence detected by the Laser Interferometer Gravitational-wave Observatory (LIGO) occurred $\sim$1.7s before the  {\it Fermi}/GBM trigger \citep{GW170817} and the sky position determined by the LIGO GW observation is consistent with the  {\it Fermi}/GBM one. The gamma-ray spectrum for the initial hard pulse is well fitted by a power-law function with a photon index  of $\alpha$ = -1.6$\pm$0.4 and an exponential cutoff at $E_{\rm peak}$ = 185 $\pm$ 62keV. For the soft tail emission, a blackbody with $k_{\rm B}T$ = 10.3 $\pm$ 1.5keV well represents the spectrum.  The fluence over the total duration is $\sim$3 $\times$ 10$^{-7}$erg$\,$cm$^{-2}$, which falls in the $\sim$50th percentile of the fluence distribution obtained by  {\it Fermi}/GBM  \citep{Goldstein:2017mmi}. Assuming a distance to the host galaxy NGC 4993 as 43Mpc,  the isotropic equivalent energy and peak luminosity in the 1keV -- 10MeV band are $\sim$3 $\times$ 10$^{46}$erg and $\sim$2 $\times$ 10$^{47}$erg$\,$s$^{-1}$, respectively, which are by 3--6 orders of magnitude smaller than those of other short GRBs detected by  {\it Fermi}/GBM. The observed features for GRB 170817A would suggest that the hard X-ray emission came from a misaligned jet  to the observer (i.e., off-axis jet).

The higher-energy gamma-ray observation of GRB 170817A by  {\it Fermi} Large Area Telescope (LAT; \cite{2009ApJ...697.1071A})
 covering the 0.1--300GeV band was unfortunately not available due to entry to the South Atlantic Anomaly  at the time of occurrence of the merger time  \cite{2018ApJ...861...85A}.   {\it Fermi}/LAT resumed scientific operation  $\sim$10$^3$s after the merger time and put a flux upper limit of 4.5 $\times$ 10$^{-10}$erg$\,$cm$^{-2}$ s$^{-1}$ (95\% confidence level) in the 0.1--1GeV band using the initial 1000s observation after the operation resumption.

Motivated by study of GRB 170817A,  investigation of previous short GRBs with a similar signature  has been performed 
\citep{2018ApJ...863L..34B,2018NatCo...9.4089T, 2019ApJ...876...89V,2020MNRAS.492.4283M}. In particular, the nearby short GRB 150101B  detected by {\it Fermi}/GBM  ({\it z} = 0.134 \cite{2015GCN.17281....1L}), which could accompany the kilonova and the off-axis jet emission suggested by the optical and X-ray observations \cite{2018NatCo...9.4089T}, has a hard spike and a soft tail. Thus, the two-component sigunature  could be a common feature for short GRBs \cite{2018ApJ...863L..34B}. The isotropic equivalent energy and  luminosity in the 1keV -- 10MeV band are $\sim$2 $\times$ 10$^{49}$erg and $\sim$4 $\times$ 10$^{50}$erg$\,$s$^{-1}$, respectively, which are in a fiducial range of short GRBs. 
One of the possible interpretation for these phenomena suggests that GRB 150101B could originate from a more on-axis jet than GRB 170817A.

%

\paragraph{MAXI}
Monitor of All-sky X-ray Image ({\it MAXI};
\cite{2009PASJ...61..999M}) is an instrument for monitoring
X-ray sky, which was launched in 2009 and has been observing from the 
International Space Station (ISS).
Because {\it MAXI} does not have a 
moving mechanism and is mounted to the ISS,
we cannot actively control the pointing direction.  
The orbital period of the ISS is 92 minutes and thus 
an X-ray source is usually observed once in the same period.
Since {\it MAXI} covers about 85\% of the sky every orbit, the 
source activity can trace back to the time before the event.

In studying X-ray counterparts of GW events, we use 
the Gas Slit Camera (GSC; \cite{2011PASJ...63S.623M})
GSC consists of 12 position sensitive proportional counters. 
Six of them covers 3 degree $\times$ 160 degree field of view (FOV)
toward the horizontal direction
and the others compose zenithal FOV with the same dimension.
These FOVs move as the ISS rotates. A scan observation (transit) for a
point source lasts 40--150 seconds depending on the source position in 
the FOV
\cite{2011PASJ...63S.635S}.

The effective area of the GSC for a source changes with  time linearly
like a triangle.
For a source with constant flux (or that can be assumed as constant in a 
scan transit), we calculate an average flux by dividing the source count 
by the effective exposure, that is a sum of the effective area along the
time.

\begin{figure}
  \begin{center}
    \includegraphics[scale=1.0]{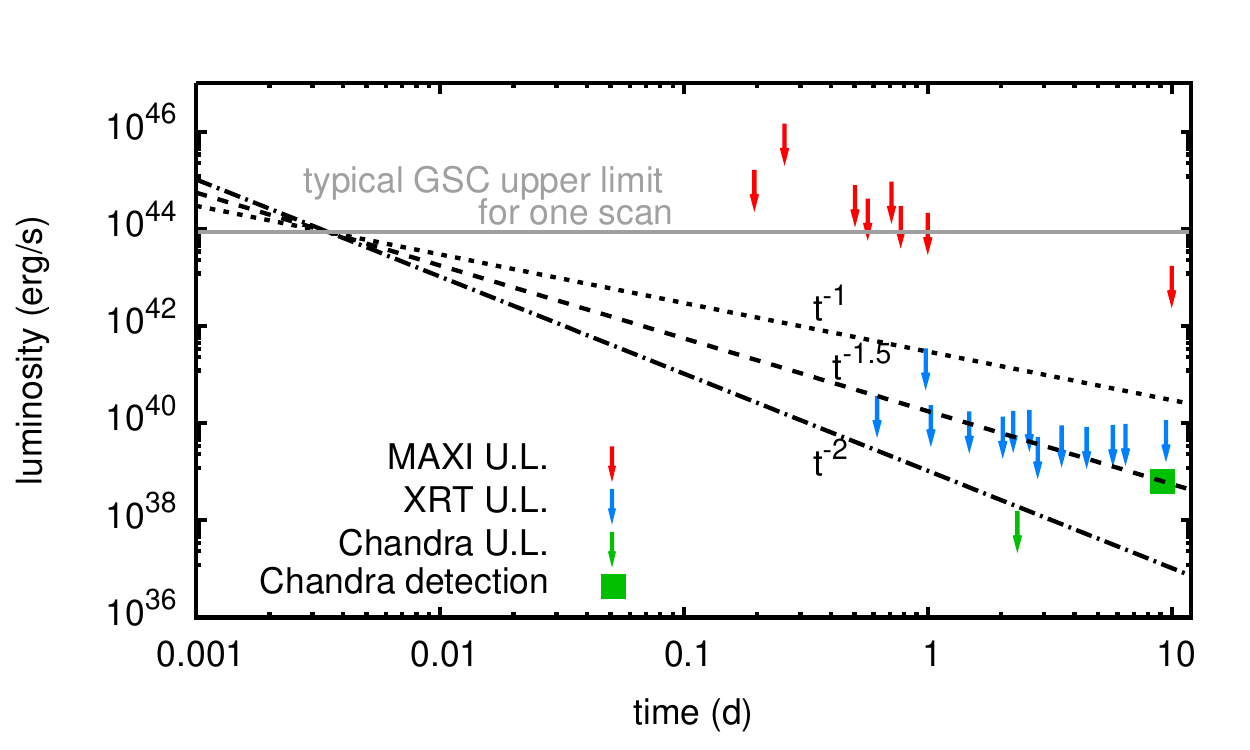}
\caption{
  \label{fig:maxi}
The observed luminosity and upper limits of the X-ray counterpart of
GW170817 in early time. {\it MAXI} upper limits are based on the
result of 
\cite{2018PASJ...70...81S}. 
The first six points are for each scan
transit and the last two are for one day and 10-day observations.
The data of {\it Swift} XRT and {\it Chandra} are converted to the luminosity
at 2-10keV assuming the spectral parameters in
their original papers
\cite{swift:phil,2019ApJ...886L..17H}. 
The dotted, dashed, and dash-dotted lines are models for luminosity $L$
as a function of the time $t$, $L \propto t^{\alpha}$, and $\alpha$ 
are -1, -1.5, and -2, respectively.
The horizontal gray line is a typical GSC upper limit for one scan
transit, scaled to the distance of GW170817 (40Mpc).
  }
\end{center}
\end{figure}

For GW170817 {\it MAXI} did not detect the X-ray emission. The upper limits
are summarized in
\cite{2018PASJ...70...81S}.
Fig.~\ref{fig:maxi} shows the luminosity and upper limits for early 
($< 10$ days) X-ray afterglows
\cite{swift:phil,2019ApJ...886L..17H}
in the case of GW170817.  The position of the EM counterpart lay at the
gap between the horizontal and the zenithal cameras for the first three 
orbits and the first scan started after 4.7 hours.
Moreover the first three observations were incomplete and 
the upper limits were 1--2 order of magnitude higher than the usual.
However, it is important to mention that this was the earliest observation
of the X-ray counterpart of the GW event.

Although the observation start time was late in this case,
if {\it MAXI's} observation starts $\sim$ 100 seconds after the trigger,
{\it MAXI} could observe an X-ray afterglow equivalent to the level
observed by Swift at several hours after the trigger. 

  The best lesson of the observation of GW170817 was that it was
  important to continuously observe the sky as long as possible.
  In O3 we have increased the observation efficiency to reduce the gap in
  the observation.

\paragraph{Swift}
The Neil Gehrels {\it Swift} Observatory ({\it Swift}; \citep{swift:neil}) consists 
of three scientific instruments.  The Burst Alert Telescope (BAT; \citep{swift:bat}) is a wide field of 
view telescope monitoring about 1/6 of the sky in the 15--150keV band.  The X-ray Telescope (XRT; \citep{swift:xrt}) 
is an X-ray telescope covering the energy range in the 0.3--10keV band in 24$^{\prime}$ field of view.  
The UV-Optical Telescope (UVOT; \citep{swift:uvot}) observes the wavelength range of 170--600nm in 
17$^{\prime}$ field of view.  

At the merger time of GW170817, the error region of GW170817 was 
occulted by the Earth and not visible by BAT \citep{swift:bat_gw170817,swift:phil}.  
Therefore, no useful information can be derived from the BAT data.  
The XRT and UVOT started their observations about an hour after the merger time 
around the center of the {\it Fermi} GBM error region ($\sim1^{\circ}$ radius).  
After the localization information became available from the LIGO and Virgo at 4.8 hours 
after the merger time, {\it Swift} started a series of 120s observations at the positions of 
known galaxies inside the 33.6 deg$^{2}$ error region of the gravitational-wave detectors.  
No new X-ray sources ($\geq$10$^{-12}$erg$\,$cm$^{-2}\,$s$^{-1}$) were found during those 
observations \citep{swift:phil}.  

{\it Swift} started the observations at the position of the optical transient of GW170817 
about 14.4 hours after the merger time.  Although the 
XRT found no X-ray from the position of the transient, the UVOT detected a fast decaying UV 
emission which is interpreted as the blue kilonova associated with GW170817.   The non-detection 
of X-ray at the early phase by XRT places a crucial difference to a typical short GRB afterglow 
which is dominated by an on-axis emission \citep{swift:phil}.  

\paragraph{CALET}
The Calorimetric Electron Telescope ({\it CALET}; \citep{calet:torii,calet:asaoka}) is 
the scientific instrument attached on the International Space Station (ISS).  The {\it CALET}  
Gamma-ray Burst Monitor (CGBM;\citep{calet:yamaoka}) is the gamma-ray burst monitor 
onboard {\it CALET} and is capable to observe emissions from 7keV to 1MeV 
(Hard X-ray Monitor; HXM, FoV of $\sim$$60^{\circ}$ from the boresight) and from 40keV to 20MeV 
(Soft Gamma-ray Monitor; SGM, FoV of $\sim$$110^{\circ}$ from the boresight) utilizing two 
different detectors.  No significant emission was observed by CGBM at the time of 
the merger of GW170817.  The 90\% upper limit is $1.3 \times 10^{-7}$erg$\,$cm$^{-2}\,$s$^{-1}$ 
in the 10-1000keV band assuming no shielding by the structure of ISS \citep{Monitor:2017mdv}.  

In addition, the main CALET instrument, the high-energy Calorimeter (CAL), observes gamma-rays 
from $\sim$1GeV up to 10TeV with a field of view of nearly 2sr. The field of view of CAL did not include 
the location of GW170817 at the time of the merger. A search for delayed emission over the 2-month period following the event resulted in 90\% C. L. upper limits on the energy flux of 
$1.2 \times 10^{-10}$erg$\,$cm$^{-2}\,$s$^{-1}$ for gamma-rays above 1GeV and 
$4.0 \times 10^{-10}$erg$\,$cm$^{-2}\,$s$^{-1}$ 
above 10GeV \citep{calet:cal}. 

\paragraph{Chandra}
The {\it Chandra} X-ray Observatory, which provides the best sensitivity in X-ray for a point source, first 
targeted to the optical transient of GW170817 at 2.2 days after the merger time.  Although no X-ray 
emission was detected in this first visit, {\it Chandra} detected the significant X-ray emission on the 2nd 
visit at 8.9 days after the merger in the isotropic luminosity of $9 \times 10^{38}$erg$\,$s$^{-1}$ at a distance 
of 40Mpc \cite{chan:nora1}.  

\begin{figure}
  \begin{center}
    \includegraphics[scale=0.4]{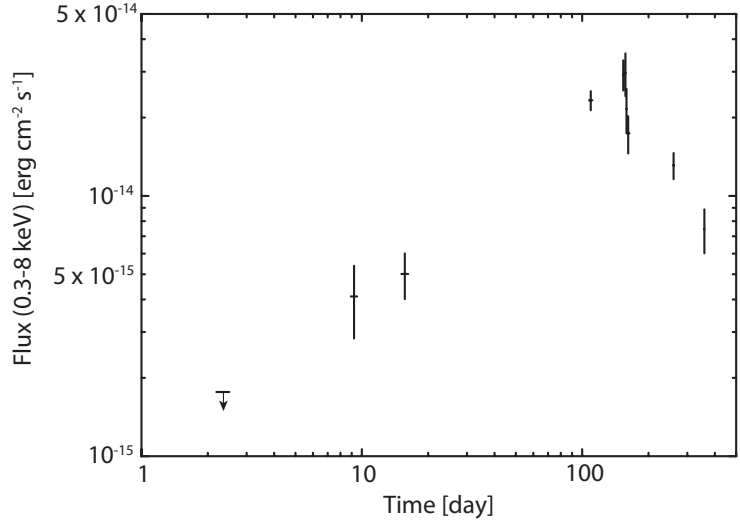}
\caption{
  \label{fig:chandra_lc}
The X-ray light curve of the counterpart of GW170817 observed by {\it Chandra}.  The source 
count-rates are converted to the unabsorbed flux in the 0.3-8keV band using the Galactic absorption 
column $N_{H} = 7.6 \times 10^{20}$cm$^{-2}$ and the photon index of 1.59 \cite{vanEerten:2018vgj}. }
\end{center}
\end{figure}

According to the year-long monitoring observations by {\it Chandra}, the X-ray flux continues to increase 
up to $\sim$160 days after the merger time by a power-law index of 0.9 \cite{vanEerten:2018vgj,chan:hajela}.  
After the peak, the flux shows the decline by a power-law index of $\sim-2$ (Fig.~\ref{fig:chandra_lc}).  There is 
a hint of an X-ray flaring feature around 155 days after the merger time \cite{chan:piro}.  {\it Chandra} has been 
collecting the data up to 2.5 years after the merger and \blue{continues} monitoring.  This unique X-ray light curve 
behavior along with the radio data will be discussed in the following section.  

\subsubsection{Theoretical Interpretation}

\begin{figure}
  \begin{center}
    \includegraphics[scale=0.4]{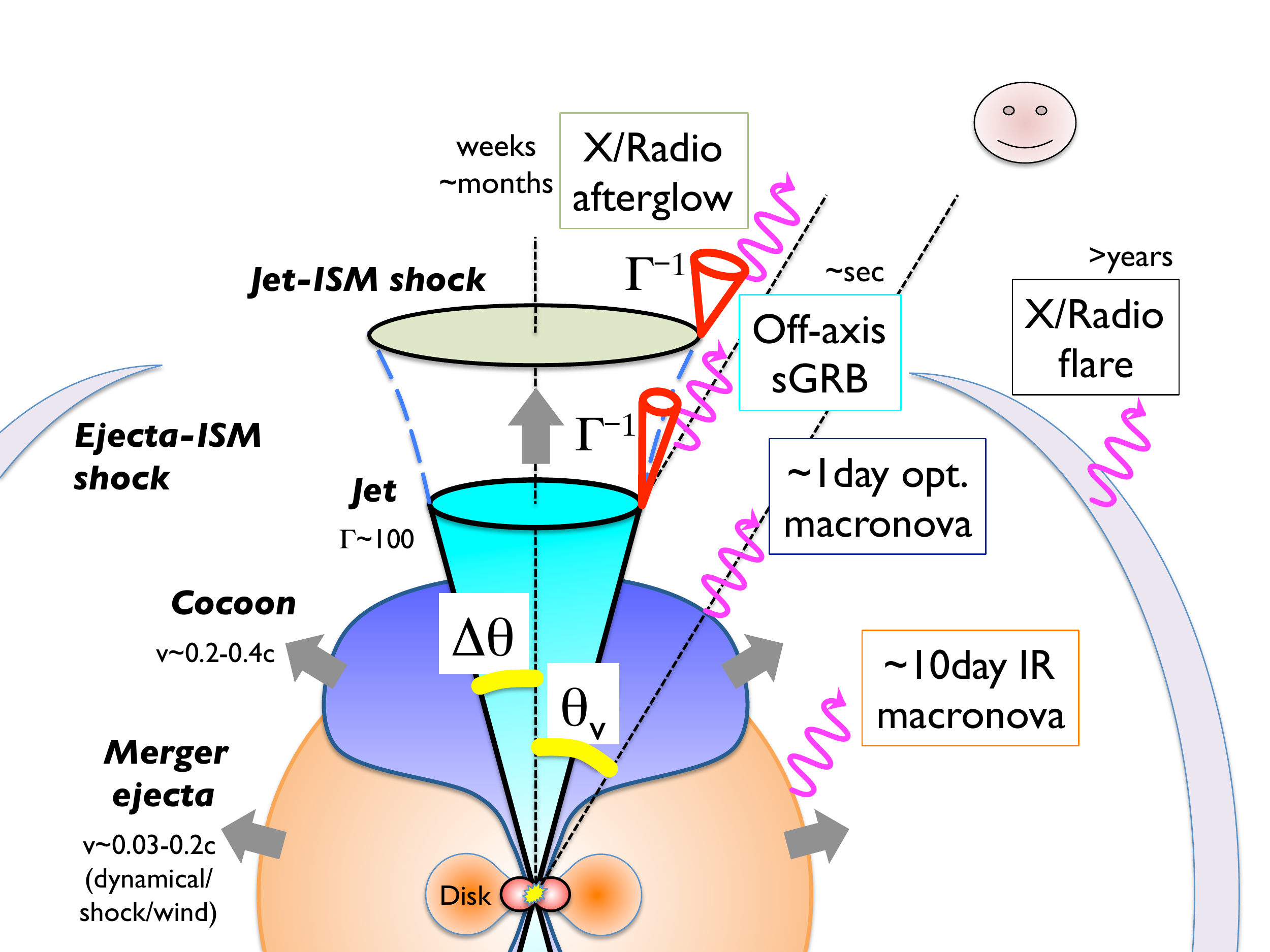}
\caption{
  \label{fig:GRBjet}
Electromagnetic counterparts associated with the off-axis jet from short GRB 170817A
(from Fig.~1 of Ref.~\citep{Ioka:2017nzl}).
  }
\end{center}
\end{figure}

The GW event GW170817 \citep{GW170817} and
the associated short GRB 170817A
\citep{Monitor:2017mdv,Goldstein:2017mmi,Savchenko:2017ffs}
have provided, for the first time,
clear evidence for a relativistic jet from a binary neutron star merger
observed from off-axis (as schematically shown in Fig.~\ref{fig:GRBjet}).
The observations also show that the jet is structured,
excluding a top-hat distribution of energy and Lorentz factor of the jet.
We also discuss the current issues about the exact structure of the jet,
the origin of the gamma-ray emission, and future prospects.

\paragraph A Evidence of an off-axis jet

The afterglow observations of GW170817A
verify the existence of a relativistic jet:
the superluminal motion of the radio image
indicates a relativistic collimated source
\citep{Mooley:2018qfh,Ghirlanda:2018uyx}
and the closure relation between the spectral index
and the light curve slope after the luminosity peak
is consistent with the standard afterglow model of a relativistic jet
\citep{vanEerten:2018vgj,Mooley:2018clx,Lamb:2018qfn},
excluding a failed-jet scenario
that the jet is choked by the ejecta from the neutron star merger
\citep{Kasliwal:2017ngb,Gottlieb:2017pju,Nakar:2018cbe}.

The jet should be strong enough for 
successfully breaking out the merger ejecta in this event
\citep{Nagakura:2014hza,Murguia-Berthier:2014pta}.
The breakout time since the jet launch ($t=t_0$)
is approximately estimated as
\begin{eqnarray}
t_b-t_0 \sim
0.17\,{\rm s}
\left(\frac{t_0-t_m}{0.1\,{\rm s}}\right)^{1/2}
\left(\frac{L_{{\rm iso},0}}{10^{51}\,{\rm erg}\,{\rm s}^{-1}}\right)^{-1/2}
+0.078\,{\rm s}
\left(\frac{L_{{\rm iso},0}}{10^{51}\,{\rm erg}\,{\rm s}^{-1}}\right)^{-1},
\end{eqnarray}
where $L_{{\rm iso},0}$ is the isotropic luminosity at the base of the jet
and $t_0-t_m$ is the time delay of the jet launch since the merger
(modifying Eq.~(32) of Ref.~\citep{Hamidani:2019qyx}).
Here the second term is necessary for this event because
the ejecta expansion velocity is comparable to the jet head velocity,
in contrast to collapsars with a static envelope.
In order to have the short GRB 170817A $\sim 1.7$s after the merger,
the jet luminosity should be large enough $L_{{\rm iso},0} \gtrsim 3 \times 10^{49}$erg$\,$s$^{-1}$ and the delay time should be short enough $t_0-t_m \lesssim 1.3$ s
\citep{Hamidani:2019qyx}.

The observed isotropic energy of the gamma-ray emission is 
very small, several orders of magnitude smaller than typical
\citep{Monitor:2017mdv,Goldstein:2017mmi,Savchenko:2017ffs}.
Therefore the jet should be off-axis for a reasonable radiative efficiency
\citep{Ioka:2017nzl,Ioka:2019jljb}.
An off-axis observer receives photons emitted
outside the relativistic beaming cone,
and the apparent energy of the off-axis jet becomes faint
(see also below).

\paragraph B Diverse jet structures

The afterglow observations,
in particular the slowly rising light curves from radio to X-ray,
are not consistent with a top-hat jet \citep{Mooley:2017enz},
but indicate a structured jet
\citep{DAvanzo:2018zyz,Ghirlanda:2018uyx,Lazzati:2017zsj,Lyman:2018qjg,Margutti:2018xqd,Ruan:2017bha,Troja:2018ruz,vanEerten:2018vgj,Lamb:2018qfn}.
These are the first confirmation of previous suspicion that
realistic GRB jets are structured 
\citep{Meszaros:1997je,Zhang:2001qt}.

However, there remains confusion about the jet structure
as different authors give different structures,
such as Gaussian \citep{Lyman:2018qjg,vanEerten:2018vgj,Lamb:2018qfn}
and power-law profiles \citep{DAvanzo:2018zyz,Ghirlanda:2018uyx}
(see Fig.~1 of Ref.~\citep{Ioka:2017nzl}).
These structures are obtained by assuming a functional form
of the jet structure with model parameters,
and adjusting the parameters
through fitting the theoretical light curves to the observations.

\begin{figure}
  \begin{center}
    \includegraphics[scale=0.35]{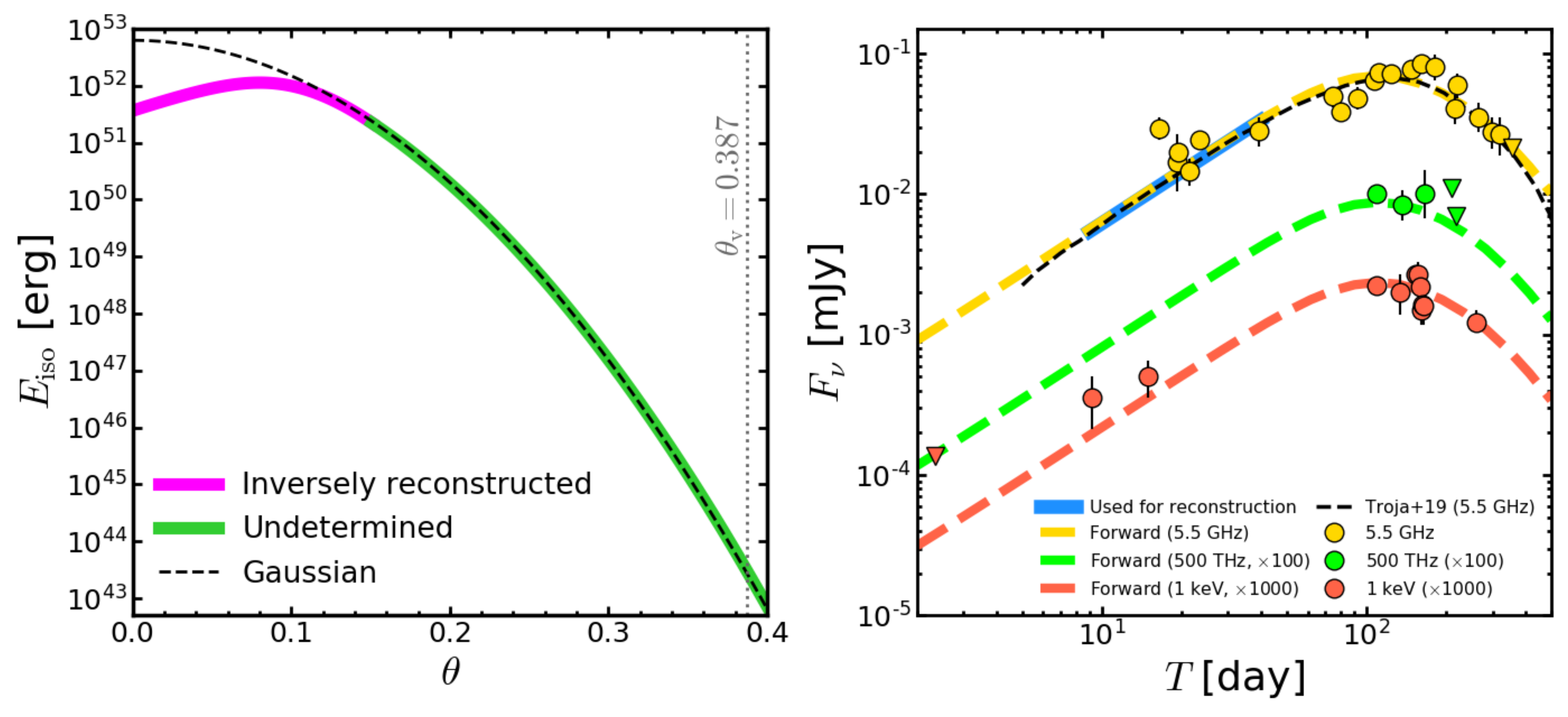}
\caption{
  \label{fig:JetStructure}
Left: Examples of the jet structure that are consistent with
the afterglow observations of GRB 170817A:
a hollow-cone (magenta line) and Gaussian jet (dashed line).
The green region is not in principle determined by the current data.
Right: The corresponding light curves for a hollow-cone (thick dashed lines)
and Gaussian jet (thin dashed line) with the data.
The blue line is used for the inverse reconstruction
(modifying Fig.~8 of Ref.~\citep{Takahashi:2019otc}).
  }
\end{center}
\end{figure}

We instead invent a novel method to determine the functional form
of the jet structure itself \citep{Takahashi:2019otc}.
This method solves an inverse problem, uniquely reconstructing
a jet structure from a given off-axis GRB afterglow
by integrating an ordinary differerential equation,
which is formulated based on
the standard theory of GRB afterglows.
Applying the inversion method to GRB 170817A, we clarify that
the current observational errors still allow various jet structures,
discovering by-product that
a hollow-cone jet is also consistent with the observations
as shown in Fig.~\ref{fig:JetStructure}.

An important caveat is that
the current afterglow observations only constrain the jet core
(magenta in Fig.~\ref{fig:JetStructure}),
not in principle the outer jet structure,
(green; which is crucial for the gamma-ray emission as discussed below).
Early afterglow observations are required to obtain
the outer jet structure \citep{Takahashi:2019otc}.

The origin of the jet structure is still unclear.
First, the jet may have intrinsic structure from the beginning of the launch.
Second, the jet may be structured during the propagation
if the outer part loads baryon from the ejecta.
Third, the cocoon component
(the collided jet and ejecta)
may contribute to the outer structure
since a part of the cocoon is relativistic
due to partial mixing of the baryon.
Early macronova/kilonova observations are important
for probing the cocoon, which has crucial information on the jet.

\paragraph C Origin of the gamma-ray emission
\label{sec:gamma}

\begin{figure}
  \begin{center}
    \includegraphics[scale=1.3]{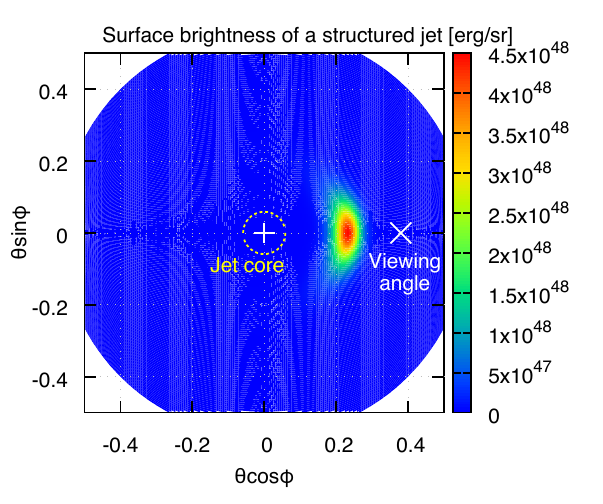}
\caption{
  \label{fig:JetSurface}
The surface brightness distribution of the prompt emission from a Gaussian jet
(from Fig.~3 of Ref.~\citep{Ioka:2019jljb})
  }
\end{center}
\end{figure}

The prompt gamma-ray emission of GRB 170817A
generally comes from an off-center jet,
neither the jet core nor the line of sight but the middle,
as in Fig.~\ref{fig:JetSurface} \citep{Ioka:2019jljb},
given the jet structure with a rapidly decaying tail
as in Fig.~\ref{fig:JetStructure}.
This is because the emission from the jet core is suppressed
by the relativistic de-beaming,
while the line-of-sight emission becomes too faint to 
dominate the off-axis emission from the inner jet.
The off-center emission solves seemingly puzzles that
the observed prompt spectrum is inconsistent
with the spectral Amati relation \citep{Amati:2002ny,Yonetoku:2003gi}
(because the off-center is different from the usually-observed jet core)
and causes the compactness problem
\citep{Matsumoto:2018xnd,Matsumoto:2019sor}
(because the off-centre jet is much less energetic
and much closer to the line of sight than the jet core).

The off-center structure still has large uncertainties
because it is not constrained by the observed afterglows
(as shown by the green region in Fig.~\ref{fig:JetStructure}).
Whether it is a jet or cocoon is also unknown.
The observed gamma-rays might be the jet emission scattered by the cocoon
\citep{Kisaka:2017mgz}.
Future observations of off-axis events will reveal the jet structure, where 
roughly $\sim 10\%$ events are expected to be brighter
at smaller viewing angles than GRB 170817A.

%% file: ptepGW_sectionB03.tex

\begin{figure}
  \begin{center}
    \includegraphics[scale=0.5]{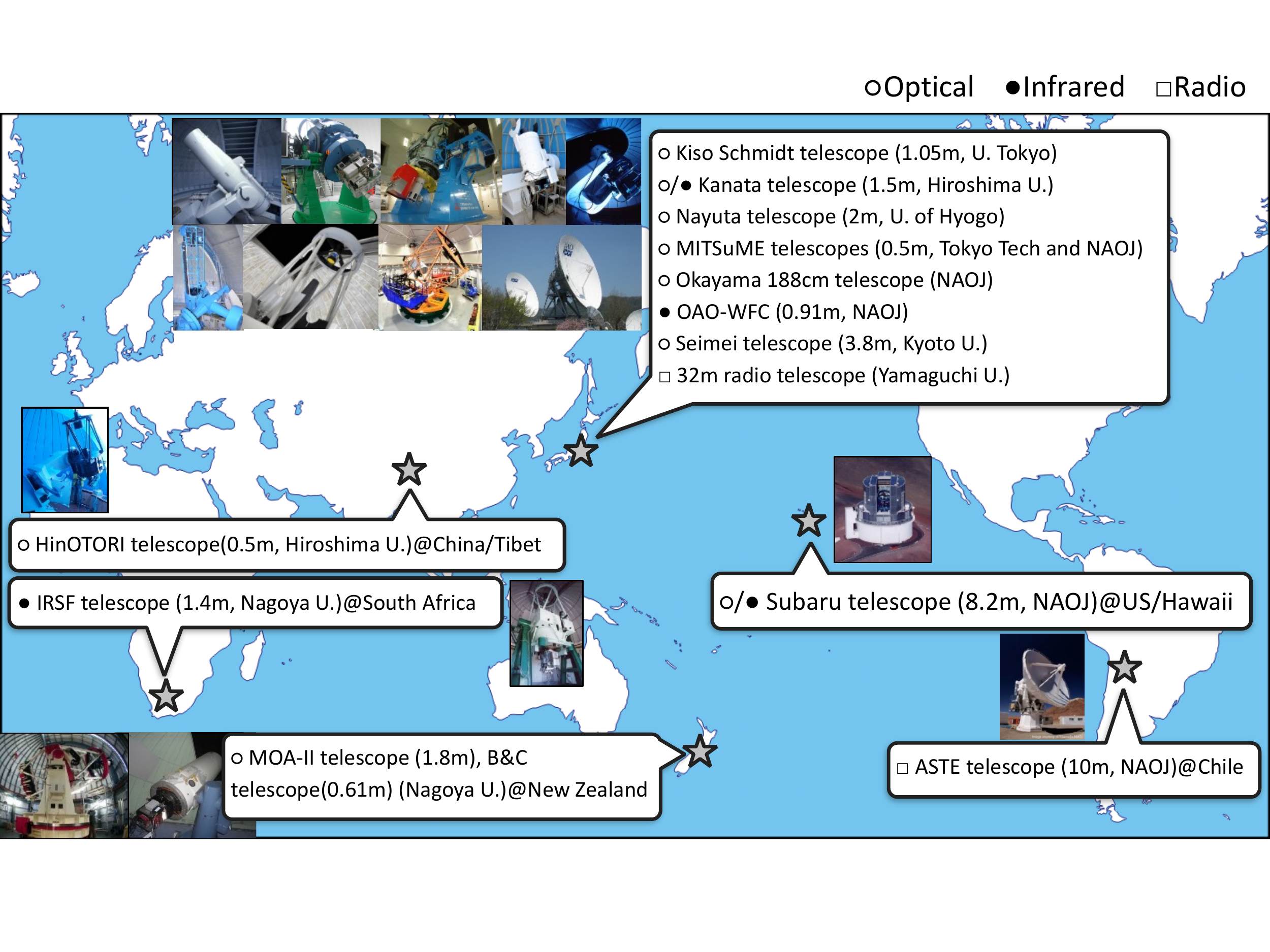}
\caption{
  \label{fig:B03_jgem}
  Optical, infrared, and radio telescopes in the J-GEM network.
  }
\end{center}
\end{figure}

The origin of the elements in the Universe is a subject of
interest in astronomy and astrophysics.
In particular, the origin of the elements synthesized by rapid neutron capture,
or so-called $r$-process, is one of the long-standing, unsolved problems.
Theoretically, NS mergers are one of the promising
candidate sites of $r$-process nucleosynthesis \cite{lattimer74,lattimer76,eichler89,freiburghaus99,roberts11,goriely11,korobkin12,bauswein13,wanajo14}.
However, there has been no observational evidence of
$r$-process by NS mergers.

When NS mergers occur, a small fraction \blue{of mass of NSs is} ejected
into interstellar space \cite{rosswog99,rosswog05,goriely11,rosswog13a,hotokezaka13,bauswein13}.
In the ejected material, $r$-process nucleosynthesis
is expected to take place.
Then, we expect optical and infrared emission
powered by radioactive decays of freshly synthesized nuclei \cite{li98,kulkarni05,metzger10,kasen13,barnes13,tanaka13}.
This phenomenon is called ``kilonova'' or ``macronova'' (hereafter we call it kilonova).
In other words, if we can detect kilonova after detection
of GWs from a NS merger,
we can test $r$-process nucleosynthesis by the NS merger.

The event rate of NS mergers can be measured from GW observations.
The ejection of $r$-process elements per event can be estimated from electromagnetic (optical/infrared) observations.
Therefore, by these multi-messenger observations,
we can study the production rate of $r$-process elements by NS mergers
and test if NS mergers can be the origin of $r$-process elements in the Universe.

To achieve observations of kilonova, we coordinated the Japanese collaboration for Gravitational wave ElectroMagnetic follow-up (J-GEM, Section \ref{sec:B03_jgem}).
Thanks to this observational network, we succeeded in observations of
the electromagnetic counterpart of GW170817 (Section \ref{sec:B03_gw170817}).
By combining our state-of-art numerical simulations (Section \ref{sec:B03_models}), we advance our knowledge about the physics of NS mergers and the origin of $r$-process elements in the Universe.

\subsubsection{J-GEM} 
\label{sec:B03_jgem}

J-GEM is a network of optical, infrared, and radio telescopes for EM follow-up observations of GW sources.
Figure \ref{fig:B03_jgem} shows a summary of telescopes in the J-GEM.
The roles of the optical telescopes can be divided into two categories: one is the wide-field surveys (Kiso, MOA-II, and Subaru Hyper Suprime-Cam (HSC)) and galaxy targeted surveys (other telescopes).
For more details of each telescope, see Morokuma {\it et al.} (2016) \cite{morokuma16}.
J-GEM signed Memorandum of Understanding with LIGO/Virgo collaboration in 2014,
and started to receive the alert of GW detection from the first observing run of Advanced LIGO (O1).

We demonstrated coordinated observations using this observing network
for the first GW source GW150914 (BBH, \cite{GW150914}).
Within 4 days from the GW detection (2 days from the initial alert),
about 24deg$^2$ area has been observed with Kiso Schmidt telescopes.
Also, within 6-7 days from the GW detection,
18 nearby galaxies were observed with the B\&C 61-cm telescope.
Although no EM counterpart has been identified, this was the first attempt of multi-messenger observations of GW sources \cite{morokuma16}.

More intensive observing campaign has been performed for
the second GW source GW151226 (BBH, \cite{GW151226PRL}).
For this event, 238 nearby galaxies were observed by galaxy targeted surveys \cite{yoshida17}, which is great improvement compared with the observations of GW150914.
Also, Kiso Schmidt telescope and MOA-II telescopes covered
778deg$^2$ and 145deg$^2$ area, respectively \cite{yoshida17}.
Furthermore, Subaru/HSC covered 63.5deg$^2$ area with unprecedented sensitivity (limiting magnitudes of 24.6 and 23.8 for the $i$ and $z$ bands, respectively, \cite{utsumi18}). The total area covered by the wide-field surveys was 986.5deg$^2$, which corresponds to $\sim$29\% of the probability map of GW151226.
It is worth emphasizing that the sensitivity of Subaru/HSC observations
was deep enough to detect expected kilonova emission even at about 200Mpc,
which is about the design sensitivity of Advanced LIGO, Advanced Virgo, and KAGRA.

Subaru/HSC is one of the most powerful wide-field imager:
its field-of-view (1.8deg$^2$) is the largest among 8-10m class telescopes.
Table \ref{tab:B03_HSC} summarized our follow-up observations using Subaru/HSC.
Details of follow-up observations of   S190510g are presented by Ohgami {\it et al.} (2021) \cite{ohgami21}.

\begin{table}[!ht]
  \caption{Subaru/HSC observations for GW sources
  \newline 
  \small{$^1$ Detected with low-letency pipeline but not recovered with the offline analysis.
  \newline 
  $^2$ Initially reported as BNS with a high probability but can be terrestrial noise.
  \newline 
  $^3$ Initially reported as Mas Gap object but later classified as BBH.} }
\label{tab:B03_HSC}
\centering
\begin{tabular}{cccccccc}
\hline
Target    & Type     & Disc. date  &  Obs. date & Filter    & Area      & Probability & Lim. mag \\
          &          &   (UUT)     &   (UT)     &           & (deg$^2$) &   (\%)      &          \\ 
\hline                                                             
GW151226  &   BBH    & 2015-12-26  & 2016-01-07   & HSC-i  & 63.1      &   12       & 24.4  \\
          &          &             &  2016-01-07  & HSC-z  & 63.3      &   12       & 24.0  \\
          &          &             & 2016-01-13   & HSC-i2  & 63.1      &   12       & 24.4  \\
          &          &             &  2016-01-13  & HSC-z  & 63.2      &   12       & 24.0  \\
          &          &             &  2016-02-06  & HSC-i2 & 65.0      &   12       & 24.5  \\
          &          &             &  2016-02-06  & HSC-z  & 65.0      &   12       & 24.0  \\
G275404$^1$ &  BH-NS & 2017-02-25  &  2017-02-27  & HSC-z  & 26.0      &   0.5      & 23.2  \\
GW170814  &   BBH    & 2017-08-14  & 2017-08-16  & HSC-Y   & 32.5      &   35       & 22.2  \\
          &          &             & 2017-08-17  & HSC-Y   & 32.6      &    35       & 22.2 \\
GW170817  &  BNS     & 2017-08-17  & 2017-08-18  & HSC-z   & 26.2      &  61       & 21.7 \\
          &          &             & 2017-08-19  & HSC-z   & 27.3      &  61       & 21.6 \\
S190510g  &  BNS$^2$ & 2019-05-10  & 2019-05-10  & HSC-Y   & 118.8     &   1.1       & 22.9 \\
S191216ap &  BBH$^3$ & 2019-12-16  & 2019-12-20  &  HSC-z  & 2.3      & 0.035      & 25.2  \\
S200224ca &  BBH & 2020-02-24  & 2020-02-25  &  HSC-r2  & 57.8      & 80.8      & 25.3 \\
         &       &             &              &  HSC-z  & 57.8      & 80.8      & 23.6  \\

          &      &             & 2020-02-28  &  HSC-r2  & 57.8      & 80.8     & 24.6  \\
          &      &             &             &  HSC-z  & 57.8      & 80.8     & 23.2  \\

     &      &                  & 2020-03-23  &  HSC-r2  & 57.8      & 80.8      & 25.5  \\
     &      &                  &             &  HSC-z  & 57.8      & 80.8      &  23.8  \\

\hline
\end{tabular}
\end{table}

\begin{figure}
  \begin{center}
    \includegraphics[scale=0.4]{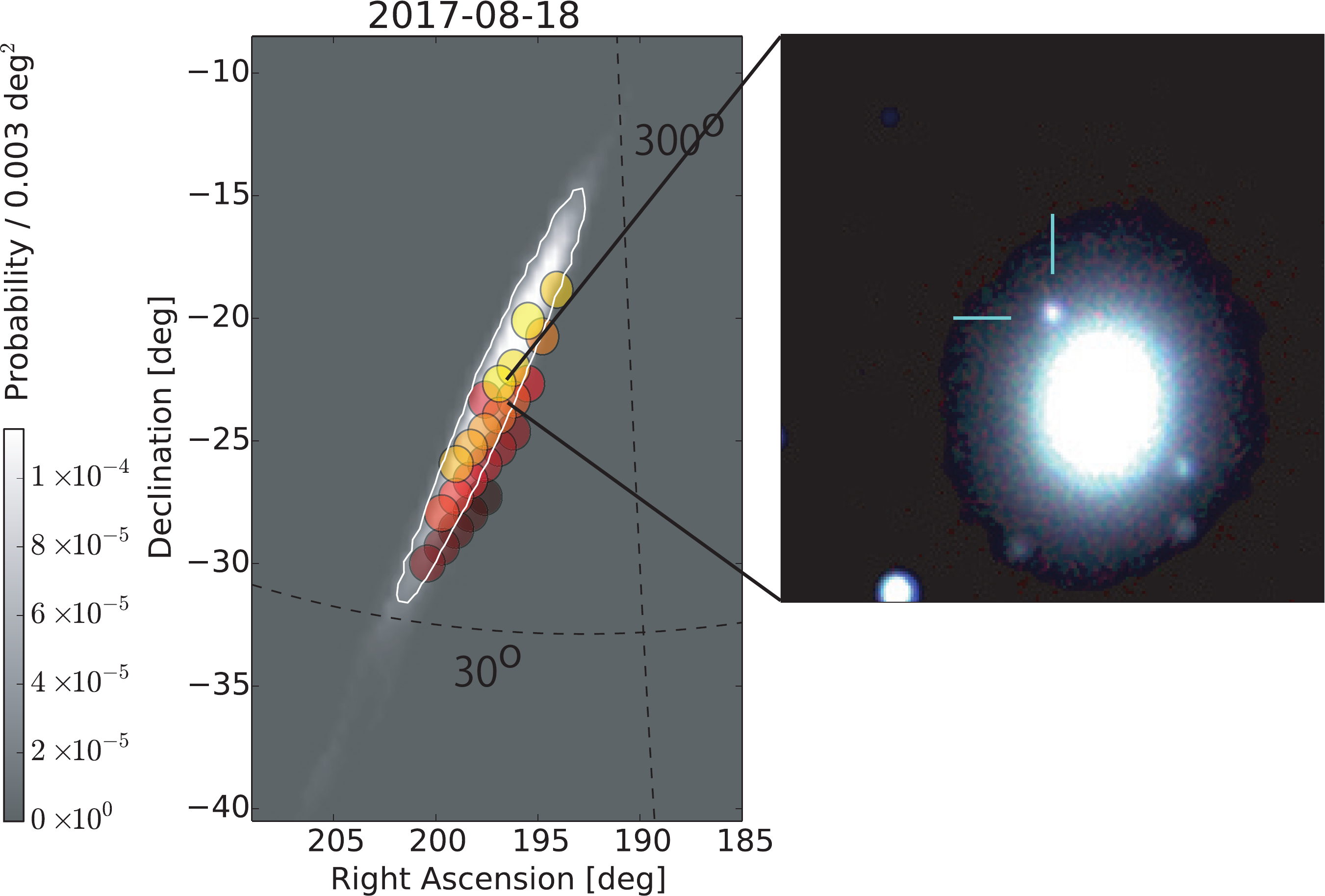}
\caption{
  \label{fig:B03_gw170817}
  Left: Localization map of GW170817 and Subaru/HSC pointing \cite{tominaga18}. Right: three-color image of the counterpart of GW170817 \cite{utsumi17}.
  }
\end{center}
\end{figure}

\subsubsection{GW170817}
\label{sec:B03_gw170817}

The first GW detection from a NS merger has been achieved
for GW170817 \cite{GW170817}.
The detection triggered EM follow-up observations
covering entire wavelength range \cite{abbott17MMA}.
Thanks to the observations with three GW detectors,
the position of the GW source is localized to $\sim 30$deg$^2$.
In optical/infrared wavelengths, a counterpart AT2017gfo
was identified by several groups \cite{andreoni17,arcavi17,chornock17,coulter17,cowperthwaite17,diaz17,drout17,evans17,kasliwal17,kilpatrick17,lipunov17,muccully17,nicholl17,pian17,shappee17,siebert17,smartt17,soares-santos17,tanvir17,tominaga18,troja17,utsumi17,valenti17}.

We have performed wide-field survey using Subaru/HSC,
covering 23.6deg$^2$ \cite{tominaga18}.
The observations recovered the detection of AT2017gfo.
Furthermore, thanks to the wide coverage and good sensitivity,
we statistically ruled out that other transients in the localization
area are associated with GW170817, or in other words, 
we concluded that AT2017gfo is the most likely counterpart.

Intensive optical and infrared observations of AT2017gfo have been performed
in the framework of J-GEM network \cite{utsumi17}.
Observations with Subaru/HSC \blue{show} that the $z$-band brightness
quickly declines.
On the other hand, the observations with IRSF/SIRIUS
shows that near-infrared light curves evolve more slowly.
This is fully consistent with the expected behavior of kilonova.

Using radiative transfer simulations, we find that the observed light curves
can be explained by $\sim 0.03 M_{\odot}$ of ejecta including lanthanide elements, which have a high opacity \cite{tanaka17a}.
Also, the blue optical component, revealed by observations with Subaru/HSC,
B\&C/Tripol5, and MOA-II/MOA-cam3, requires an ejecta component with
a smaller fraction of lanthanide.
Since the estimated total ejecta mass is higher than the expected ejection
by the first dynamical ejection from the NS merger, we concluded that the observed signals are mainly produced by post-merger mass ejection.

\begin{figure}
  \begin{center}
        \begin{tabular}{cc}
    \includegraphics[scale=0.5,angle=270]{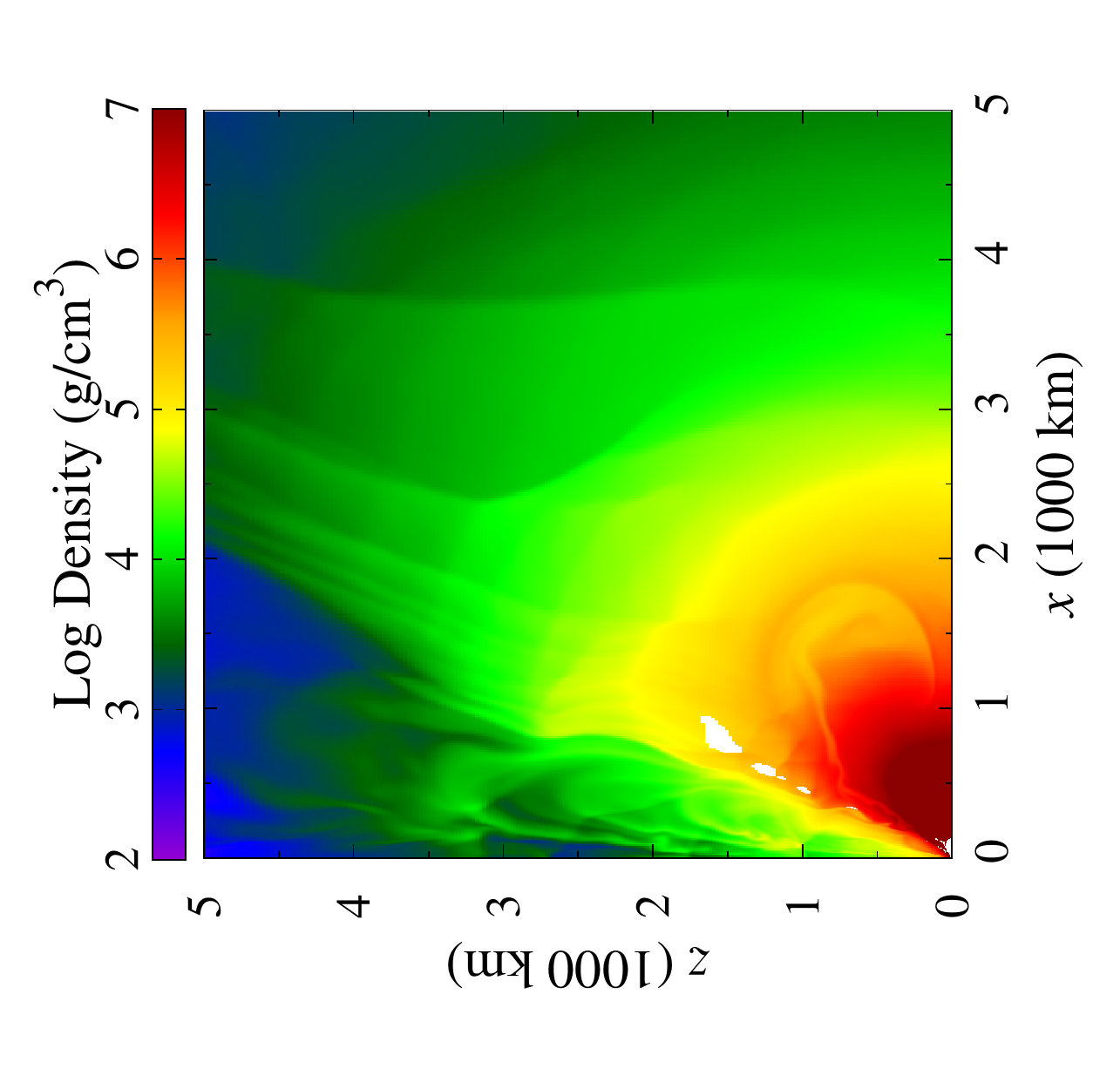} &
    \includegraphics[scale=0.5,angle=270]{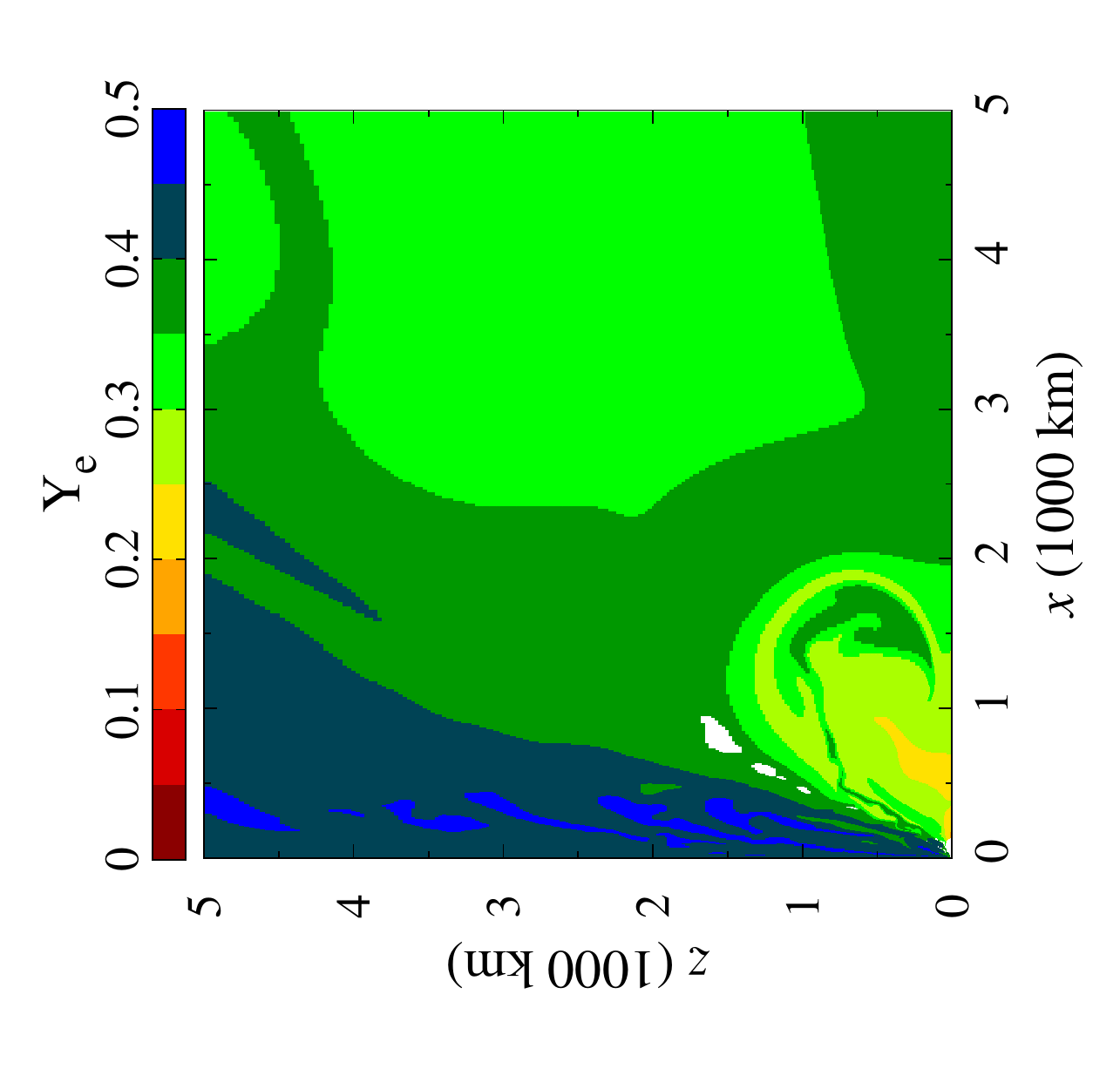}\\
        \end{tabular}    
\caption{
  \label{fig:B03_models}
  Density \blue{distribution} and electron fraction distribution in the post-merger ejecta, obtained by
  viscous-radiation hydrodynamics simulations \cite{fujibayashi18}.
  }
\end{center}
\end{figure}

\begin{figure}
  \begin{center}
    \includegraphics[scale=0.8]{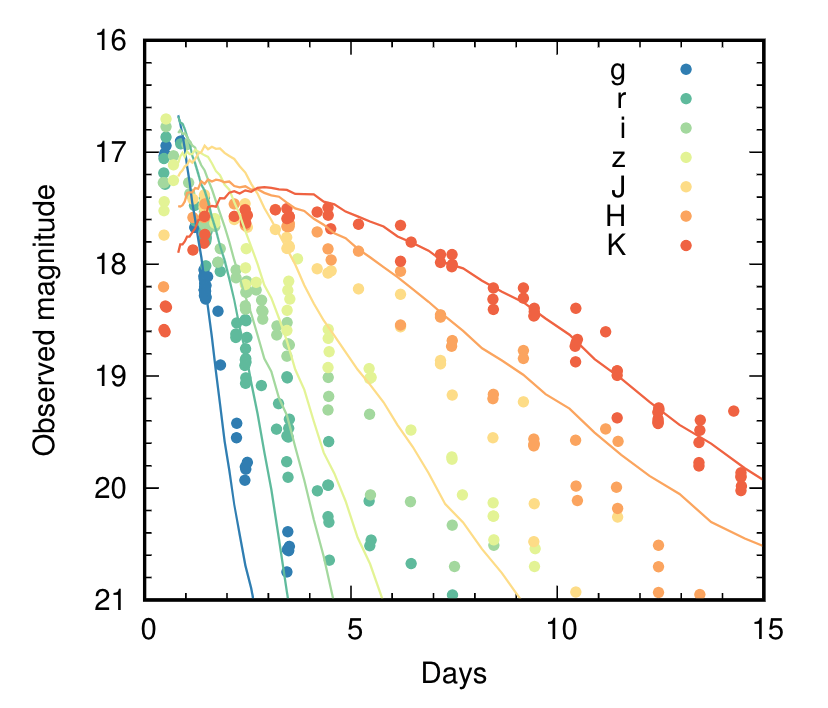}
\caption{
  \label{fig:B03_lc}
Multi-color light curves of GW170817/AT2017gfo (circles) compared with the results of two-dimensional radiative transfer simulations (solid lines) \cite{kawaguchi18,kawaguchi20}
  }
\end{center}
\end{figure}

\subsubsection{Advances of Theoretical Models}
\label{sec:B03_models}

The observed properties of AT2017gfo are consistent with
the expectation from numerical relativity simulations \cite{shibata17}.
In particular, to have an ejecta component with a moderate lanthanide fraction,
\blue{at least temporal} presence of hypermassive neutron star is suggested.
If the merger results in the prompt collapse to \blue{a BH},
there is no source of neutrino radiation,
and thus, almost entire ejecta are expected to be lanthanide-rich,
which is not consistent with the observations.
In this way, numerical relativity simulations play crucial roles
to connect the initial conditions estimated from GWs (mass and mass ratio)
with the final outcome of kilonova.

\blue{Mass ejection from NS mergers can be roughly divided into two phases: (1) dynamical mass ejection mainly by the tidal force and shock interaction in the dynamical time scale and (2) subsequent post-merger mass ejection from the accretion torus.}
Observations of GW170817/AT2017gfo highlighted the importance of
post-merger mass ejection from NS mergers.
However, the detailed simulations have been difficult.
We have performed long-term general relativistic neutrino radiation hydrodynamics simulations, by taking into account the effects of viscosity \cite{fujibayashi18}. We find that post-merger ejecta ($> 10^{-2} M_{\odot}$) dominate over the dynamical ejecta ($< 10^{-2} M_{\odot}$). Thanks to the inclusion of neutrino transfer, we could also provide the distribution of electron fraction in the ejecta. We showed that the ejecta have high electron fraction ($> 0.25$), which means that post-merger ejecta is lanthanide-poor (Fig.~\ref{fig:B03_models}).

Depending on the mass and mass ratios of the NS merger, various ejecta components with different properties (mass, velocity, and electron fractions) are expected. Therefore, it is important to study the observable outcome using the results of numerical relativity.
We have performed multi-dimensional radiative transfer simulations
by consistently taking the interplay of multiple ejecta components into account \cite{kawaguchi18}.
We showed that AT2017gfo can be naturally explained by the geometry predicted by numerical relativity simulations (Figure \ref{fig:B03_models}).

We then extended our radiative transfer simulations for various initial conditions. Using the updated atomic data \cite{tanaka20}, Kawaguchi {\it et al.} (2020) \cite{kawaguchi20} showed that the properties of kilonova should have diversity depending mainly on the total mass of the system. Since this is multi-dimensional simulations, we can also predict the variety of the luminosity for different viewing angles. It is shown that optical emission is greatly suppressed when NS merger is observed from the equatorial plane while infrared emission is almost unchanged. These simulations will be useful to facilitate future EM follow-up observations.

\begin{figure}
  \begin{center}
    \includegraphics[scale=1.0]{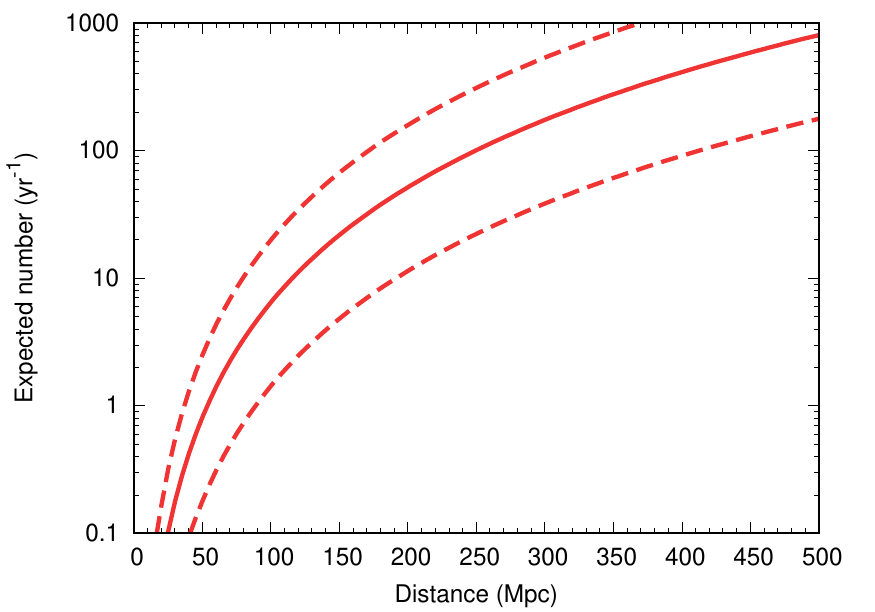} 
\caption{
  \label{fig:B03_future}
  Expected number of NS merger as a function of distance assuming the event rate estimated from the O2 run \cite{LIGOScientific:2018mvr}. The dashed lines show the range for 90\% confidence level.
  }
\end{center}
\end{figure}

\subsubsection{Future Prospects}
\label{sec:B03_future}

Our observational and theoretical results indicate that NS mergers synthesize a wide range of $r$-process elements. With the current estimate of the event rate and mass ejection, NS merger can explain the total amount of $r$-process elements in our Galaxy \cite{rosswog17,hotokezaka18}. However, the event rate is still largely uncertain. Also, we still do not fully understand that the mass ejection is universal or not. Varieties in the mass ejection and nucleosynthesis are expected by numerical simulations. Therefore, it is important to observe more NS mergers to accurately derive the event rate and to obtain the general picture of the mass ejection and nucleosynthesis.

The sensitivity of GW detectors are still improving.
In the third observing run (O3), a typical sensitivity corresponds to the range of the NS merger up to 130Mpc (LIGO Livingstone), 110Mpc (LIGO Hanford), and 45Mpc (Virgo). Thanks to these high sensitivity, two reliable GW events including at least one NS have been reported. However, due to the poor positional localization or large distances, any EM counterpart has not been identified.
We refer the reader to Sasada {\it et al.} (2021) \cite{sasada21} and Ohgami {\it et al.} (2021) \cite{ohgami21} for our observing effort in O3.
The situation will be improved when the sensitivity of Virgo becomes comparable. Furthermore, when KAGRA joins the network with a comparable sensitivity, positional localization will be greatly improved (down to a few deg$^2$ for the signals such as GW150914, \cite{gaebel17}).


Figure \ref{fig:B03_future} shows the expected event number per year as a function of the distance. Once all the sensitivity of GW detectors reach $\sim 200$Mpc, we expect $> 10$ events every year. Kilonova associated with such events is expected to be faint, $\sim 22$ mag in optical within 2-3 days after the merger. Therefore, observations with Subaru/HSC will play important roles. Also, deep near-infrared imaging and spectroscopic observations with large-aperture telescopes, such as Subaru, Keck, Gemini, and upcoming TAO 6.5m telescope \cite{doi18}, will be important for firm identification of kilonova. Such multi-wavelength observations will provide us with unique information of \blue{$r$-process} nucleosynthesis.


%% file: ptepGW_sectionC01.tex

\subsubsection{The CCSN GW signatures}

 Unlike the GW signals from compact binary coalescence where a template-based search is best suited (e.g., \cite{rasio99}), 
 gravitational waveforms from CCSNe are essentially of stochastic nature. This is because the waveforms are affected by
turbulence in the postbounce core, which is governed by
the multi-D, non-linear hydrodynamics.
In order to clarify the GW emission mechanisms, extensive numerical simulations have been done so far 
in different contexts (e.g., \cite{Dimmelmeier08,Scheidegger10,CerdaDuran13,Ott13,Yakunin15,KurodaT14,pan18,viktoriya18} and
 \cite{Kotake13} for a review). 
For canonical supernova progenitors \cite{Heger05}, core rotation is generally
 too slow to affect the dynamics (e.g., \cite{takiwaki16,summa18}).
For such progenitors, the GW emission takes place 
 in the postbounce phase, 
 which is characterized by prompt convection, neutrino-driven 
convection, proto-neutron star (PNS) convection, the standing accretion shock instability (SASI, \cite{thierry15}), and the $g$(/$f$)-mode 
oscillation of the PNS (e.g., \cite{EMuller97,EMuller04,Murphy09,Kotake09,BMuller13,viktoriya18}).
 
 The most generic GW emission feature seen in recent 
self-consistent 3D CCSN models 
 is the one from the PNS oscillation \cite{kuroda16,andresen17,radice19,alex19}.
 The characteristic GW frequency increases almost monotonically 
with time due to an accumulating accretion to the PNS, which ranges 
approximately from $\sim 100$ to $1000$Hz.
On the other hand, the typical 
frequency of the SASI-induced GW signals is concentrated in 
 the lower frequency range of $\sim 100$ to $260$Hz and persists when the SASI 
dominates over neutrino-driven convection \cite{kuroda16,andresen17,andresen18,radice19}.
 The detection of these distinct GW features  is considered as the key to infer which one is working more predominantly
 in the preexplosion supernova core, namely neutrino-driven 
convection or the SASI \cite{andresen17}. 

\begin{figure*}[htbp]
\begin{center}
  \includegraphics[width=1.0\linewidth]{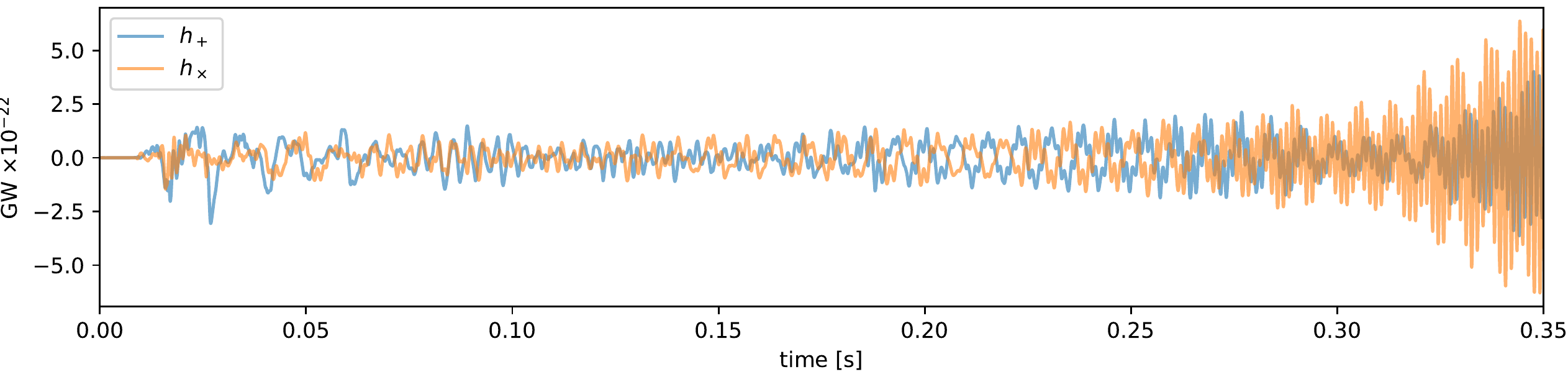}
  \caption{An example gravitational waveform of the $+$ (blue line) and $\times$ (orange line) mode from 
  a 3D-GR CCSN model of $15 M_{\odot} $ star (model SFHx in \cite{kuroda16}). 
   The time is measured after core bounce (:$t=0$). A source distance of $10$kpc is assumed. The waveform was extracted 
via a standard quadrupole formula with GR corrections \cite{KurodaT14}. Since the waveform is weakly dependent
 on the source orientation, an unbiased direction is chosen (e.g., from the north pole ($(\theta,\phi)=(0,0)$). This figure is taken from \cite{kawahara18}.
    \label{f1}}
\end{center}
\end{figure*}
Shown Fig.~\ref{f1} is an example waveform taken from a 3D CCSN simulation in \cite{kuroda16}. For the model, the hydrodynamics evolution is self-consistently 
followed in full general relativity (GR), starting 
from the onset of 
core-collapse of a $15M_\odot$ star \cite{WW95}, through core bounce, up to $\sim$ 
350ms after bounce. As consistent with the outcomes from recent 3D models (e.g., \cite{andresen17,radice19}), the 
hydrodynamic evolution and the associated GW waveform is characterized by the prompt convection phase shortly after bounce
($T_{\rm pb}\lesssim 20$ms with $T_{\rm pb}$ the postbounce time, shown simply as "time" in Fig.~\ref{f1}),
 then the linear (or quiescent) phase,
 which is followed by the non-linear phase ($T_{\rm pb}\gtrsim 140$ms) when the vigorous SASI activity (as well as the growing GW amplitudes) was observed for the model  (see \cite{kuroda16} for more details). The dominance of the SASI over neutrino-driven convection 
   persists over $140\mbox{ms}\lesssim T_{\rm pb} \lesssim 300$ms, after which neutrino-driven convection dominates over the SASI.
  
In order to discuss the detectability
 of the CCSN GW signals (like of Fig.~\ref{f1}), previous 
 studies have traditionally relied on a GW spectrogram analysis, aiming to specify the GW feature by the excess power in the time-frequency domain (by taking the square norm of the short-time Fourier transform). However, the spectrogram analysis has a trade-off relation between the time and frequency resolution, leading to a limited time-frequency localization. Because of this drawback, some features could have been 
potentially overlooked in the previous spectrogram analysis in the context of CCSN GWs. Recent developments in the time-frequency analysis (TFA) have yielded various time-frequency representation alternatives to the spectrogram. In particular, quadratic time-frequency representations, such as the Wigner--Ville distribution and its modified forms, enable us to perform high-resolution TFA (e.g., \cite{cohen95,stankovic2013time,Boashash2015}). 

\begin{figure*}[htbp]
\begin{center}
  \includegraphics[width=0.49\linewidth]{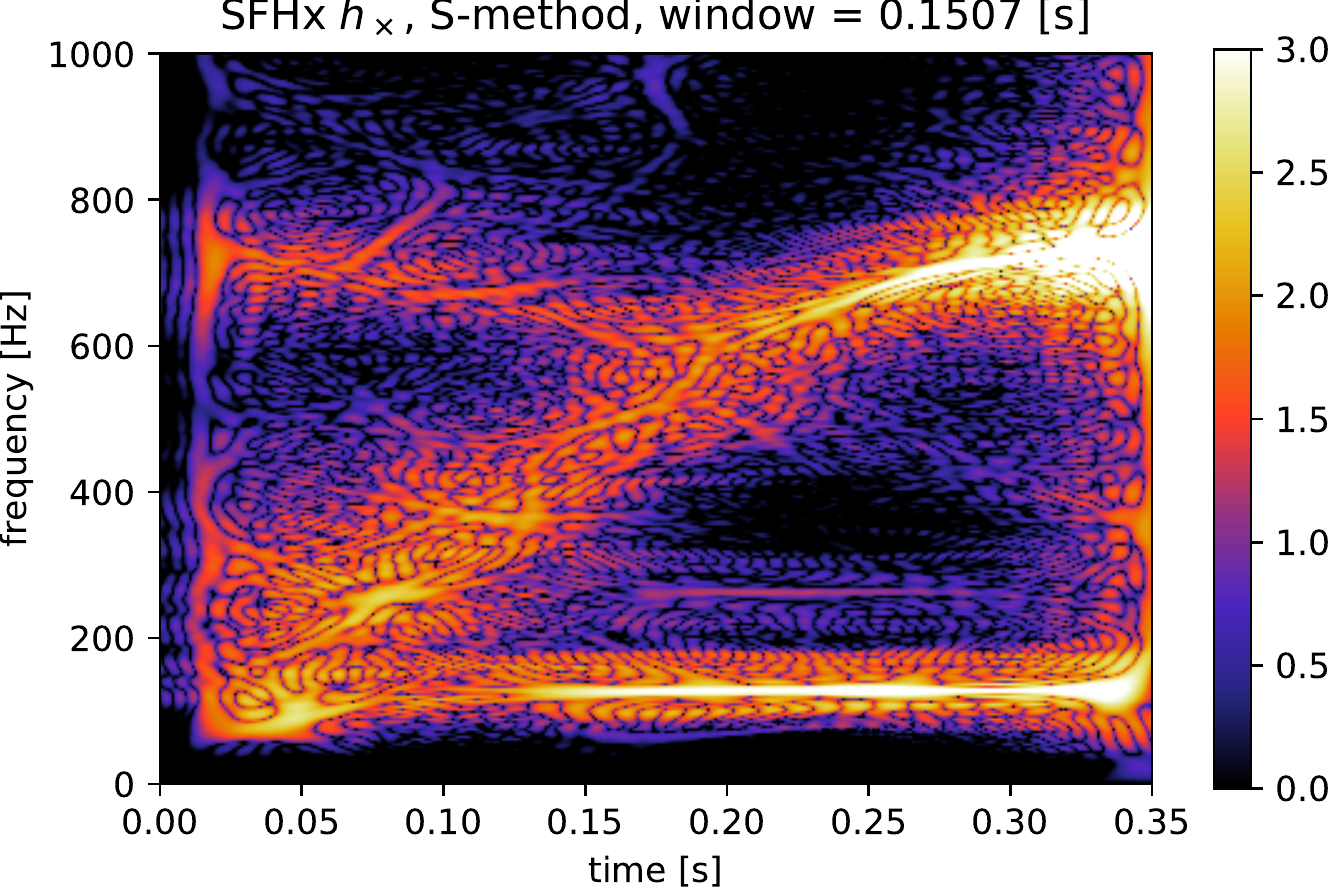}
   \includegraphics[width=0.49\linewidth]{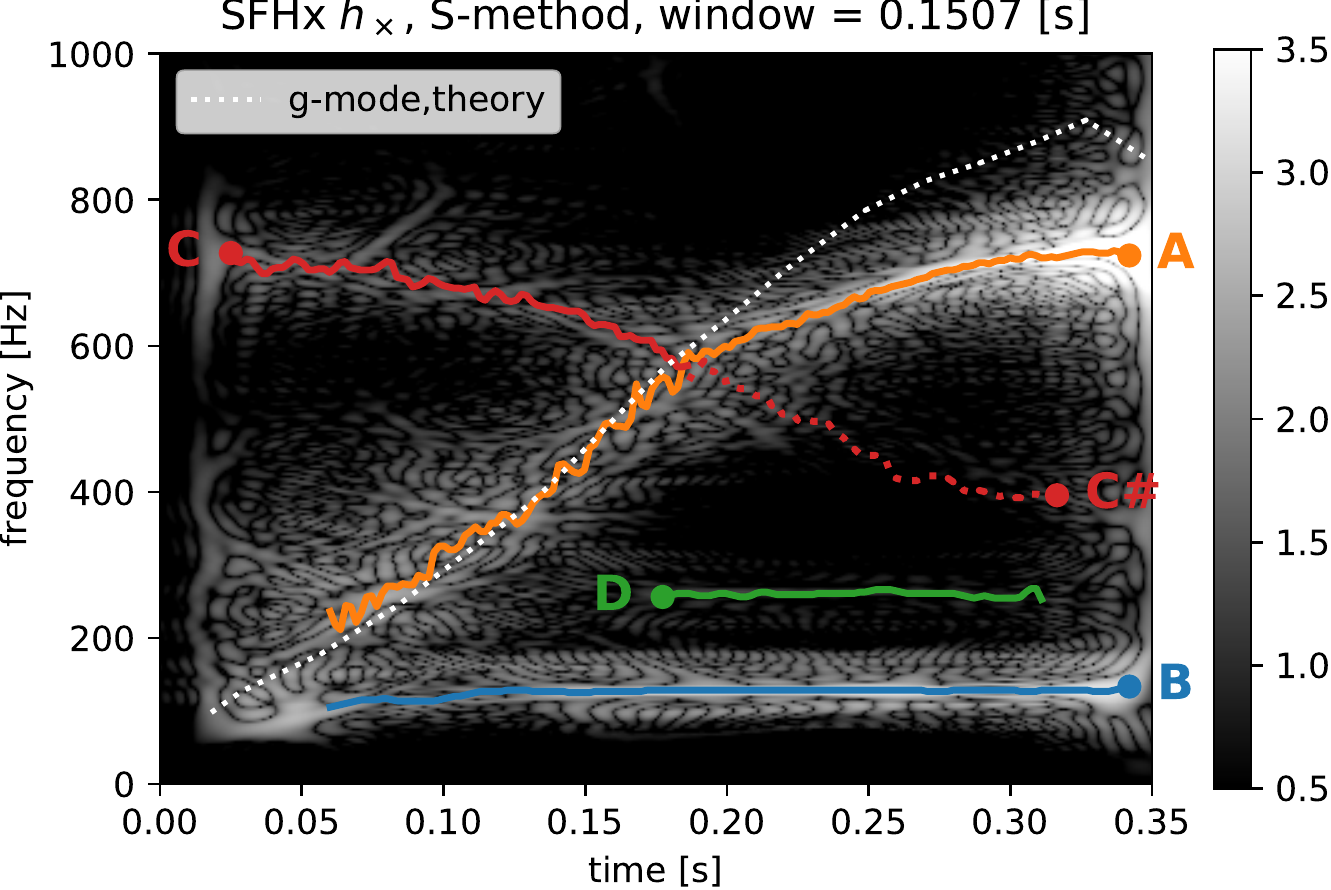}
  \caption{The left panel is the time-frequency representation of the sample waveform (Fig.~\ref{f1}),
   which is extracted by the S-method. The right panel shows the GW mode identification of the spectrogram based on the instantaneous frequency identification method via the S-method (see \cite{kawahara18} for more details). In the right panel, the white dashed line represents the theoretical prediction of the peak GW frequency of the PNS $g$-mode oscillation \cite{BMuller13}. Note that the deviation of mode A from the white dashed line (especially in the later postbounce phase) was previously observed in the literature \cite{andresen17,viktoriya18}. These figures are taken from \cite{kawahara18}.
   \label{f2}}
\end{center}
\end{figure*}

Recently, the S-method utilizing the Wigner--Ville  distribution was applied in \cite{kawahara18}, which  successfully extracted several key (CCSN) GW features in the time-frequency domain (left panel of Fig.~\ref{f2}, where the waveform (Fig.~\ref{f1}) was used). In comparison with the conventional TFA 
based \blue{on} the Fourier transform (e.g., Fig.~1 in \cite{kuroda16}), the S-method significantly improves the sharpness of the modes. Furthermore by defining the instantaneous frequency (IF) of the excess in the time-frequency domain (corresponding to the bright-color regions in the 
left panel of Fig.~\ref{f2}, see \cite{kawahara18} for more details),  five modes (A B, C, $\mathrm{C}^\#$, and D, the right panel of Fig.~\ref{f2}) were identified. 

Mode A in the right panel of Fig.~\ref{f2} corresponds to the PNS $g$(/$f$)-mode oscillation, which is excited in the vicinity of the PNS surface by the downflows and/or directly originates from the deceleration of infalling convective plums \citep{BMuller13}. Modes B and C are quasi-static modes, in the sense that these two modes barely change with time in the time-frequency domain. Mode B ($f_B\sim$ 130Hz) is originated from SASI, the frequency of which is twice of the SASI frequency ($f_{\rm SASI} \sim$ 65Hz) because of the quadrupole nature of the GW emission (see also \cite{andresen18}). Mode D is the overtone of the mode B ($f_D \approx 2 f_B$), which is most likely to come from the PNS core oscillation (see Fig.~8 in \cite{kawahara18}). Mode C intersects mode A at $T_{\rm pb} \sim 180$ms, after which this mode is denoted as $\mathrm{C}^\#$. Although yet to be investigated elaborately, modes C and $\mathrm{C}^\#$ are likely to arise in the innermost region ($\sim10$km) \cite{alex18}. Note that the typical frequency of the SASI-induced GWs is in the range of $\sim$100 -- 260Hz (e.g., modes B and D in the left panel of Fig.~\ref{f2}),
 which is in the best sensitivity range of the currently running interferometers.
For a Galactic event, the signal-to-noise ratio (SNR) of the GW signals predicted
in the most recent 3D CCSN models was estimated in the range of $\sim$ 4 -- 10 for the advanced
LIGO \cite{andresen17}. With the third-generation detectors on line
(e.g., Cosmic Explorer and Einstein Telescope), these signals would be never missed
for the CCSN events throughout the Milky way.

\subsubsection{Circular Polarization of CCSN GWs}
Recently, yet another GW feature has been reported, which is circular polarization of the CCSN GWs. The importance of detecting the GW circular polarization was first pointed out 
 by \cite{Hayama16} in the context of rapidly rotating core-collapse. More recently, Hayama {\it et al.} (2018) \cite{Hayama18} presented analysis of the GW
 circular polarization using results from core-collapse of a non-rotating $15$$M_{\odot}$ star \cite{kuroda16}, the waveform of which is already shown in Fig.~\ref{f1}.
 
 \begin{figure}
\begin{center}
\includegraphics[width=1.0\linewidth]{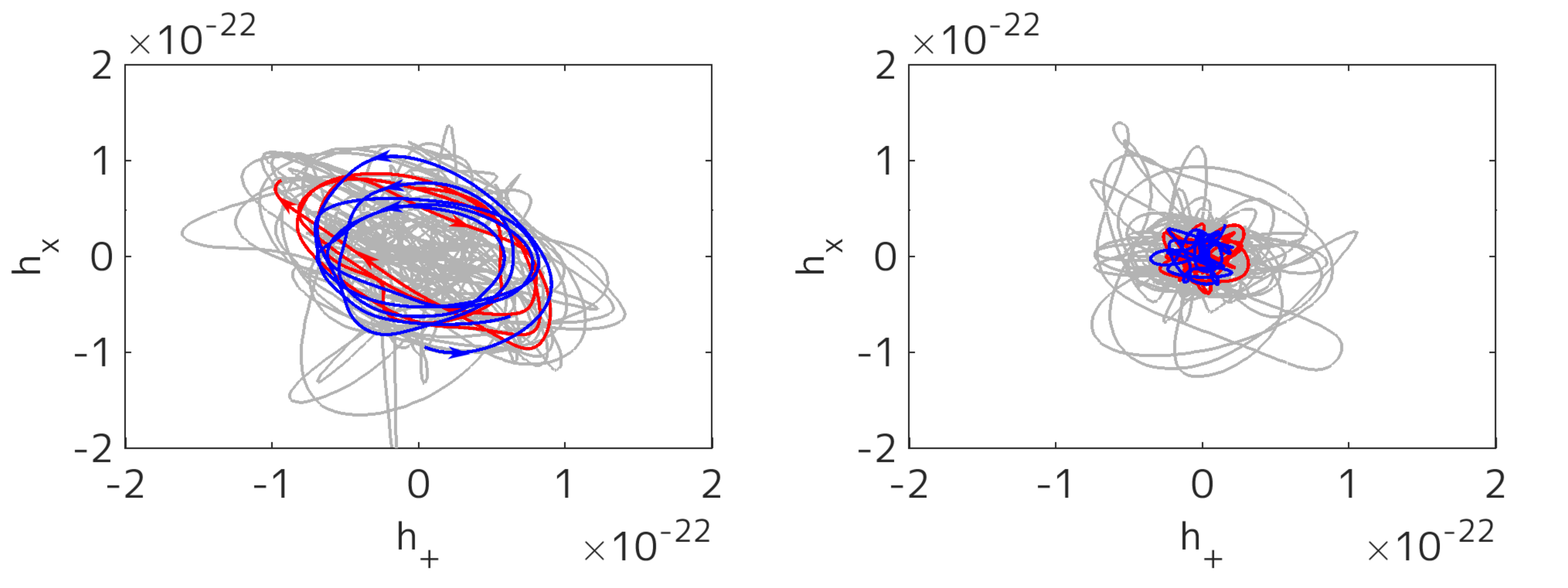}
\caption{Shown are the trajectory of the GW circular polarization on the 
$h_{+}-h_{\times}$ plane of models SFHx (left panel) and TM1 (right panel) of 3D-GR simulations using different equations of state (see text). The two characteristic epoch with the right-handed 
($140\mbox{ms}\lesssim T_{\rm pb} \lesssim 200$ms) and left-handed 
($220\mbox{ms}\lesssim T_{\rm pb} \lesssim 320$ms) polarization are highlighted with 
the red and blue color. The gray line denotes the whole 
trajectory during the simulation time. The source distance is assumed at 10kpc. These figures are taken from \cite{Hayama18}.
}
\label{f3}
\end{center}
\end{figure}
 
 Figure \ref{f3} visualizes the time evolution of the GW circular polarization (hereafter, CP) of models SFHx (left panel) and TM1 (right panel) from \cite{Hayama18}. Note that SFHx \cite{SFH} and TM1 denotes the name of the nuclear equation of state (EOS) employed in the 3D-GR simulation \cite{kuroda16}. The key difference is that the softer EOS (SFHx) makes the PNS radius and the shock radius at the shock-stall more compact than those of TM1. This leads to \blue{much} stronger activity of the SASI
  for SFHx compared to TM1.
  
  In fact, one can see a clearer
 polarization signature for SFHx (left panel of Fig.~\ref{f3}) characterized by the bigger
 GW amplitude with the right-handed (red line) and left-handed mode (blue line)
 than those for TM1 (right panel).
 Note in each panel that the two SASI-dominant phases of SFHx are colored by red or blue (see also bottom panels  
of Fig.~1 in \cite{Hayama18}). The non-axisymmetric flows associated with spiral SASI give rise to the circular polarization.
 Before $T_{\rm pb} \sim 140$ms, the CP is small because  the SASI activity is still weak. But, after $T_{\rm pb} \sim 140$ms when the (spiral) SASI activity begins to be vigorous (e.g., \cite{kuroda16}), the right-handed CP firstly emerged, which is followed by the left-handed CP 
(the blue line) until $T_{\rm pb} \sim 320$ms, after which
neutrino-driven convection dominates over the SASI.


To explore the detectability of the CP, Monte Carlo simulations of the coherent network analysis \cite{hayama15} were performed for the signal reconstruction, where the network of LIGO Hanford, LIGO Livingston, VIRGO and KAGRA was considered.
 They reported the enhancement of the signal-to-noise ratio of the GW circular polarization (${\rm SNR}_{\rm CP}$) relative to that of the GW signal itself (${\rm SNR}_{\rm TF}$). Although this is likely because the Gaussian noise
 has little component of the circular polarization, more detailed analysis is required to 
  draw a robust conclusion whether or not
the GW CP could provide a new probe to 
clarifying the CCSN inner-workings such as the SASI and the \blue{PNS oscillations} (see \cite{Hayama18} for more detail).
 

\subsubsection{Rapid Rotation: GW and Neutrino signals from an exploding $27M_{\odot}$ star}

Rapid rotation in the iron core
(with the initial rotation frequency typically greater than $0.5$rad/s)
leads to significant rotational flattening of the collapsing and bouncing core,
 which produces a time-dependent quadrupole (or higher) GW emission (e.g.,
 \cite{ott_rev} for a review).
For the bounce signals having a strong and characteristic 
signature, the iron core must rotate enough rapidly\footnote{Although rapid rotation
  and the strong magnetic fields in the core
 \blue{are} attracting great attention as the key to solve
  the dynamics of collapsars and magnetars, one should keep in mind that recent
 stellar evolution calculations predict that such an extreme condition can be
  realized only in a special case \cite{woos06} ($\lesssim$ 1\% of massive star population).}. 
 The GW frequency associated with the rapidly rotating collapse and bounce is in
 the range of $\sim 600 - 1000$Hz \cite{Dimmelmeier08}.
 A current estimate based on a coherent network analysis using predictions from a set of 3D models
 shows that these GW signals could be detectable up to
 about $\sim 20$kpc for the most rapidly rotating model (\cite{hayama15}, see
 also \cite{powell19,gossan16}).

\begin{figure}
\begin{center}
\includegraphics[width=0.48\linewidth]{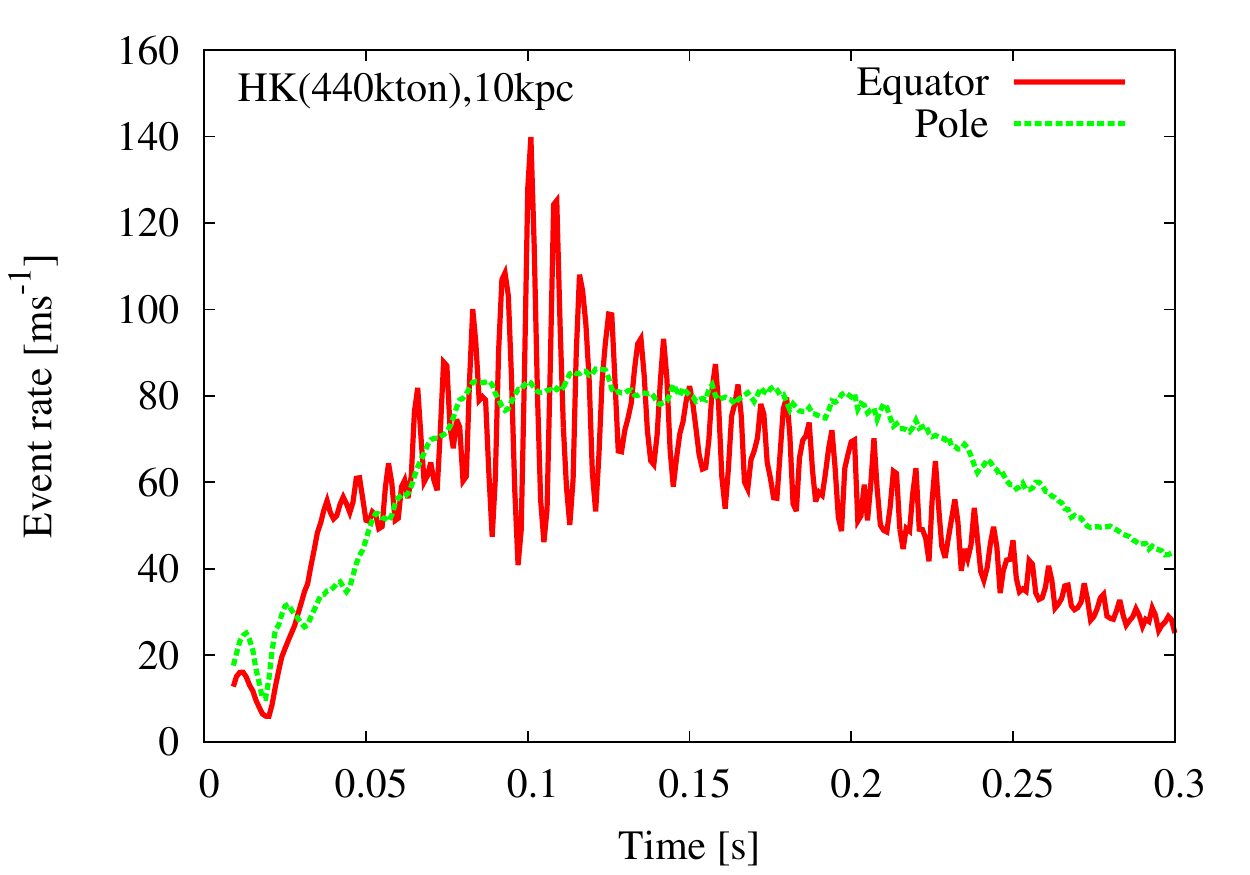}
\includegraphics[width=0.48\linewidth]{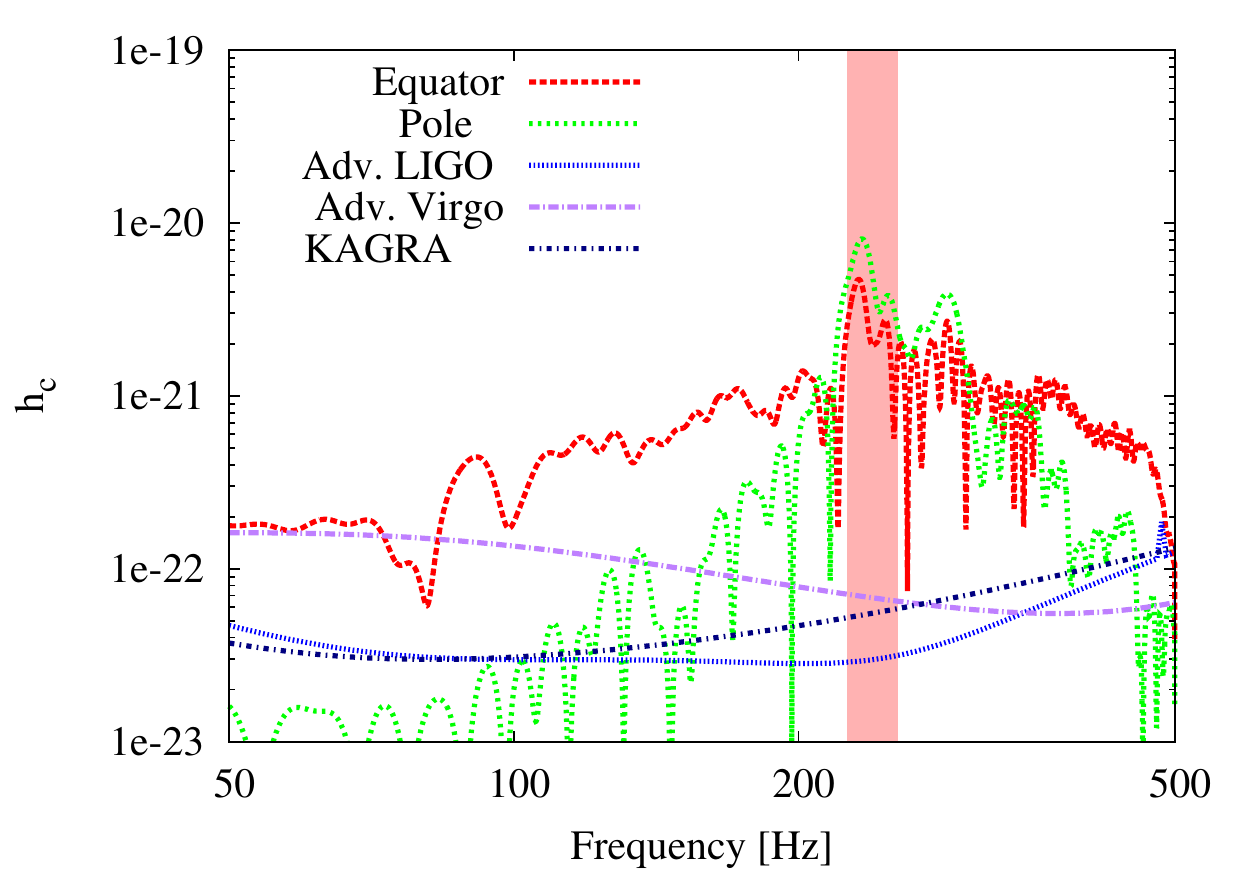}\\
\end{center}
\vspace{1cm}
 \caption{The left panel shows detection rates of $\bar{\nu}_e$ at 10kpc for a rapidly rotating
 $27 M_{\odot}$ model as a function of time after bounce \cite{takikotake}. 
The red and green lines correspond to the 
 event rates (per 1ms bin) for an observer along the equator 
 and the pole (e.g., parallel to the rotation axis), respectively.
A quasi-periodic modulation (red line) corresponds to $\sim 120$Hz (red lines)  seen for the observer along the equator.
The right panel shows the 
characteristic GW spectra relative to the sensitivity curves of advanced LIGO,
 advanced Virgo (\cite{advv})
and KAGRA (\cite{aso13}). The peak 
GW frequency (the vertical red band)
is about twice of the neutrino modulation frequency.
}\label{f5}
\end{figure}

Figure \ref{f5} shows the neutrino and GW signals   \cite{takikotake} obtained 
in 3D core-collapse supernova simulation 
of a rapidly rotating 27$M_{\odot}$ star that is exploding due to the growth
of the so-called low-T/$|W|$ instability \cite{takiwaki16}. The time modulation seen in the left panel corresponds
 to the neutrino light-house effect where the spinning of strong neutrino emission
 regions around the rotational axis leads to quasi-periodic 
modulation in the neutrino signal. 
 Depending on the observer's viewing angle,
 the time modulation will be clearly detectable in 
 IceCube and 
the future Hyper-Kamiokande. The GW emission is also anisotropic 
where the GW signal is emitted, as previously identified (see \cite{Kotake13} for
a review), 
most strongly toward the equator at rotating core-collapse 
and bounce, and the non-axisymmetric instabilities 
in the postbounce phase lead to stronger GW emission 
toward the spin axis. The right panel in Fig.~\ref{f5} shows
that these GW signals can be a target of LIGO-class detectors for a Galactic event.
  The origin of the postbounce GW emission (e.g., 
  the bar-mode ($m=2$) deformation of the PNS due to the low T/$|W|$ instability), naturally explains 
 the reason that the peak GW frequency is about twice of the neutrino 
modulation frequency. These results demonstrate that the simultaneous 
detection of the rotation-induced neutrino and GW signatures 
could provide a smoking-gun signature of a rapidly 
rotating PNS at the birth (see also \cite{walk18,walk19}).

\subsubsection{Hydrodynamics and GW signals from a BH-forming massive star}

Successful observations of 
GWs from binary black holes (BBHs) by the LIGO-VIRGO collaborations are now posing a new challenge; what is the origin of the massive BHs ?
One of the most plausible scenarios is a binary stellar evolution in a low-metallicity environment 
(see \cite{abbott16} for a review).
 It has been proposed that two massive stars 
in the approximate range of $40$ to $100 M_\odot$ 
lead to the formation 
of a massive helium core after experiencing the Roche-lobe 
overflow and common envelope phase 
(e.g., \cite{bel10,Langer12,Kinugawa16} for collective 
references therein).
The gravitational collapse of the massive core ($\sim30M_\odot$)
 could account for some of the relevant BH mass ranges (at least in 
 the high-mass end) in the GW events, although the formation path to
 the massive core and further to the BH 
is still very uncertain due to the complexity of the binary 
evolution and the fallback dynamics (e.g., \cite{Fryer99}).

In order to clarify the formation process of the BH,
one requires GR neutrino 
radiation-hydrodynamics core-collapse simulations of such massive stars in 3D.
 Due to the high numerical cost,
 most of the previous studies with BH formation have been done assuming
 spherical symmetry (1D) (e.g., \cite{janka16,Kotake12} and collective references
 therein).
 In the context of multi-D simulations with multi-energy neutrino transport, Ref. \cite{Pan17} reported 1D and two-dimensional (2D) core-collapse simulations of a solar-metallicity 40$M_\odot$ star
 using two-flavor IDSA scheme (\cite{idsa}) and 
a post-Newtonian gravity to include GR effects. 
 More recently,  a BH-forming 3D-GR simulation of a 
zero-metallicity 40$M_\odot$ star with an approximate neutrino transport (FMT) scheme was reported in \cite{chan18}.
It is only recently that the first 3D-GR simulation (using a 70$M_{\odot}$ star \cite{Takahashi14}) with detailed neutrino transport \cite{shibata11} following the dynamics up to BH formation was published \cite{Kuroda18}. In the simulation, the evolution equations of metric, hydrodynamics are solved
 based on the Baumgarte-Shibata-Shapiro-Nakamura formalism \cite{Shibata95,Baumgarte99}
 and the evolution of 
neutrino radiation field based on the multi-energy M1 scheme 
\cite{Kuroda18}. Casting the helium core of $\sim 31$$M_{\odot}$,  the zero-metallicity 70$M_{\odot}$ star was chosen as one of the possible progenitor candidates of the BBHs.

\begin{figure}
\begin{center}
\includegraphics[width=120mm]{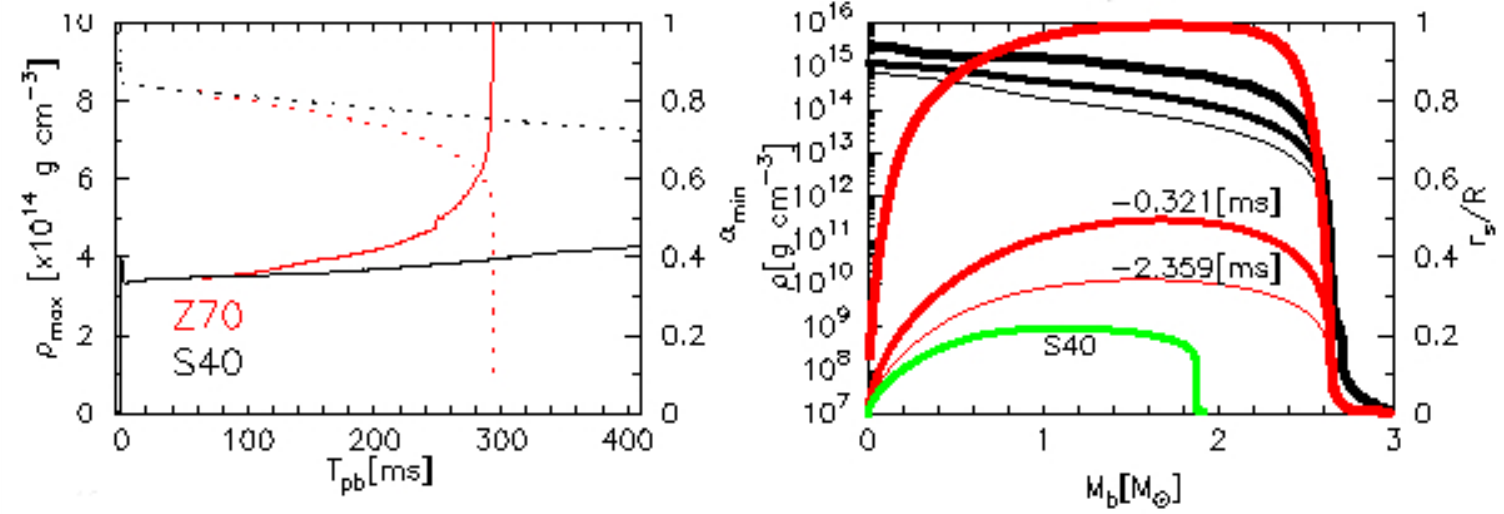}
\caption{The left panel shows time evolution of the maximum rest-mass density
  (solid line) and minimum lapse function (dotted line) for a 70$M_\odot$ star (denoted as Z70, red line) and 
  a 40$M_\odot$ star (as S40, black line), respectively.
The right panel shows angle-averaged profiles of the (rest-mass) density (black line) 
and the ratio $r_{\rm s}/R$ (red lines, see text) for Z70 as a function of the enclosed baryon mass for three representative time slices near the final simulation time.  "-0.321" and "-2.359"ms denote the 
time before the final simulation time 
($T_{\rm fin} = 293$ms after bounce). The green line corresponds 
to $r_{\rm s}/R$ for S40. These figures are from \cite{Kuroda18}. \label{f6}}
\end{center}
\end{figure}

The left panel of Fig.~\ref{f6} shows temporal evolution of the maximum 
 density $\rho_{\rm max}$ (solid lines) and the minimum lapse 
$\alpha_{\rm min}$ (dotted lines) for the 70$M_\odot$ star (denoted as Z70, red line) and the 40$M_\odot$ star (as S40, black line), respectively. Note that S40 is shown for comparing with previous results. The compactness parameter
 of S40 is much smaller ($\xi_{2.5}\sim0.26$) than that of Z70. The maximum density at bounce 
($\rho_{\rm max}\sim 4 \times 10^{14}$g$\,$cm$^{-3}$) is quite similar
between Z70 and S40. After bounce, the increase of 
the maximum density of Z70 (red solid line) is significantly faster 
than that of S40 (black solid line). 
In both Z70 and S40, the minimum lapse 
(dotted lines) shows a gradual decrease after bounce. At around 
$T_{\rm fin} = 293$ms after bounce, it shows a drastic drop to $\alpha_{\rm min}=0.0645$ 
for Z70, indicating that the PNS core starts to collapse
rapidly toward a BH formation. 

The right panel of Fig.~\ref{f6} explains
 how $T_{\rm fin}$ is related to the 
BH formation time. Shown in the panel is the profiles of 
 (angle-averaged) density (black lines) and a diagnostics to
 measure the BH formation as 
a function of the enclosed baryon mass $M_{\rm b}$ at some 
representative snapshots near $T_{\rm fin}$
 for Z70 (red lines) and at $T_{\rm pb} = 400$ms 
for S40 (green line). As a BH-formation diagnostics, 
 the ratio of $r_{\rm s}/R$ is shown, where $r_{\rm s}$ and $R$ denotes the Schwarzschild radius and the radial coordinate, respectively.
 One can see that the maximum $r_{\rm s}/R$ 
is $\sim0.3$ at 2.359ms before $T_{\rm fin}$ (the thin red line labelled 
 by "-2.359 [ms]") and rapidly increases with time, approaching to unity (precisely,
 $0.932$) at $T_{\rm fin}$ (thickest red line), which was
 judged as the epoch of the BH formation in \cite{Kuroda18}. 
It should be noted that for the unambiguous definition of the BH formation, one 
 requires the implementation of the so-called apparent
 horizon finder (e.g., \cite{thorn04}) in numerical relativity simulation, which is still left as a future work.
 
 At the (fiducial) BH formation time, the mass and the radius is 
$M_{\rm b(g),BH}\sim2.60(2.51)$$M_\odot$ and 
$R_{\rm iso}\sim4$km, respectively.
By contrast, S40 shows significantly less compact structure (green line)
 at the final simulation time ($T_{\rm pb} = 400$ms).
 The BH formation should occur much later, 
possibly when the mass shell at $R(M_{\rm b}=2.6 M_\odot)\sim10^9$cm 
accretes to the stalled shock. 
Using the same EOS (LS220 \cite{LS}), this expectation is in line 
with \cite{Pan17}, 
\blue{which} reported the BH formation at $T_{\rm pb}\sim 700$ms 
and with \cite{chan18} at $T_{\rm pb}\sim 1$ s.

 The time evolution of the (angle-averaged) shock radii $R_{\rm s}$, the gain radius $R_{\rm g}$,
the ratio of the advection timescale to the neutrino-heating timescale
in the gain region $\tau_{\rm adv}/\tau_{\rm heat}$, and the mass in the gain region $M_{\rm gain}$ \blue{are} presented in Fig.~2 of \cite{Kuroda18},
respectively.
For model Z70, the shock revival was obtained after $T_{\rm pb}\, \gtrsim \, 260$ms.
At this time, the maximum temperature in the core 
 becomes as high as $T\sim100$ MeV at a slightly off-center region 
at $R_{\rm iso}\sim10$km (equivalently 
at $M_{\rm b}\sim1.0 M_\odot$).
Subsequently the high temperature region propagates outward in the mass 
coordinate, although spatially inward, due to the continuous mass 
accretion. The maximum temperature reaches $\sim 170$ MeV at $R_{\rm iso}\sim1$km 
($M_{\rm b}\sim1.4 M_\odot$) just before $T_{\rm fin}$. 
 In this second collapse phase to the forming BH, 
the high neutrino emission makes the heating timescale shorter
 than the competing advection timescale in the gain region. 
Aided by strong convection behind the shock, the stalled shock is revived
  at $T_{\rm pb}\,\gtrsim \,260$ms
 ($\tau_{\rm adv}/\tau_{\rm heat} \geq 1$). This also results in the increase 
in the gain mass (see the blue line) due to the shock expansion.

\begin{figure}
  \begin{center}
          \includegraphics[clip,width=70mm]{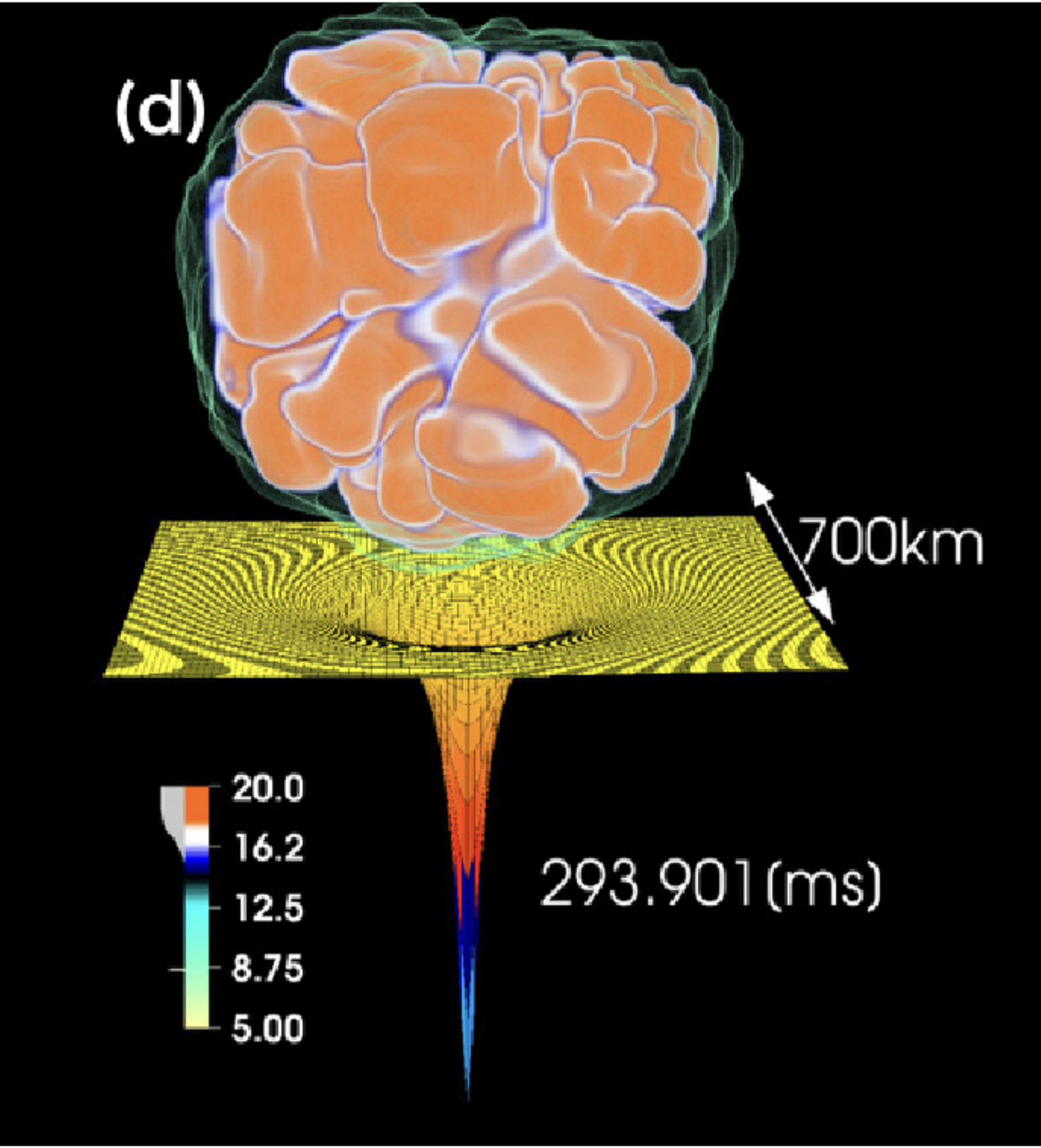}
          \caption{A snapshot of the entropy distribution 
            ($k_{\rm B}$ baryon$^{-1}$) for Z70 at $T_{\rm pb} \sim 294$ms just before
            the BH formation (\cite{Kuroda18}).
The sheet represents the lapse function ($\alpha$) on the $z=0$ plane. This figure is taken from \cite{Kuroda18}.
  \label{f7}}
  \end{center}
\end{figure}

Figure \ref{f7} visualizes a 3D hydrodynamics feature near at the
BH formation.
During the first $\sim160$ms after bounce, the neutrino heating is still weak 
and high entropy bubbles are yet to appear.
Only after $T_{\rm pb}\gtrsim 230$ms, the formation of high entropy plumes
($s\gtrsim 15 k_{\rm B}$ baryon$^{-1}$) was seen due to the intense neutrino heating from the hot PNS.
At this time, the mass in the gain region $M_{\rm gain}$ also starts to 
increase. The expansion of the 
(merging) high entropy plumes was observed, leading to the shock revival.
 The lapse function shows a steepest drop in the center 
(see the cusp in the plane of Fig.~\ref{f7}), which corresponds to 
 the BH formation. By expanding the shock radius into the spherical harmonics,
 we find that the deviation of the shock from spherical symmetry 
(in the low-modes $\ell = 1,2$) is less than $\sim2\%$. 
This clearly indicates that neutrino-driven convection dominates over the SASI in this case.

\begin{figure}
  \vspace{0.5cm}
  \begin{center}
          \includegraphics[width=40mm,angle=-90]{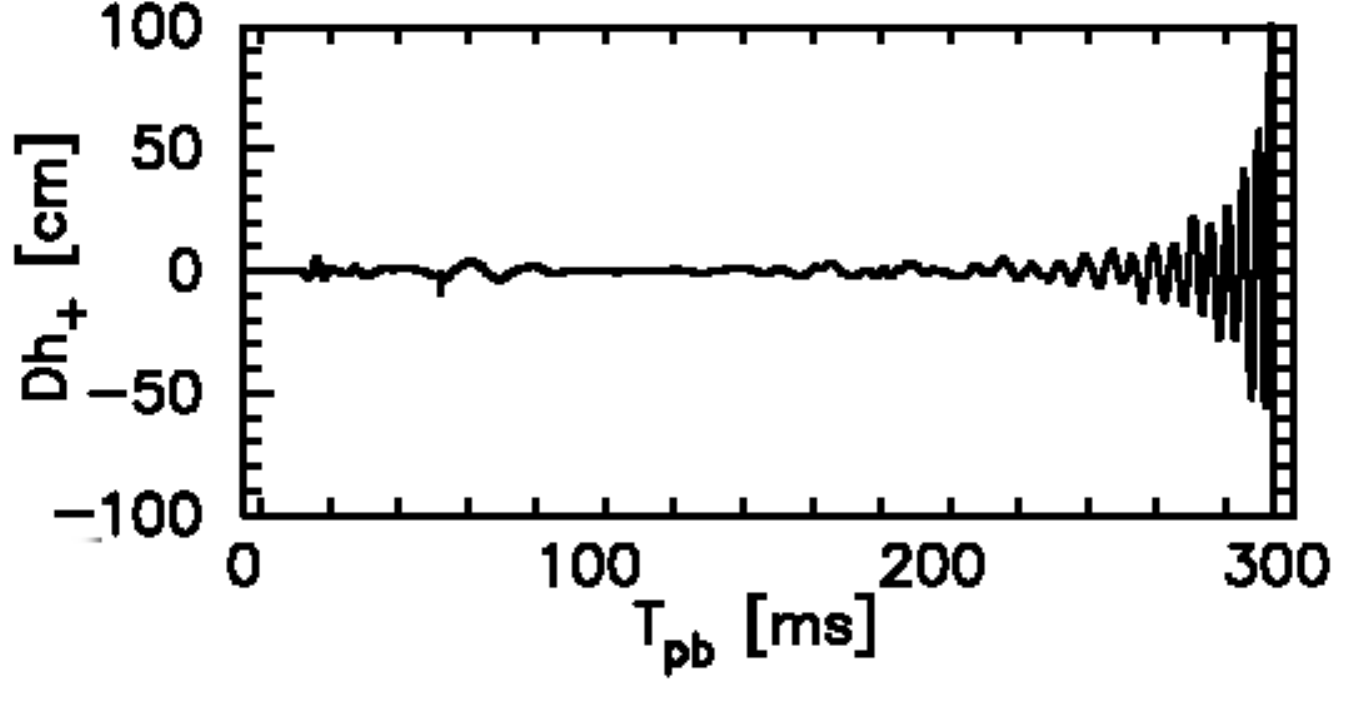}
          \caption{A gravitational waveform for the BH-forming core-collapse of a 70$M_{\odot}$ star. Note that $h_{+}$ and $D$ denote the GW amplitude of the $+$ polarization
 and the distance to the source, respectively. 
  \label{f8}}
  \end{center}
\end{figure}

Finally, Fig.~\ref{f8} shows the GW prediction for the 
Z70 star.
The waveform is extracted along positive $z$-axis via the quadrupole formula.
The GW amplitude \blue{(multiplied by the distance to the source)} stays at small value $\lesssim 10$cm before 
  the shock expansion occurs ($T_{\rm pb} \sim 260$ms).
The strong GW emission thereafter mainly originates from strong
 convection motion 
behind the shock.
Just before the final simulation time, the GW amplitude reaches $\sim100$cm, though only for a short duration.
Although more quantitative discussion is apparently needed,
 a rough estimate shows that the GW signal could be targeted 
  by third-generation GW detectors such as
the Einstein Telescope (ET) and the Cosmic Explorer (CE) (\cite{ET,CE}) for the source distance less than $\sim$ Mpc.
Regarding the neutrino signals, both electron type (anti-)neutrinos show decreasing
and plateau trend for $T_{\rm pb}\gtrsim 260$ms, whereas heavy-lepton neutrinos show a rapid increase both in
 the luminosity and energy. These features are consistent
  with those previously identified in 1D full-GR simulations with Boltzmann neutrino 
transport (\cite{matthias04}), which is due to 
rapid contraction of the PNS to the forming BH
(see also, \cite{Sumiyoshi07,tobias09}
). The detection
 of the short-live ($\sim 300$ms after bounce) neutrino signals are basically 
 limited to Galactic events (see \cite{mirizzi} for a review). However,
 further study would be needed to clarify the contribution of these BH forming massive stars to the prediction of diffuse neutrino supernova 
background (e.g., \cite{Lunardini09,horiuchi18}). 

\subsubsection{Short Summary and Future Prospects}
To summarize, each phase of a CCSN has a range of characteristic GW signatures and
 the GW polarization that can provide diagnostic constraints on the evolution
 and physical parameters of a CCSN and on the explosion hydrodynamics.
 The GW signatures described in this section commonly appear in the frequency range of $100$ - $1000$Hz in recent simulations of the CCSN.
 Among them, the PNS oscillatory modes are currently recognized as the model-independent
 GW signature.
The characteristic frequency ($f_p \propto M/R^2$) of the fundamental ($f/g$) modes is predominantly dependent
on the PNS mass ($M$) and the radius ($R$) \cite{bernhard13}.
In order to break the degeneracy, the detection of other eigen-modes ($p,w$ modes)
of the PNS oscillations is mandatory. In fact,  the GW asteroseismology of the PNS
has just started \cite{forne19,morozova18,sotani17}, the outcomes of which should reveal
the requirement of the next-generation detectors to detect the whole of the
eigen-modes (presumably extending up to several kHz) for the future CCSN event. 

 \blue{Regarding the SASI-induced GWs for a Galactic supernova source, they are predicted to be detectable by the advanced LIGO with the singal-to-noise ratio (SNR) in the range of $\sim 4$--$10$ by the most recent 3D CCSN models \cite{andresen17}. With the third-generation detectors (e.g. Cosmic Explorer and Einstein Telescope) online, these signals would never be missed for CCSN events throughout the Milky Way. A current estimate of SNR for GWs from rapidly rotating collapse and bounce based on predictions from a set of 3D models using a coherent network analysis
 shows that these GW signals could be detectable up to
 about $\sim 20$kpc for the most rapidly rotating model (\cite{hayama15}; see also Refs.~\cite{powell19,gossan16}).
 See a recent review by Ref. \cite{ernazar} for a more comprehensive review regarding the detectability of the CCSN GW signals by the future detectors. }

 After decades of progress, the CCSN theory is finally converging, but not without detailed numerical effort and much theoretical sweat. But at last not in the least, we have to draw a caution that the 
 current numerical results and the associated GW predictions that we have reported in this Section should depend
 on the next-generation calculations by which more sophistication can be made 
not only in the neutrino transport schemes (such as Boltzmann transport \cite{sumi12,harada19,nagakura19}), but also in the treatment of GR with the BH formation. Therefore we provide here only a snapshot of the moving (long-run) documentary film that records our endeavours for making our dream of "GW astronomy of the next Galactic CCSN" come true.

%% file: ptepGW_sectionC02.tex
One of the primary sources of near-field GWs, core collapse supernova explosions are among the most dramatic and important events to take place in the Universe. Agents of great destruction -- anything within tens of light-years is utterly annihilated by the dying star -- they are nevertheless responsible for the existence of life itself, for they (and the collisions of neutron stars they produce) are the sole source of all elements heavier than helium. Therefore, understanding these complex explosive events is necessary if we are to understand why we are here, and why the Universe looks the way it does today.  

Supernova neutrinos, famously observed from SN1987A, provide a unique and vital probe into the inner dynamics of these events. Released together with GWs during the initial stellar collapse, neutrinos and GWs are both certain to travel through any obscuring dust or gas and remain undiminished upon their arrival at \blue{the} Earth. Neutrinos also carry information regarding the end state of the star: for explosions within our galaxy, collapses into neutron stars or black holes, the eventual sources of far-field GWs, can be differentiated via observations of neutrino emissions.

Our goal is to make theory and experiment work together so \blue{that} we will be ready to make the best possible observations of the next galactic explosion and maximize our extraction of information on the explosion mechanism, progenitor, and nuclear physics ingredients.

In order to both predict and make sense of the neutrino signals from the next explosion in the Milky Way galaxy, the world's most advanced supernova numerical simulations are needed. Using nuclear data such as equation of state and neutrino reactions, and by solving the 6D Boltzmann equation in 2D/3D, we can provide detailed predictions of the time distribution and energy spectra of supernova neutrinos. 

To verify these complex computational models, we will enrich the famous \blue{Super-Kamiokande (SK)} detector with gadolinium [Gd] salt. Doing so will turn it into the world's most advanced supernova neutrino detector, capable of real-time tagging and identification of individual supernova neutrino interactions with nanosecond-scale time resolution. This will make the diffuse supernova neutrino flux from all past supernova explosions visible for the first time. Starting around the middle of 2020, a gadolinium-loaded SK will collect a steady stream of supernova neutrino data, the first such new data in over 30 years.

The new experimental measurements made possible by gadolinium will then be fed into the theoretical models, testing their predictions using real supernova neutrino data and allowing us to better prepare for the next nearby supernova explosion.  The models will be refined as needed and as indicated by the past supernova data. The gadolinium-loading of SK will also greatly improve the detector's response to a nearby supernova.  Therefore, both experimentally and theoretically, we will be well-prepared and ready for it.

\subsubsection{SK-Gd}

Water Cherenkov detectors such as Kamiokande, IMB, and SK, have been used for decades as effective detectors for neutrino interactions and nucleon decay searches.  While many important measurements have been made with these very large detectors, a major drawback has been their inability to efficiently detect the presence of thermal neutrons. 

If a water Cherenkov detector could be improved to observe the neutrons produced by the inverse beta process ($\bar{\nu}_e p \rightarrow e^+ n$), then backgrounds would be greatly reduced.  As a result, supernova neutrinos from explosions 35,000 times more distant than SN1987A could be seen by an improved SK, covering 50\% of the entire Universe. Perhaps more important for less distant explosions, neutron tagging of inverse beta events would facilitate the de-convolution of a galactic supernova's various signals, allowing a much more complete interpretation of the physics of the burst. In coincidence with a GW signal, detailed information concerning the explosion's dynamics would become even more valuable.

The key is to add 0.2\% by mass of a soluble gadolinium compound like gadolinium sulfate,  \mbox{Gd$_2$(SO$_4$)$_3$}, to the water. Doing so will make $>$90\% of the neutrons visible as a consequence of the gamma rays released by gadolinium's capture of thermalized neutrons.  This technique and the various new scientific advances it would make possible was first proposed in the Physical Review Letters article, ``GADZOOKS! Anti-neutrino spectroscopy with large water Cherenkov detectors''\cite{gad}. The publication of this paper introduced the concept of gadolinium-enhanced water Cherenkov detectors, a transformational technology with a strong impact on the physics community.  It provided a clear and cost-effective path to extend and build upon the pioneering Kamioka work of years past.  

In order to test this new technology, a dedicated gadolinium R\&D facility was built in 2009 in the Kamioka mine near SK. Called EGADS (Evaluating Gadolinium's Action on Detector Systems)~\cite{egads}, it consists of a gadolinium-loaded 200-ton scale model of SK, complete with 240 50-cm photomultiplier tubes, its own DAQ and readout electronics, a novel selective water filtration system, and water transparency evaluation equipment. Based on its successful operation, in 2015 the SK Collaboration formally approved the plan to add gadolinium to SK, designating this new phase of the experiment ``SK-Gd''.  The T2K Collaboration, a long-baseline neutrino oscillation experiment based at J-PARC which uses SK as their far detector, formally approved the gadolinium-loading plan in 2016.

In order to prepare SK for the addition of gadolinium, the detector had to be 
opened and refurbished for the first time in twelve years.  There were four main tasks:

\begin{enumerate}
    \item Replace the photomultiplier tubes that had failed (a few hundred out of 13,000) since the previous in-tank refurbishment in 2006.
    \item Fix a small water leak in the SK tank.
    \item Clean any rust and other dirt that had accumulated in the detector since its original completion in 1996.
    \item Install additional water piping to increase the total water flow for increased water purification, and to enable better control of the flow direction in the tank.
\end{enumerate}

This in-tank work, which required over 3000 person-days of effort, was successfully carried out between May 2018 and January 2019.  By February 2019 the detector had been refilled with pure water, and was ready for the addition of \mbox{Gd$_2$(SO$_4$)$_3$}.  After taking the T2K beam schedule into account, the first gadolinium is now expected to go into the tank in mid-2020.  By searching for the small but constant, diffuse flux of neutrinos produced by all core collapse explosions since the onset of stellar formation in the early Universe, it is expected that by 2022 we will have collected the world's first additional supernova neutrinos since SN1987A.

\subsubsection{Theoretical studies of supernova explosion and neutrino emission}

In the rest of this section, we report the progress of theoretical study of supernova neutrinos for the detection of SK with the emphasis on detailed description of neutrino transfer and microphysics.  In order to provide the prediction of supernova neutrino detections at SK with gadolinium loading, it is mandatory to get rid of the uncertainties in numerical simulations to describe the explosion mechanism and neutrino emission.  One of the remaining uncertainties of supernova simulations is the neutrino transfer, which has been routinely treated with approximate ways in multi-dimensional simulations.  We perform the sophisticated numerical simulations of neutrino-radiation hydrodynamics based on the direct solver of Boltzmann equation.  The first principle-type calculations enable us to determine the final outcome of explosion, provide the solid prediction of neutrino emissions and evaluate the uncertainties of microphysics such as equation of state.  

\subsubsection{Boltzmann-radiation-hydrodynamics simulation of the core-collapse supernovae}

We have performed several simulations using the Boltzmann-radiation-hydrodynamics code \blue{under} 2D/3D. The details of the code are illustrated in~\cite{Sumi12,Naga14,Naga17}. Various nuclear equations of states (EOSs), progenitor models, and the rotational velocities are employed. Here, we show the important features of the obtained results.

\begin{figure}[ht]
\centering
\begin{minipage}{0.4\hsize}
\centering
\includegraphics[width=\hsize]{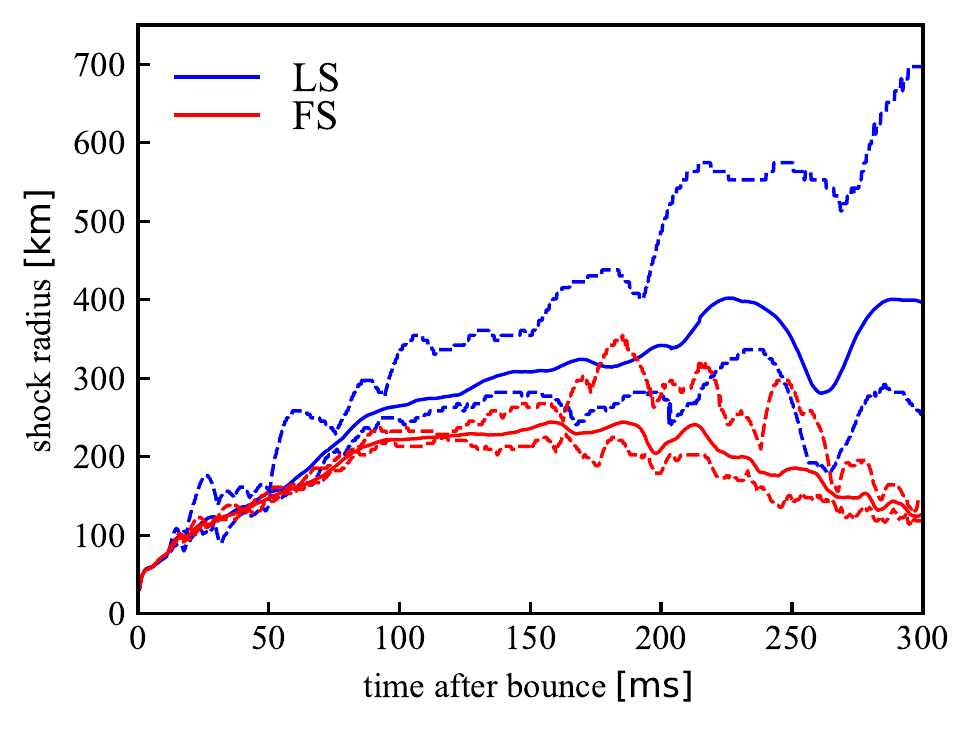}
\end{minipage}
\begin{minipage}{0.4\hsize}
\centering
\includegraphics[width=\hsize]{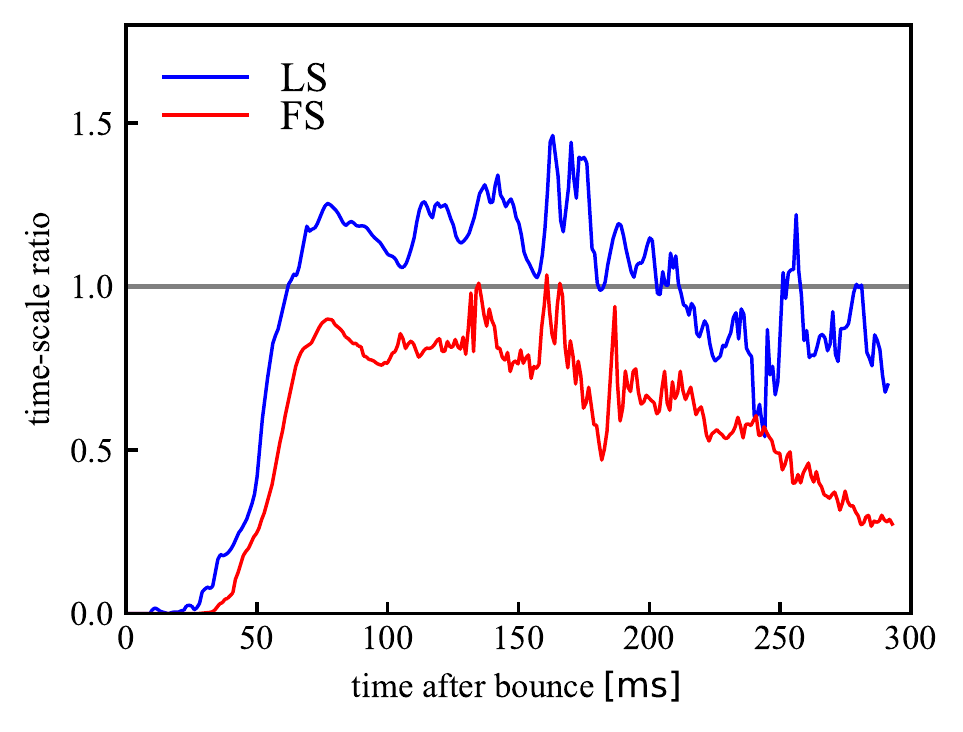}
\end{minipage}
\caption{(Left) The time evolution of the shock radii for the LS (red) and FS (blue) models. The thick solid lines are the angular averaged shock radii, and the dashed lines show the ranges between the minimum and the maximum shock radii. (Right) The time evolution of the timescale ratio for the LS (red) and FS (blue) models. \label{fig:LSFS}}
\end{figure}

First, we examined the effect of the nuclear EOSs~\cite{Naga18}. We employ the Lattimer-Swesty (LS) EOS~\cite{LS91} and the Furusawa-Shen (FS) EOS~\cite{FS13}. The former (latter) EOS model employs the soft (hard) nuclear force model and the single nuclear approximation (nuclear statistical equilibrium treatment) for the nuclear composition. The progenitor model is the $11.2\,M_\odot$ progenitor model taken from~\cite{whw02}.  The simulations with the different EOS models show a significant difference: the LS model shows shock revival while the FS model not (the left panel of Fig.~\ref{fig:LSFS}). A useful quantity to understand the difference is the timescale ratio, which is defined as $\tau_{\rm adv}/\tau_{\rm heat}$, where $\tau_{\rm adv} := M_{\rm gain}/\dot{M}$ and $\tau_{\rm heat} := E_{\rm gain}/Q_{\rm gain}$ are the advection timescale and the heating timescale, respectively. The advection timescale is the timescale with which a fluid element flows through the gain region. The heating timescale is the timescale with which a fluid element is heated and become gravitationally unbound. The gain region is the region where the neutrino heating exceeds cooling. The symbols $M_{\rm gain}$, $\dot{M}$, $E_{\rm gain}$, and $Q_{\rm gain}$ are the mass in the gain region, the mass accretion rate, the total energy including the gravitational binding energy in the gain region, and the net neutrino heating rate in the gain region, respectively. If the timescale ratio exceeds unity, the neutrino heating proceeds sufficiently fast and the explosion succeeds. The right panel of Fig.~\ref{fig:LSFS} shows that the timescale ratio of the LS model exceeds unity, while that of the FS model not. Although $\dot{M}$, $E_{\rm gain}$, and $Q_{\rm gain}$ are similar between the LS and FS models, $M_{\rm gain}$ for the LS model is larger than that of the FS model. This is because the turbulence is stronger for the LS model. Indeed, the prompt convection is stronger for the LS model, and it enhances the neutrino driven convection at the later stage. The stronger prompt convection in the LS model is originated from the stronger photodissociation of the accreting nuclei: the accretion flow outside the shock in the LS model contains more heavy nuclei, and hence the energy consumed by the photodissociation of these nuclei is larger for the LS model. Therefore the proper treatment of the nuclear composition of an EOS is important to assess the influence of EOSs on the CCSNe.

\begin{figure}[ht]
\centering
\includegraphics[width=0.8\hsize, bb= 0 0 632.032479 324.016651]{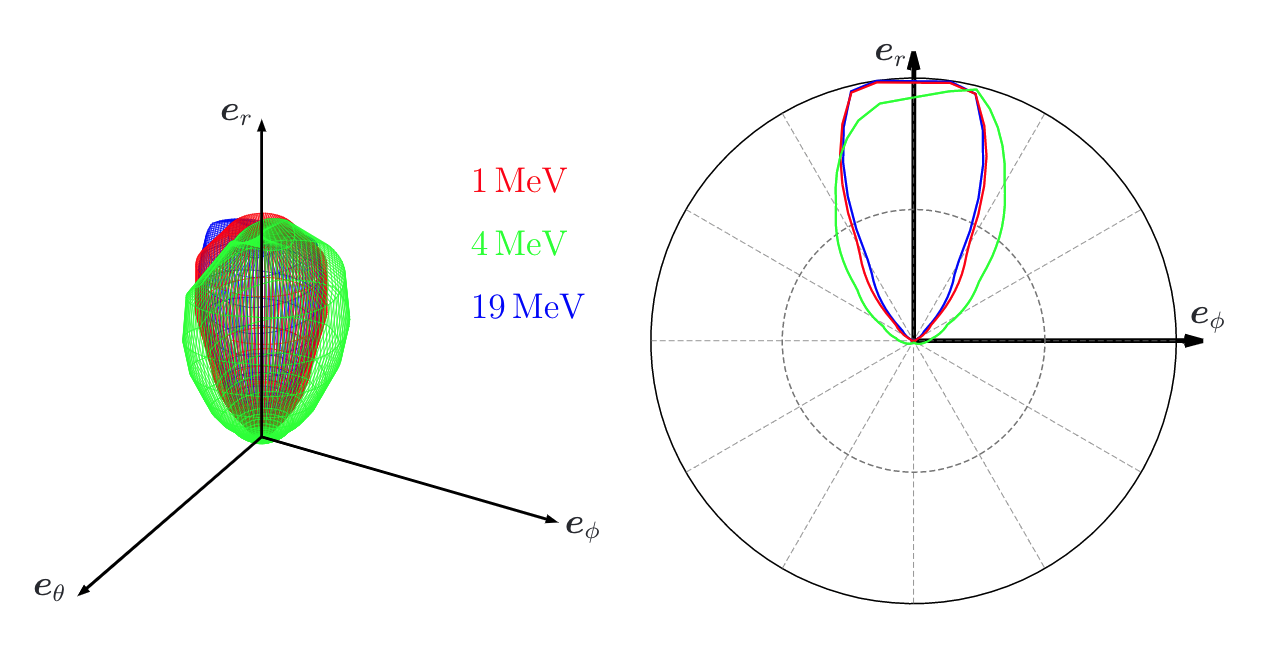}
\caption{The angular distributions of neutrinos outside the shock on the equator at $12\,{\rm ms}$ after bounce. The left panel shows the 3D distributions, and the right panel shows the sections by the $r$-$\phi$ plane. The different colors correspond to the different energies as indicated in the left panel. \label{fig:distri}}
\end{figure}
\begin{figure}[ht]
\centering
\includegraphics[width=0.8\hsize, bb=0 0 649.033353 274.014081]{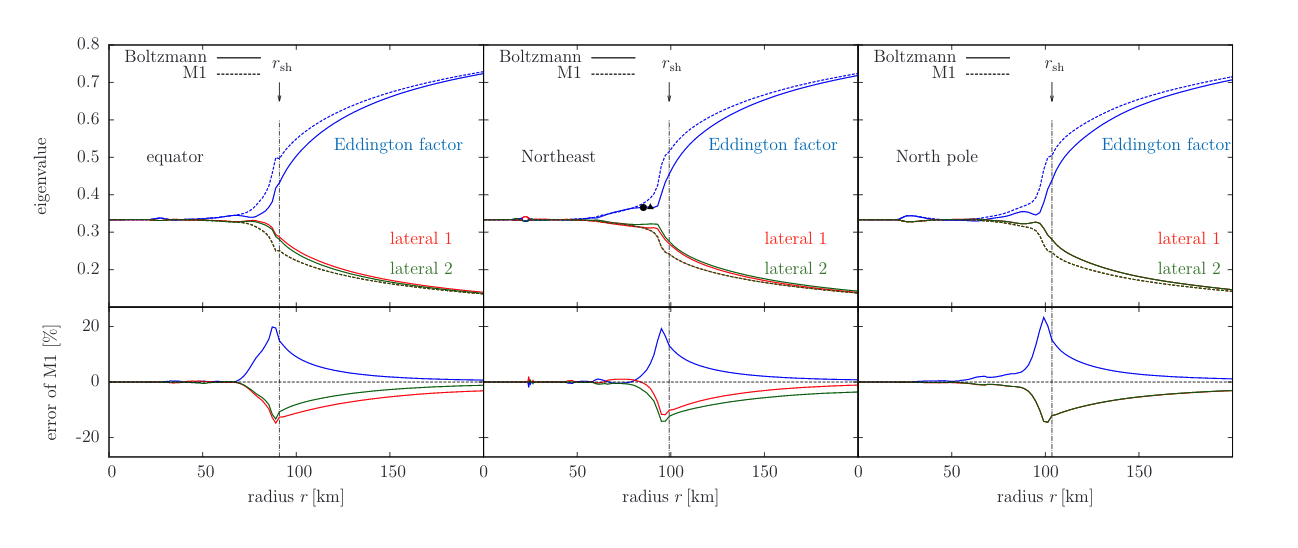}
\caption{The eigenvalues of the Eddington tensors calculated from the distribution function (solid) and the M1-closure scheme (dashed) in different directions at $12\,{\rm ms}$ after bounce. The largest eigenvalue is called the Eddington factor (red), and the other eigenvalues are named lateral 1 and 2 (blue and green). The upper row shows the eigenvalues itself, while the lower row displays the fractional difference between the solid and dashed lines. \label{fig:eigen}}
\end{figure}

Second, we investigated the effect of rotation~\cite{Hara19}. The employed EOS and progenitor model are the FS EOS and the $11.2\,M_\odot$ model from~\cite{whw02}. The rotational velocity profile is so-called the shelluler rotation profile: $\Omega(r) = 1\,{\rm rad/s}/(1+(r/1000\,{\rm km})^2)$. However, the employed rotational velocity is too slow to affect the postbounce dynamics even though the velocity is almost the highest according to the current stellar evolution theory~\cite{Heg00}: the shock radii and other dynamical features are similar between the rotating and non-rotating models. What is more interesting here is the momentum space distributions of neutrinos. Figure \ref{fig:distri} shows the neutrino angular distributions in the laboratory frame outside the shock on the equator. Especially, the distribution of the high-energy neutrinos is tilted to the $\phi$-direction. This is because matter rotates and drags the neutrino distribution to the rotational direction. This detailed angular distribution is only accessible by the Boltzmann solver. Since we have such information, we can assess the accuracy of the M1-closure scheme, one of the approximate neutrino transport method. The second angular moment of the distribution divided by the zero-th moment is called the Eddington tensor. The M1-closure scheme estimates the Eddington tensor from the energy flux and density of neutrinos. In Fig.~\ref{fig:eigen}, we compare the eigenvalues of the Eddington tensor calculated from the distribution function directly and the M1-closure scheme. The largest eigenvalue, or the Eddington factor, is $\sim 20\%$ larger for the M1-closure scheme. This difference originates from so-called ray-collision between outward and inward rays. If we can get some information about these rays from neighboring matter, we may improve the accuracy of the M1-closure scheme.

Third, we suggested a new mechanism for the proto-neutron star kick motion~\cite{Naga19}. We performed the core-collapse simulation with the $15\,M_\odot$ model of~\cite{whw02} and \blue{Togashi-Furusawa} (TF) EOS~\cite{TF17}. The TF EOS is one of the most realistic EOSs. The most interesting feature of this model is the proto-neutron star kick. Thanks to the exact treatment of the proto-neutron star in our code, the proto-neutron star moves with the velocity of $\mathcal{O}(10)\,{\rm km/s}$. So far, the driving force of the proto-neutron star motion is considered to be the gravity of the asymmetric ejecta~\cite{Sc06}. However, the driving force observed in our simulation seems the recoil of the asymmetric neutrino emission. With the proto-neutron star motion, the asymmetric distribution of neutrinos are sustained, and hence the driving force by the neutrinos persists. This result suggests the different mechanism of the proto-neutron star motion.

Finally, we have performed the 3D simulation with the Boltzmann-radiation-hydrodynamics code. Due to the limited computational resources, the dynamics and the neutrino distributions until $\sim 20\,{\rm ms}$ after the core bounce are investigated. With this simulation, the neutrino distribution function with no spatial symmetry is obtained for the first time in the world. This result provides us the important clues to understand the behavior of the neutrino in the supernovae.

\subsubsection{Light curve of supernova neutrinos}

In order to extract the information as much as possible from the detection of supernova neutrinos from the next supernova in future, it is necessary to prepare the templates of neutrino signals by systematically covering the variation of progenitors and microphysics.  One of such systematic sets is the supernova neutrino \blue{database} provided by~\cite{nak13a}.  This data is now routinely used to evaluate the event rates at SK, replacing the rather classic data set by~\cite{tot98}.  

The supernova neutrino \blue{database} is constructed by the combination of supernova simulations for the set of progenitors with different masses and metallicities.  The numerical code of general relativistic neutrino-radiation hydrodynamics under the spherical symmetry~\cite{Yamada:1996uh,sum05}, which is the first-principle calculation, is utilized to study the series of phenomena; gravitational collapse, core bounce, formation of central object, the shock propagation and the associated neutrino emission.  The numerical code for the thermal evolution of proto-neutron star with the flux-limited neutrino diffusion in general relativity~\cite{suz94,nak18} is utilized to study the neutrino emission from the cooling phase of the proto-neutron star over 20$\,$s.  In order to connect the two stages, we take out the profile of a central object from the core-collapse simulation and continue the cooling simulation to simulate the shock revival and the cease of accretion in the explosion.  

We demonstrated the method to extract the physics information from neutrino emissions of the Galactic supernova based on the supernova neutrino database and additional simulations~\cite{suw19}.  We explored the time profile of neutrino detection events at the SK for the set of progenitors with different masses and shock revival time.  We provided the basic feature of the rise and fall of neutrino burst to extract the information of the core bounce of massive stars in the early phase within 0.3$\,$s.  We explored also the long term behavior of neutrino burst over 20$\,$s up to the phase of fading out.  We performed additional long simulations of the proto-neutron star cooling in the case of light and massive neutron stars.  We found that the SK and Hyper-Kamiokande are able to detect the long term evolution over 100$\,$s for the massive neutron stars in Galactic supernovae.  We proposed a method to determine the mass of central object by plotting the cumulative event number summed from the last event detection in time-backward manner.  These results are used as the basis to prepare for the future detection of the next supernova in order to determine the details of supernova mechanism and the neutron star.  The dependence of the neutrino signals on the equation of state (EOS) is now in progress under the development of the EOS table for supernova simulations as described below.  

\subsubsection{Information of supernova mechanism from supernova neutrino detectors}

\blue{
The accumulation of predicted signals of supernova neutrinos will enable us to extract the information of supernovae from the detection of neutrino bursts in the future~\cite{janka2017,BMueller2019}. SK will detect $\sim$10,000 neutrino events from a supernovae in our Galaxy and provide valuable and unprecedentedly detailed information regarding the supernova mechanism. For nearby explosions, pre-supernova neutrinos from the silicon burning stage just before explosion will be studied by both KamLAND and the newly gadolinium-loaded SK.

Events from the dynamical phase will reveal hydrodynamical instabilities and rotation, which are augmented by the detection at IceCube~\cite{Marek2009,takikotake}. Observation of events from the cooling phase will clarify the birth of compact object with mass and radius, which may constrain the equation of state of dense matter~\cite{suw19}.

A combined detection of GWs and neutrinos at KAGRA and SK would serve to unveil the supernova mechanism by the correlation between them~\cite{Yokozawa2015,naka016}. The planned next generation of big detectors such as DUNE, JUNO, and Hyper-Kamiokande~\cite{Hyper-K-report,Hyper-K-SN} with their extended supernova neutrino detection ranges can be expected to provide additional information (See~\cite{SNEWS2} for further references).

Regarding multimessenger astronomy, in the case of a galactic supernova an early alert is possible since neutrinos are generated earlier in the explosion and therefore arrive before the electromagnetic radiation. Such an alert could be generated and disseminated either through large, directional, high-confidence detectors like the gadolinium-loaded SK acting independently to announce a burst, or a network of detectors which detect lower-confidence signals acting in coincidence~\cite{SNEWS2} to reduce the chance of individual false alerts.
 }
\subsubsection{Neutrino bursts from black hole formation}

Studying the neutrino burst from the black hole formation is an interesting target of the SK among the variety of gravitational collapse of massive stars~\cite{sum07}.  In the case of stars of more massive than those for ordinary supernovae, 40--50$M_{\odot}$ for example, with the intense accretion of matter from outer layers, the retreat of shock wave is inevitable and there is no chance to have the explosion.  The mass of central object increases monotonically and attains the maximum mass, which can be supported by the equation of state.  The black hole is formed due to the re-collapse of proto-neutron star at the critical mass.  The neutrino signal in this case has a characteristic signature with a short duration, typically about 1$\,$s, and increasing average energies and luminosities.  This is cased by the increasing mass due to the accretion and the associated increase of density and temperature.  This type of short neutrino burst can be a hallmark of black hole formation and can be used to constrain the equation of state of hot and dense matter~\cite{sum06,nak10}.  

The detection of GWs and neutrino bursts from the black hole formation may provide the information on the central object and the equation of state in addition to the case of core-collapse supernovae 
(see sec.~\ref{ptepGW_secC01}).  We revealed the characteristic feature in the frequencies of GWs from the dynamical evolution of proto-neutron star toward the black hole formation.  We analyzed the time evolution of accreting proto-neutron stars to determine the fundamental and gravity modes of GWs~\cite{sot19} by utilizing the results of the numerical simulations of massive stars with a set of equation of state~\cite{sum07}.  The density increase toward the black hole mass leads to the rise of frequencies of GWs as a function of average density of the proto-neutron star.  The ratio of the frequencies of the two modes can be characterized by the compactness of the proto-neutron star, therefore, the information of mass and radius.  Moreover, the termination of neutrino signal can provide the timing of the black hole formation and the information of maximum density through the combined analysis of GWs and neutrino signals.  

\subsubsection{Diffuse supernova neutrino background}
The numerical simulations of the supernova explosions and the black hole formation from various massive stars are essential for the detection of diffuse supernova neutrino background, which is the main target of the SK with gadolinium loading.  The numerical simulations of the core-collapse supernovae in multi-dimensions (see also sec.~\ref{ptepGW_secC01}) and the black hole formation as discussed above have been applied to provide the integrated energy spectra of neutrino emission from the progenitors in the wide range of stellar mass.  The feature of neutrino burst depends on the compactness of massive stars, which in turn determines the accretion rate of matter and affects the neutrino emission.  In the wide coverage of massive stars with different compactness, the case of black hole formation contribute to hard energy spectra due to the energetic bursts.  Hence the contribution of black hole formation for large compactness increases the event rates of diffuse supernova neutrino background.  The long term observation at SK and Hyper-Kamiokande can provide the constraint on the critical compactness and the ratio of black hole formation in various stellar collapse~\cite{hor18}.  

\subsubsection{Equation of state for supernovae and neutron star mergers}

The equation of state is one of the important ingredients in core-collapse supernovae as well as neutron star mergers.  It is also one of remaining uncertainties in nuclear physics to determine the outcome of explosions, the birth of compact object and the neutrino emission.  In addition to the progress of numerical simulations of supernovae, developments of the data table of supernova EOS for simulations have been made over the decades.  One of the popular sets of supernova EOS is the Shen EOS based on the relativistic mean field theory~\cite{she98a,she98b,she11}, which has been applied to provide extended versions of data table of supernova EOS~\cite{oer17}.   The Shen EOS table is publicly available on the web and has set the standard format to provide thermodynamical quantities for the usage of numerical simulations.  The Shen EOS has been used in the numerical simulation for the supernova neutrino database.  

Recently, the Shen EOS has been revised to improve the density dependence of the symmetry energy~\cite{she20}.  This modification is motivated by the recent progress of observational data on neutron stars from the GW detection and the X-ray observations.  The symmetry energy of the original Shen EOS, which was determined by the fitting to the nuclear masses and radii, has been claimed to be rather large as compared with other frameworks.  Recent nuclear experiments also provide constraints on the behavior of symmetry energy and help to improve the description of symmetry energy.  A new term of density-dependent iso-vector interaction in \blue{the relativistic mean field theory} provides a smaller symmetry energy while keeping good properties of symmetric nuclei and matter.  It provides also smaller radii of neutron stars within the observational constraints.  The updated Shen EOS is now applied to numerical simulations of core-collapse supernovae and proto-neutron star cooling.  It has been recently shown that the influence of the updated Shen EOS at high densities mainly appears in the evolution of proto-neutron stars when the matter becomes neutron-rich and remains minor in the dynamics around the core bounce.  It is interesting to see the influence of the full data table of the updated Shen EOS in core-collapse supernovae and neutron star mergers.  

In addition to the revisions of the nuclear interaction,  the improvement of the supernova EOS of hot and dense matter has been made to describe the mixture of nuclei under the nuclear statistical equilibrium~\cite{fur17a,fur17b}.  There is recent progress of the construction of EOS tables by the microscopic nuclear many body theories such as the variational method and the Dirac Brueckner Hartree-Fock theory~\cite{fur20,tog17} based on the nucleon-nucleon interactions, being different from the effective nuclear many body theories.  In addition, the sets of EOS tables with systematic coverage of EOS parameters are under the development and will be applied to the study of core-collapse supernovae and neutron star mergers.  These development will help to study the signal of neutrinos and GWs and to probe the dense matter in these astrophysical events.

\subsubsection{Supernovae simulations on GPUs}

The dynamics of the supernovae is described by hydrodynamic equations for dense matter and the Boltzmann equation for neutrino transport under the gravitational effect described by the theory of general relativity. In addition to these coupled equations, data of physics processes such as sets of equation of state and reaction rates for neutrinos are needed to be implemented in the numerical simulations. Since the neutrino distribution is a function in six dimensions (three dimensions in space and three dimensions in neutrino momentum space), the required computational resources quickly increases with grid resolution of these degrees of freedom. Therefore the numerical simulations of supernovae tend to be challenges in large-scale computational techniques and algorithms.

So far most of large-scale simulations have been performed on massively parallel clusters, such as the K computer. Recently another type of architecture, which makes use of arithmetic accelerators such as GPUs, has become popular in high performance computing. Although the accelerator device has an advantage in cost performance, the code implementation becomes much more involved to achieve desired performance. Furthermore, whether accelerators work efficiently depends strongly on the structure of numerical algorithms.

Presumably for these reasons, application of GPUs to the simulations of core-collapse supernovae has been restrictive. As for the simulation code including the neutrino transport, the VERTEX code was ported to GPUs by employing CUDA~\cite{Dannert2013}. In the VERTEX code, its most time consuming part is calculation of the collision term of the Boltzmann equation that exhausts almost half the simulation time. Offloading one dominant reaction process to GPUs achieved 1.8 times acceleration of whole simulation time.

We develop an efficient scheme to exploit accelerator devices such as GPUs in the numerical simulations of the core-collapse supernovae~\cite{Matsufuru2018,Matsufuru2018b}. As the first step, we apply the offloading of simulation bottlenecks to the computations of neutrino-radiation hydrodynamics under spherical symmetry~\cite{Yamada:1996uh}. By adopting the implicit scheme for the evolution equation, an iterative linear equation solver for the coefficient matrix is the most time consuming part. To offload this part to the GPU devices, we employ OpenACC as a framework of implementation as well as make use of cuBLAS library that is available on NVIDIA's CUDA environment. We change the data layout to maximize the efficiency of data transfer between the device memory and cores, in accord with so-called coalesced access. With both the OpenACC and cuBLAS, significant acceleration is achieved so that the linear equation solver is no longer a primary bottleneck~\cite{Matsufuru2018}.

The secondary bottlenecks of the original code are the computation of collision term in the Boltzmann equation and the inversion of block diagonal part of the coefficient matrix. The latter is required in the weighted Jacobi-type preconditioner for the iterative linear equation solver~\cite{Imakura:2012aa}. Offloading to GPU devices successfully accelerates these tasks~\cite{Matsufuru2018b}. For the inversion of block matrices, we employed a blocked version of the Gauss-Jordan algorithms~\cite{Quintana:2001} that is suitable to the many-core architecture. With these improvements, systematic simulations with better resolution have become feasible. Further optimization as well as extension to 2D and 3D simulations are underway. We are also implementing a code for PEZY-SC processor, another kind of accelerator device, by applying the technique developed for GPUs.

%% file: ptepGW_section5.tex
\subsubsection{Varieties of gravitational waveform within general relativity}
The GW signals from coalescing binaries 
can be decomposed into inspiral, merger and ringdown waveforms. 

Here we focus on the inspiral phase, in which the orbital frequency and hence the 
GW frequency gradually increases as the binary separation decreases. 
In the case that the binary with aligned spins evolves along a sequence of quasi-circular orbit, 
the inspiral waveform in the frequency domain can be approximately represented as
\begin{equation}
  \tilde h(f)={\cal A} f^{-7/6} e^{i\psi(f)}\,,  
  \label{eq:GRwaveform}
\end{equation}
where the overall amplitude ${\cal A}$ is 
$\propto {\cal M}^{5/6}$, and 
\begin{eqnarray}
 \psi(f)=2\pi f t_c -\phi_c +\frac{3}{128}\left(\pi{\cal M}f\right)^{-5/3}
  \Biggl[1+\frac{20}{9}\left(\frac{743}{336}+\frac{11}{4}\eta\right)v^2
   -4\left(4\pi -\beta\right)v^3+\cdots
  \Biggr]\,,
  \label{simplewaveform}
\end{eqnarray}
in the post-Newtonian (PN) approximation~\cite{Blanchet:2013haa}.
Here, $\eta=\mu/M$ is the symmetric mass ratio, $v=(\pi M f)^{1/3}$ is
the orbital velocity, and 
\begin{equation}
    \beta=\left(\frac{113}{12} - \frac{19}{3}\eta\right)\chi_{\rm eff}
  + \frac{19}{6} \eta \sqrt{1-4\eta}\left(\frac{\mbox{\boldmath{$S$}}_1}{m_1^2}
   -\frac{\mbox{\boldmath{$S$}}_2}{m_2^2}\right)\cdot 
     \hat{\mbox{\boldmath $L$}} \,. 
\end{equation}
The second term in the \blue{right} hand side of the above equation
becomes a small contribution because
$\eta \sqrt{1-4\eta} \leq \sqrt{3}/18 = 0.096\cdots$,
but we should note that this term can be dominant
for spins antialigned with each other.

The first term in the square brackets of Eq.~\eqref{simplewaveform} is 
the term obtained by using the Newtonian dynamics and the 
quadrupole energy loss formula due to GW emission. 
So, when we discuss the phase evolution, we count the PN order 
relative to this leading term. The second term is relatively $O(v^2)$, 
which we refer to as $1\,$PN order. In the following,
we summarize additional effects which \blue{should be carefully discriminated from modifications due to extension of general relativity (GR).}

\begin{description}
\item[The effect of spins:]~\\
The lowest PN effect of spins 
on the phase evolution 
appears at the $1.5\,$PN order. The effect 
depends only on the spin components projected to the direction 
parallel to the orbital angular momentum at this order, 
which is encapsulated in 
the coefficient $\beta$. 
When the spins are not aligned with the direction of the 
orbital angular momentum, spin precession occurs. 
As a result, the amplitude modulation occurs, which 
helps to solve the degeneracy between spins and other parameters~\cite{Yagi:2009zm}. 
The GW templates for precessing 
binaries are ready (see, e.g., Ref.~\cite{Khan:2018fmp} and references therein).

\item[Eccentric binaries:]~\\
The orbital eccentricity decays like
$e\propto a^{19/12}$ at the late stage of binary evolution, 
where $a$ is the orbital semi-major axis~\cite{Peters:1964zz}. 
Therefore, binaries are circularized well before they merge 
unless they are borne with a close separation and a large eccentricity. 
One possible scenario to form binaries that maintain a sufficiently 
large eccentricity at the coalescence time is the binary formation 
by capture in star clusters. Hence, detection of eccentricity would 
be useful to distinguish the formation channels of binaries that 
become GW sources. 
In the GW waveform, the effect appears most significantly in the 
phase evolution, unless the eccentricity is not extremely large. 
The correction appears in the phase evolution at $-19/6\,$PN order. 
The minus sign of the PN order reflects the fact that the 
orbital eccentricity decays rapidly. 


\item[Dark matter cloud:]~\\
In the presence of dark matter, the waveform of GWs can be affected~\cite{Macedo:2013qea}.
In the scenario of the existence of the dark matter mini-spike around an IMBH~\cite{Zhao:2005zr,Bertone:2005xz}, the waveform of the GWs from an EMRI/IMRI system is modified due to the effects of the gravitational pull, the dynamical friction, and the accretion of dark matter~\cite{Eda:2013gg,Eda:2014kra,Yue:2017iwc}.
They have shown a possibility of the measurement of the power-law index of the dark matter mini-spike radial profile with space-borne GW detectors such as LISA. (See also Ref.~\cite{Hannuksela:2019vip}.)
On the other hand, when the mass ratio is around $O(10^3)\sim O(10^4)$, energy dissipation due to the dynamical friction is comparable to the total binding energy of dark matter. 
If one considers a modification of dark matter distribution due to the energy dissipation, the differences of the waveform from the vacuum case becomes much smaller, even the spike has a large power-law index~\cite{Kavanagh:2020cfn}.
\end{description}

\subsubsection{Model independent 
test of \blue{waveform} consistency with general relativity}
From the inspiral waveform, we can determine the 
binary parameters. At the same time, we can test if 
there is no deviation from the standard waveform predicted for 
non-spinning quasi-circular binaries in GR. 
We can test the presence of spin and eccentricity of each binary component 
BH. At the same time we can also test the deviation from the 
prediction of GR. 

Here we summarize the test of GR performed by LIGO/Virgo collaborations~\cite{LIGOScientific:2019fpa}. 
The first test is the one to test if the residual data which is 
obtained by subtracting the best-fit template is consistent with 
the Gaussian noise. There is no inconsistency observed. 

They also have made a consistency test between the estimations of parameters obtained from 
the inspiral part and from the merger-ring down part. 
Small perturbation modes of a BH have its own frequencies and damping rates determined by its mass and spin. Those damped oscillation modes, which are called quasinormal modes (QNMs), are excited at the epoch of formation of the remnant BH after coalescence. See, e.g., Ref.~\cite{Kokkotas:1999bd} as a review of QNMs including the history. Among various QNMs, 
the dominant mode is the fundamental ($n=0$) mode with the harmonic $(\ell,\,m)=(2,\,2)$, and the GW waveform is written as
\begin{equation}
h(t) = A e^{-(t-t_0)/\tau}
\cos(2 \pi f_{\rm R} (t-t_0) - \phi_0) \,, 
\end{equation}
where $f_{\rm R}$ is the oscillation frequency,
and $\tau$ is the damping time, and 
$t_0$ and $\phi_0$ are the starting time and its phase, respectively. $A$ is the amplitude at the starting time.
We can find the information of $f_{\rm R}$ and $\tau$
which are related to the mass $M_{\rm rem}$
and spin $a_{\rm rem}$ of the remnant BH, in Refs.~\cite{Berti:2005ys,BertiQNM}.

As for the initial phase, we can maximize the SNR
with an analytical method presented
in Ref.~\cite{Nakano:2003ma,Nakano:2004ib}
which is based on Ref.~\cite{Mohanty:1997eu}.
Since it is difficult to determine the starting
time (e.g., see Ref.~\cite{Sakai:2017ckm} for the estimation),
the fractional differences 
\begin{equation}
\frac{\Delta M_{\rm rem}}{\bar M_{\rm rem}}
= \frac{2(M_{\rm rem}^{\rm insp}-M_{\rm rem}^{\rm post-insp})}{M_{\rm rem}^{\rm insp}+M_{\rm rem}^{\rm post-insp}}
\,, \quad
\frac{\Delta a_{\rm rem}}{\bar a_{\rm rem}}
= \frac{2(a_{\rm rem}^{\rm insp}-a_{\rm rem}^{\rm post-insp})}{a_{\rm rem}^{\rm insp}+a_{\rm rem}^{\rm post-insp}}
\,,
\label{eq:param_consist}
\end{equation}
between the quantities for the inspiral
and the post-inspiral phases have been used 
as an inspiral-merger-ringdown consistency test 
for BBHs in Ref.~\cite{LIGOScientific:2019fpa}.
In the GR case, we will have
\begin{equation}
\frac{\Delta M_{\rm rem}}{\bar M_{\rm rem}}=0 \,,
\quad
\frac{\Delta a_{\rm rem}}{\bar a_{\rm rem}}=0 \,,
\end{equation}
within the measurement error range.
The analysis result for GW events
in the LIGO-Virgo Catalog GWTC-1
is shown in Fig. 2 and Table III of Ref.~\cite{LIGOScientific:2019fpa}.
Here, we should also note that the SNR only for the ringdown
phase is not so high in the current observations.
Therefore, the merger-ringdown phase
with a combination of phenomenological and analytical BH
perturbation theory parameters~\cite{Meidam:2017dgf},
is treated in the parameterized test of GWs.
This analysis result for GW events in the LIGO-Virgo Catalog GWTC-1
is also shown in Fig. 3 of Ref.~\cite{LIGOScientific:2019fpa}.
On the other hand, when we estimate the mass and spin
of the remnant BH (e.g., the right panel of Fig. 4
in Ref.~\cite{LIGOScientific:2018mvr}) without testing gravity,
we use the inspiral-merger-ringdown waveform from
numerical relativity~\cite{Pretorius:2005gq,
Campanelli:2005dd, Baker:2005vv}.

Using only the ringdown phase,
one of the best way to test gravity 
is ``black hole spectroscopy'' by observing multiple QNMs~\cite{Dreyer:2003bv} (\blue{see also Refs.~\cite{Berti:2007zu,Gossan:2011ha} for testing the no-hair theorem with black hole ringdowns, and}
another simple method is
proposed by Ref.~\cite{Nakano:2015uja}).
For example, we may consider Eq.~\eqref{eq:param_consist}
with substitution,
\begin{eqnarray}
&&M_{\rm rem}^{\rm insp}=M_{\rm rem}^{(\ell=2,m=2)} \,,
\quad
M_{\rm rem}^{\rm post-insp}=M_{\rm rem}^{(\ell=3,m=3)} \,,
\cr
&&a_{\rm rem}^{\rm insp}=a_{\rm rem}^{(\ell=2,m=2)} \,,
\quad
a_{\rm rem}^{\rm post-insp}=a_{\rm rem}^{(\ell=3,m=3)} \,,
\end{eqnarray}
if the $(\ell=3,m=3)$ QNM is the next dominant mode.
In order to do so, it is necessary to extract 
$f_{\rm R}$ and $\tau$ of the weak ringdown signal from
the noisy data accurately for each QNM. 
In Ref.~\cite{Nakano:2018vay}. 
As the first step, 
mock data of GWs
which include some deviation from the GR prediction was prepared,
and it was applied to the following five methods
to extract the dominant QNM:
(1) plain matched filtering with ringdown part method, 
(2) matched filtering with both merger and ringdown parts method,
(3) Hilbert-Huang transformation method, 
(4) autoregressive modeling method, and 
(5) neural network method.
It was found that the determination of $f_{\rm R}$ 
is much easier than that of $\tau$
although the accuracy depends on the analysis methods.
Interestingly, it turned out that the standard matched filtering does not give the 
optimal inference in this problem. 


As a unified framework to test possible modifications of gravity, one can 
use parametrized post-Einstein (PPE) approach~\cite{Yunes:2009ke}. 
In this approach, we consider the modification of 
the functions that appear in the GW 
waveform in GR~\eqref{eq:GRwaveform} 
without specifying the model as 
\begin{equation}
  A(f)\to (1+\sum_i \alpha_i v^{2a_i})A_{GR}(f)\,, 
  \qquad
   \psi(f) \to \psi_{GR}(f)+\sum_i\beta_i v^{2b_i-5/2}\,, 
\end{equation}
where $\alpha_i$ and $\beta_i$ specify the amplitude of 
modification, while $a_i$ and $b_i$ specify the PN order 
of modification. The point is that the leading order 
correction tends to be given by a particular PN order. 
In many cases the leading order effects appear in the 
phase evolution. 
Of course, the test targeting at a particular modification can 
be more sensitive. 

LIGO/Virgo collaborations did not give the constraints on the PPE parameters directly. 
Instead, they examined the constraint on 
the modification to the model parameters contained in the waveform model 
IMRP{\small HENOM}Pv2, \blue{following the method proposed in 
Ref.\cite{Chandra:2010} and implemented by Ref.\cite{Agathos:2013upa}. }
For the inspiral part, they test the 
modification to the existing 
PN coefficients in the phase evolution up to $3.5\,$PN order in GR as well as the 
$-1\,$PN and $0.5\,$PN coefficients which are not present in GR templates. 
Also, the modification of the parameters that characterize the merger 
and ring-down phases is also tested. 

For the modification of GW propagation, what they did is similar to PPE, but the 
constraints are derived for the amplitude of modification per propagation 
distance by stacking the data. 
In all cases mentioned above, significant deviation from GR prediction 
has not been obtained.

\subsubsection{\blue{Model independent 
test of extra polarizations}}
The GW interferometers basically detect the tidal deformation of the 
physical distances among nearby test masses, which 
can be described by the geodesic deviation equation for 
slowly moving objects 
\begin{equation}
  \ddot x^i \approx -R^i{}_{0j0}x^j\,.  
\end{equation}
Namely, what we can observe is the rank 2 tidal 
tensor $E_{ij}=R_{i0j0}$. 
The absolute distance cannot be measured by the current
GW interferometer, and they measure the difference of the distance changes 
in different directions. 
Hence, the trace part of $E_{ij}$ is irrelevant. 
In generic dynamics of geometry, therefore, 
there are five components which can be decomposed 
into three dimensional 1 scalar, 2 transverse vector, 2 transverse traceless 
tensor components. 

It is not easy to think of gravity models in which 
scalar or vector components of the tidal waves are emitted without 
modifying the GW waveform significantly. 
For example, in the case of scalar-tensor theories, the effect of 
scalar wave emission, which also contributes to the excitation of 
the scalar component of the tidal tensor, appears most significantly 
in the orbital evolution of the binary due to the radiation reaction 
that it causes. If we consider such models that can excite larger 
magnitude of tidal tensor in the scalar or vector mode 
without losing much energy, the coupling of matter field 
to those modes tends to be too large to hide the influence on  
the tests of gravity in solar system. 

For the reason mentioned above, sensible models would be a 
conversion of the excitation energy from the tensor modes, i.e., standard 
GW modes, to the other exotic modes during the propagation. 
In that case, the waveform can be identical to the original 
GW waveform although the phase velocity of the scalar 
and vector waves can be different from the speed of light.  
In that case, it is difficult to put some constraint to the 
conversion of the energy during the propagation. 
So far, only comparison of likeliness between pure tensor modes 
and pure scalar mode or between pure tensor modes and pure 
vector modes has been done, finding that pure tensor modes are 
largely preferred~\cite{Abbott:2018lct,LIGOScientific:2019fpa,Takeda:2020tjj}. 

In a general framework, a GW signal can be a superposition of scalar and/or vector modes in addition to the ordinary tensor modes:
\begin{equation}
 S_a(t,\hat{\Omega})=\sum_a F_a^{P}(t,\hat{\Omega})h^P(t).
 \label{eq:GW-pol-signal}
\end{equation}
\blue{Here $F_a^{P}$ is the antenna pattern function of the $a$-th detector for the polarization $P$, and $h_P$ is the GW waveform for the polarization $P$, where $P=+$, $\times$, $V$, $W$, $S$ and $L$~\cite{Poisson:2014,Nishizawa:2009bf,Schutz:1987}. 
 $V$ and $W$ modes are vector modes and $S$ and $L$ modes are scalar modes, 
 which satisfy $F_a^{S} = - F_a^{L}$~\cite{Nishizawa:2009bf}. 
The antenna pattern functions are dependent on 
a GW source location  
$\theta$ and $\phi$, which are the latitude and longitude, respectively. }
 The separability of the polarization modes of GWs from a point source has been investigated by solving an inverse problem for the first time in~\cite{Hayama:2012au}. \blue{It has been shown that in a nontensorial polarization search, the necessary number of detectors is at least the same as the number of polarization modes to be searched, e.g., at least three detectors are necessary to search for a superposition of two tensor and one scalar modes.} The mode separability in the case of compact binary coalescences is subtle because their GW waveforms are well modeled with multiple source parameters. It is shown in~\cite{Takeda:2018uai} that even with correlations and degeneracies among the parameters, a mixture of the polarization modes are separable with the same number of detectors as the number of the modes. Then the scalar or vector modes with similar amplitudes to the tensor modes can be detected. For a binary neutron stars (BNS) observed with the third-generation detectors such as Einstein telescope\cite{ET} or cosmic explorer\cite{CE}, the mode separation with smaller number of detectors is possible due to the long duration of a signal ($\gtrsim 1$ hour) and the time-evolving detector antenna patterns~\cite{Takeda:2019gwk}. 
 
 With the help of an electromagnetic counterpart, more polarization modes than mentioned above can be searched once the sky location is fixed. \blue{We will discuss this possibility in the next.}

\begin{figure}
  \begin{center}
    \includegraphics[width=115mm]{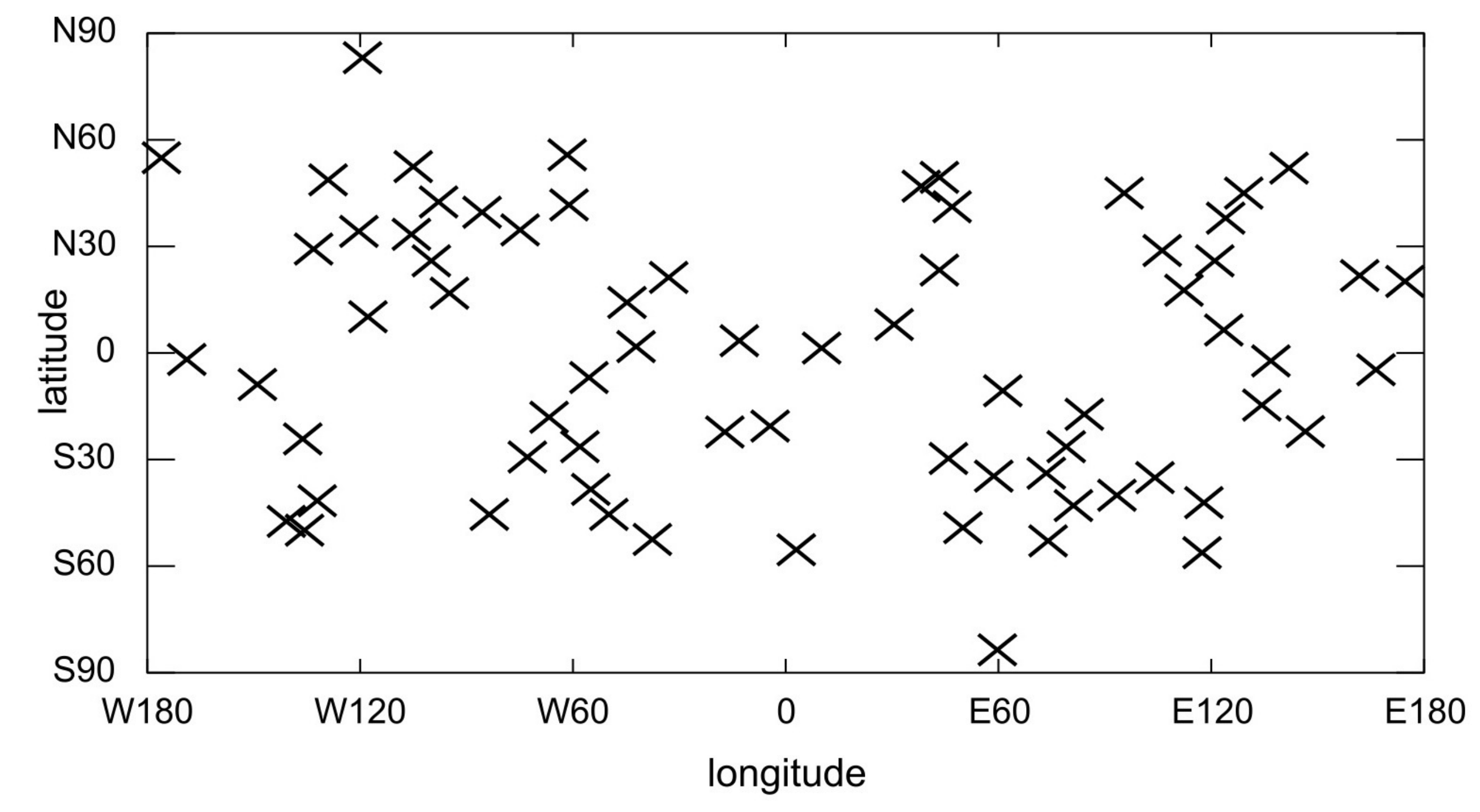}
    \caption{Seventy sky positions in the earth-centered coordinates. At these points for a GW source, the spin-1 test can be done in principle, because the spin-0 and spin-2 modes can be perfectly eliminated. The numerical calculations were done in Reference~\cite{Hagihara:2018azu}. }
    \label{figure-70}
  \end{center}
\end{figure}

%
\paragraph{\blue{Direct polarization test with the help of an electromagnetic counterpart}}
\blue{We consider to analyse the data from a GW source, 
$S_a(t,\hat{\Omega})+ n_a(t)$, 
where $n_a$ is the detector noise.  }
\blue{Equation (\ref{eq:GW-pol-signal})} means that tests of all polarization modes 
with a network of interferometers require at least five detectors. 
However, Hagihara {\it et al.} 
pointed out that there exist particular sky 
positions that allow for a vector mode test, 
because scalar modes, even if they exist, can be 
perfectly eliminated from a certain combination of 
the strain outputs at the four detectors 
only for these sky directions~\cite{Hagihara:2018azu}
\blue{(See Fig.~\ref{figure-70}).}  
They also found that a vector mode test can be done 
by using four detectors~\cite{Hagihara:2019ihn}. 
They put also a direct upper bound on 
a vector component from GW170817 event, 
since the scalar modes can be largely suppressed 
for this event~\cite{Hagihara:2019ihn,Hagihara:2019rny}. 

Very recently, they proposed a general formulation 
for directly testing extra GW polarization 
and a certain condition that allows for a scalar test. 
First, we focus on the network composed of the four ground-based interferometers. 
For a given source,  we know its sky position, 
as is the case of GW events with an electromagnetic counterpart such as GW170817. 
Then, one knows exactly how to shift the arrival time 
of the GW from detector to detector. 

For example, 
the projection operator is defined to 
eliminate $W$, $+$ and $\times$ modes. 
It is denoted as $\Pi^{aW+\times}$. 
For this example, $h_W$, $h_+$ and $h_{\times}$ in the strain outputs $\{S_a\}$ are eliminated as 
\begin{align}
\Pi^{aW+\times} S_a 
= 
&\left(\varepsilon^{abcd}F_a^SF_b^WF_c^+F_d^{\times}\right) (h_S-h_L)
+ 
\left(\varepsilon^{abcd}F_a^VF_b^WF_c^+F_d^{\times}\right) h_V 
+ \Pi^{aW+\times} n_a .
\label{4null-wpc}
\end{align}

If the coefficient of $h_V$ in Eq. (\ref{4null-wpc}) 
vanishes in a certain sky region, 
there remains only the spin-0 part in the null steam.  
Therefore, the spin-0 polarization test is possible, 
if a GW source is found in this sky region. 
The vanishing coefficient condition is 
\begin{align}
D_4 
&\equiv
\left|
\begin{array}{cccc}
F_1^V & F_1^W & F_1^+ & F_1^{\times} \\
F_2^V & F_2^W & F_2^+ & F_2^{\times} \\
F_3^V & F_3^W & F_3^+ & F_3^{\times} \\
F_4^V & F_4^W & F_4^+ & F_4^{\times} \\
\end{array}
\right| = 0 . 
\label{D4}
\end{align}
Note that this is invariant 
for choosing a reference axis 
for the polarization angle. 
$D_4 =0$ (or sufficiently small $D_4$ practically) is a condition for directly testing scalar modes separately from the other modes only by four detectors 
\cite{Hagihara:2019rny}. 

A possible straightforward procedure of the GW data analysis along this direction is as follows. 
At first, we determine the sky location of the GW/EM source from multimessenger astronomy, especially by optical and VLBI observations.  Secondly, by using the sky location, the arrival time shift for each GW detector is taken into account. Next, the (time-shifted) strain output at each detector is substituted into $S_a$ in the left-hand side of Eq. (\ref{4null-wpc}). If the left-hand side of Eq. (\ref{4null-wpc}) is above the noise level (the last term of the right-hand side), then, the existence of extra GW polarizations could be suggested. If it is comparable to (or lower than) the  noise level, a certain {\it direct} upper bound can be placed on extra GW polarizations (through the first and second terms the right-hand side). In this procedure, explicit templates of GW waveforms are not needed. In this sense, this test of extra GW polarizations is {\it robust}. Further study along this course is left for future. For instance, sophisticated algorithms and pipelines would be important for real data analysis, because the nature of detector noises is quite complicated.

\begin{figure}
  \begin{center}
    \includegraphics[width=115mm]{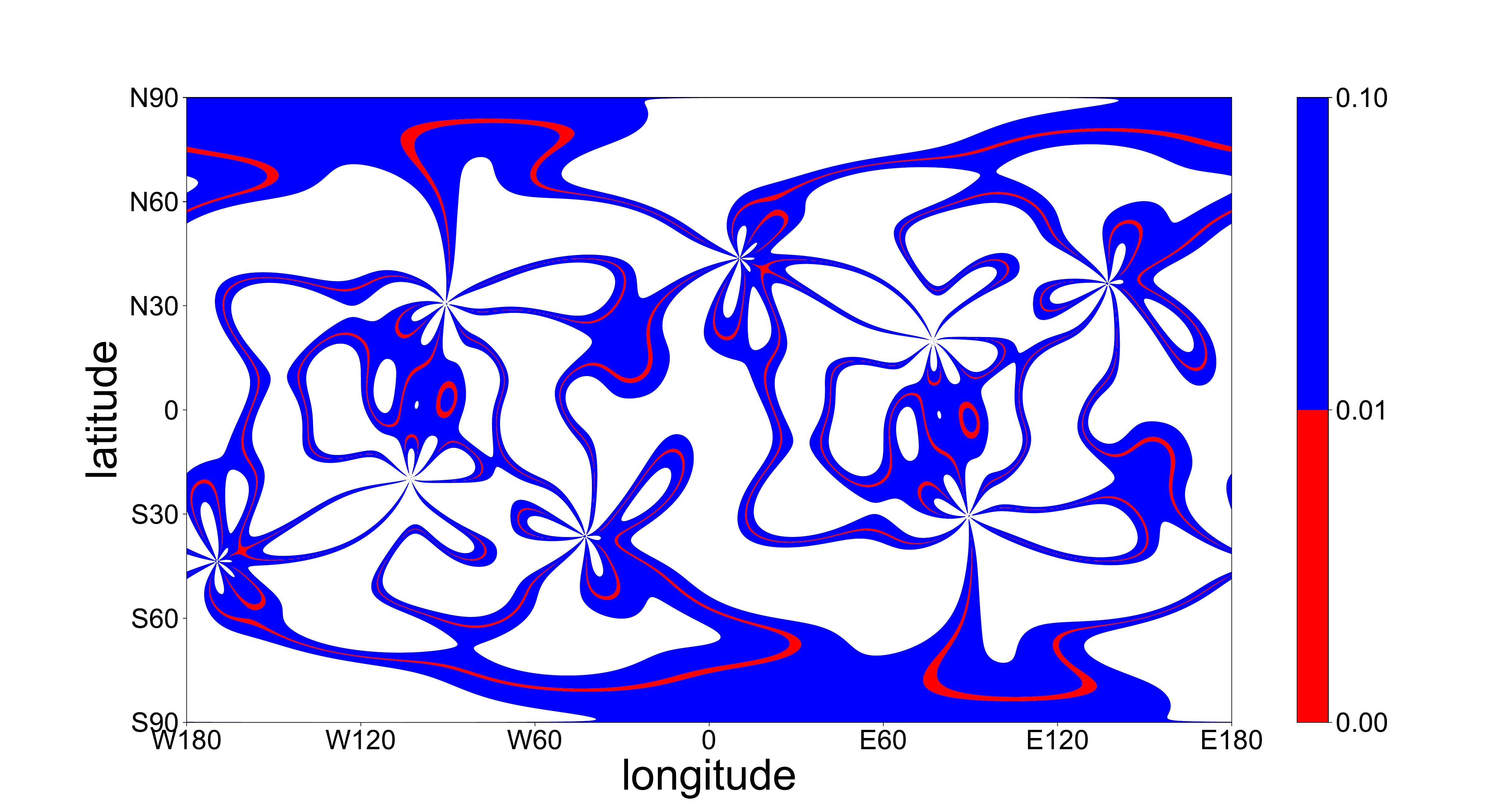}
    \caption{Contour plot of $D_4$ in the earth-centered coordinates. 
    The numerical calculations were done in Reference~\cite{Hagihara:2019rny}. }
    \label{figure-D4contour}
  \end{center}
\end{figure}

\blue{
With four detectors, consistency tests with GR 
can be done 
by examining if the wave in the fourth detector, 
say KAGRA, is consistent with the three-detector network. 
}
\blue{Equation~(\ref{eq:GW-pol-signal})} for strain outputs 
($a = 1, \cdots, 5$) from the five detectors including  LIGO-India can be always solved for  
$h_S - h_L$, $h_V$, $h_W$, $h_+$ and $h_{\times}$. 
In principle, adding \blue{the} fifth detector will thus allow 
for a direct 
\blue{search} of \blue{extra} GW polarizations for 
{\it any} sky region. 
This would be \blue{interesting} in probing new gravitational physics \blue{beyond GR}.

\subsubsection{Gravitational waveform corresponding to various modifications of gravity}
\label{sec:modificationlist}
The sources of GWs are composed of the objects of extremely strong gravity in the sense 
that the deviation from the Minkowski spacetime is significant. The spacetime curvature 
radius is comparable to the size of the system in the case of BH and/or NS binaries. 

In the case of BBHs,
the emission of electromagnetic counterpart is not expected much. 
In fact, no detection of electromagnetic counterpart has been reported so far. 
In the case of binaries that include a NS,
we have a chance to detect the electromagnetic 
counterpart. In fact, various follow-up observations succeeded in finding the counterpart 
for GW170817, the first coalescing BNS event detected by GWs~\cite{GW170817} (see Sec.~\ref{ptepGW_sec3}).  
However, even in that case 
the emission of the electromagnetic counterpart in the region of truly strong gravity
cannot be observed directly because of the large 
opacity of the high density matter. 

In contrast, GWs are emitted from the bulk motion of the objects where the 
strong gravity is really at work. 
Hence, GWs are thought to be an 
indispensable probe of strong gravity. 
Here, we summarize various modifications on the GW waveform by considering representative extended theories of gravity. 

\paragraph{Scalar tensor gravity}
As the most typical modification of gravity, one can consider 
scalar-tensor theories. A representative theory will be 
Brans-Dicke theory~\cite{Brans:1961sx}, defined by the action 
\begin{equation}
     S=\frac{1}{16\pi}\int d^4x\sqrt{-g}
       \left(\phi R-\omega_{BD}\phi^{-1}\phi_{,\alpha}\phi^{,\alpha}\right)
       -\sum_a \int d\tau_a m_a(\phi)\,,
\end{equation}
where $\omega_{BD}$ is the so-called Brans-Dicke parameter. 
When we send $\omega_{BD}$ to $\infty$, Einstein gravity is recovered.
$m_a(\phi)$ is the effective mass of a star labeled by $a$, and it depends on 
the local value of the scalar field around the star. 
The effective gravitational coupling depends on $\omega_{BD}$ as
$G_{\rm eff}=(4+2\omega_{BD})/(\phi(3+2\omega_{BD}))$, and 
the dimensionless 
scalar charge is defined by $s_a:=\partial \ln m_a(\phi)/\partial\ln \phi$, 
which can be interpreted as the dependence of the gravitational binding 
energy of the star on $G_{\rm eff}$. 

The modification to the GW waveform appears at $-1\,$PN order 
like $\psi(f)=\cdots+(3/128)(\pi{\cal M}f)^{5/3}\left[\alpha v^{-2/3}+1+\cdots\right]$ in Eq.~\eqref{simplewaveform}, 
where the coefficient $\alpha$ is 
given by $\alpha=-5(s_1-s_2)^2/(64\omega_{BD})$~\cite{Will:1989sk}. 
This is caused by the dipole radiation of the scalar field. 
If the scalar charge of two stars are proportional to their mass, 
i.e., if the dimensionless scalar charges are identical, the 
dipole radiation does not occur. Hence, the dipole radiation 
is not expected from equal-mass BNSs.

In this model, numerical approach is also possible, but the 
effect is larger for lower frequencies.
Therefore, the template based on 
the PN approach mentioned above would be 
sufficient, unless positive detection of the presence of 
scalar charge is reported. 

One scenario to give a scalar charge to compact stars is spontaneous 
scalarization in the context of generalized Brans-Dicke model, 
in which $\omega_{BD}$ is promoted to a function of $\phi$~\cite{Damour:1992we}. 
Rather intuitive understanding of this process can be obtained 
by introducing the canonically normalized field $\varphi=\int\sqrt{2\omega_{BD}(\phi)/\phi}d\phi$. 
In terms of this new field, the gravitational action would be 
rewritten as 
\begin{equation}
     S=\frac{1}{16\pi}\int d^4x\sqrt{-g}
       \left(\phi(\varphi) R-
       \frac{1}{2}\varphi_{,\alpha}\varphi^{,\alpha}\right)\,.
\end{equation}
From this expression, one can find that 
\blue{the $\varphi$-dependent part of} the energy of a static body would be given approximately by 
\begin{equation}
    \int d^3 x \left(\varphi_{,i}\varphi^{,i}-8\pi \phi(\varphi) \rho\right)
    \approx R^3\left(\varphi/R^2-8\pi \rho \phi(\varphi)\right)\,,
\end{equation}
where, after the curvature is replaced with the energy density $\rho$ using the 
Einstein equation as an approximation, the deviation 
from the Minkowski metric is neglected. In the last line, the 
size of the system is set to $R$ and $\varphi$ is assumed to 
vanish outside the compact object. 
If the function $\phi(\varphi)$ is convex upward at the origin 
and the density becomes high enough or 
the size of the compact object is large enough, non-trivial 
configuration of the scalar \blue{field} develops inside the star, which 
leads to giving a scalar charge to the star. The point is 
that this happens only in the situation where $\rho R^2\approx 
M/R$ is sufficiently large.


Within the category of scalar-tensor theory, more generalization is possible.
The Horndeski theory~\cite{Horndeski:1974wa} is the most general scalar-tensor theory
having second-order equations of motion.
This theory can, however, be generalized further in a healthy way to
the so-called degenerate higher-order scalar-tensor (DHOST)
theories~\cite{Langlois:2015cwa,Crisostomi:2016czh,BenAchour:2016fzp},
which inevitably have higher-order equations of motion, but still
propagate the same number of degrees of freedom
as the Horndeski theory does (i.e., one scalar and two tensor modes).
The Horndeski and DHOST theories offer us a powerful unifying framework to
study the physics of dark energy and
modified gravity (see, e.g.,~\cite{Kobayashi:2019hrl} for a review).

Measuring the speed of GWs can constrain
the Horndeski and DHOST theories.
The Lagrangian for the DHOST theories satisfying $c_{\rm GW}=1$ is  given
by~\cite{Creminelli:2017sry,Sakstein:2017xjx,Ezquiaga:2017ekz,Baker:2017hug}
\begin{align}
{\cal L}&=G_2(\phi,X)-G_3(\phi,X)\Box\phi+
f(\phi, X)R +A_3(\phi,X)\Box\phi \phi^\mu\phi_{\mu\nu}\phi^\nu
\notag \\ &\quad
+A_4 \phi^\mu\phi_{\mu\rho}\phi^{\rho\nu}\phi_\nu
+A_5(\phi^\mu\phi_{\mu\nu}\phi^\nu )^2,
\label{eq:dhost_Lag}
\end{align}
where $X:=-\phi^\mu\phi_\mu/2$ and
\begin{align}
A_4=-\frac{1}{2f}\left(2A_3f-3 f_X^2-2A_3f_XX+A_3^2X^2\right),
\quad
A_5=-\frac{A_3}{f}\left(f_X+A_3X\right).
\end{align}
We thus have 4 free functions of $\phi$ and $X$.
For $A_3=0$ and $f=f(\phi)$, this reduces to a subclass of the Horndeski theory.
Solar System and astrophysical experiments/observations
are potentially able to put further bounds on
the functions in the above Lagrangian.

For Solar System and astrophysical tests of the Horndeski and DHOST theories,
it is important to understand the Vainshtein screening mechanism~\cite{Vainshtein:1972sx}, 
which suppresses the scalar-mediated force and
typically is implemented in scalar-tensor theories
whose Lagrangian depends on the second derivatives of the scalar field.
In the Horndeski subclass with $A_3=0$ and $f=f(\phi)$,
the Vainshtein mechanism is known to work perfectly
for a static spherically symmetric body, $\Phi=\Psi=-G_NM/r$,
inside a certain radius (which is sufficiently larger than
the size of stars and the Solar System), where
$\Phi$ and $\Psi$ are two metric potentials defined by
$\delta g_{00}=-2\Phi$ and $\delta g_{ij}=-2\Psi\delta_{ij}$,
$G_N:=1/16\pi f$ is the effective Newton constant,
and $M$ is the mass contained within $r$.
Therefore, it would be difficult to
place Solar System/astrophysical constraints on this subclass.

In generic DHOST theories, the behavior of weak gravitational fields
is more interesting. It was found in~\cite{Langlois:2017dyl,Crisostomi:2017lbg,Dima:2017pwp} that
\begin{align}
\Phi'=G_N\left(\frac{M}{r^2}+\frac{\Upsilon_1}{4}M''\right),
\quad
\Psi'=G_N\left(\frac{M}{r^2}-\frac{5\Upsilon_2}{4}\frac{M'}{r}
+\Upsilon_3 M''\right),\label{eq:partial_br}
\end{align}
where $G_N$ and $\Upsilon_i$ are written in terms of $f$ and $A_3$,
and the prime denotes differentiation with respect to $r$.
Since $M=$ \blue{constant} outside the source body,
Eq.~\eqref{eq:partial_br} shows that gravity is modified only inside astrophysical bodies
in generic DHOST theories.
(In other words, Vainshtein screening operates nicely in the
vacuum exterior region.)
This partial breaking of Vainshtein screening
was first discovered in~\cite{Kobayashi:2014ida}.
A number of astrophysical tests of Vainshtein-breaking theories have been
proposed so far. Going beyond the weak gravity regime,
it \blue{was} shown in~\cite{Kobayashi:2018xvr} that the structure of relativistic stars is
quite sensitive to the Vainshtein-breaking effect, which would potentially give
tight constraints on DHOST theories in the strong gravity regime.

It was pointed out in~\cite{Creminelli:2018xsv,Creminelli:2019nok} that gravitons can decay
into $\phi$ in DHOST theories. To avoid this, it is required that
$A_3=0$. Otherwise, GWs would not be observed.
Upon imposing this constraint we have $A_4=3f_X^2/2f$ and $A_5=0$.
This yields a special subclass of DHOST theories characterized by 3 free functions,
in which the behavior of weak gravitational fields is
completely different from~\eqref{eq:partial_br},
as shown in~\cite{Hirano:2019scf,Crisostomi:2019yfo}.
(In fact, Eq.~\eqref{eq:partial_br} is derived assuming $A_3\neq 0$.)
The main result of~\cite{Hirano:2019scf,Crisostomi:2019yfo} is summarized as follows:
(i) a kind of fine-tuning is required even in the vacuum exterior region
so that Solar System tests are evaded;
(ii) in the interior region the metric potentials
$\Phi$ and $\Psi$ obey the standard inverse power law, but the two do not coincide;
(iii) the interior and exterior values of the effective Newton constant are
different,
\begin{align}
    G_{\rm in}=\frac{G_{\rm ext}}{1-Xf_X/f}\,,
    \quad 
    G_{\rm ext}=\frac{1}{16\pi f(1-Xf_X/f)}\,.
\end{align}
All of these features are
in sharp contrast with the case of generic DHOST theories.

A stringent bound on this particular class of DHOST theories
with $A_3=0$ comes from the fact that
the measured value of the Newton constant
(which is supposed to be $G_{\rm ext}$)
is different from
the gravitational coupling of GWs
by a factor of $1-Xf_X/f$. This can be seen from
the quadratic Lagrangian for GWs,
\begin{align}
    {\cal L}_{\rm GW}^{(2)}=\frac{1}{16\pi G_{\rm GW}}
    \left(h^{\mu\nu}\Box h_{\mu\nu}+\cdots\right)+
    \frac{1}{2}h^{\mu\nu} T_{\mu\nu}, \quad 
    G_{\rm GW}=\frac{1}{16\pi f}.
\end{align}
Assuming that
the scalar radiation does not take part in the energy loss,
one may naively replace the Newton constant in
the standard quadrupole formula with $G_{\rm GW}$.
Then, from the orbital decay of the Hulse-Taylor pulsar we find that
$|Xf_X/f|\lesssim 10^{-3}$.
It would be interesting to explore other astrophysical constraints
on this special class of Vainshtein-breaking modified gravity.

\paragraph{Quadratic gravity}
As a low energy effective theory of the dynamics of the metric, 
the term in Lagrangian at the lowest order of mass dimension 
is the cosmological constant and the next leading order is 
the Einstein-Hilbert action. In this sense, 
possible natural extension of gravity is to consider to 
add terms quadratic in curvature to the Lagrangian. 

There are three parity preserving terms, $R^2$,  
$R_{\mu\nu}R^{\mu\nu}$ and $R_{\mu\nu\rho\sigma}R^{\mu\nu\rho\sigma}$. 
$R^2$ term is classified as an alternative representation of scalar-tensor 
theories~\cite{Maeda:1988ab}. One combination, 
$R_{GB}=R^2-4R_{\mu\nu}R^{\mu\nu}+R_{\mu\nu\rho\sigma}R^{\mu\nu\rho\sigma}$, 
is know as the Gauss-Bonnet term, which does not contribute to the 
equations of motion as it can be written as a total divergence. 
Hence, truly remaining new possibility is, e.g., 
adding $R_{\mu\nu}R^{\mu\nu}$. 
However, this theory has a ghost degree of freedom, i.e., it 
contains infinitely many negative norm/energy states, which leads 
to an absence of stable vacua. This is caused by the presence of 
higher derivative terms in $R_{\mu\nu}R^{\mu\nu}$.
Hence, we need to treat such a model 
simply as a low energy effective theory. 
Then, the higher derivative terms can be replaced by using the 
lower order equations of motion, which is now the Einstein equations, 
up to an appropriate field redefinition~\cite{Weinberg:2008hq}. 
Once we apply this procedure to replace $R_{\mu\nu}R^{\mu\nu}$ 
with $T_{\mu\nu}T^{\mu\nu}$, the model is reduced to 
just Einstein gravity with matter fields described by a little 
complicated Lagrangian. 

As a parity violating term quadratic in the curvature, 
we can think of the Chern-Simons term, 
$R \tilde R:=\epsilon^{\alpha\beta}{}_{\sigma\chi}R^{\sigma\chi}{}_{\mu\nu}
R^{\mu\nu}{}_{\alpha\beta}/2$, 
where $\varepsilon^{\mu\nu\rho\sigma}$ is the Levi-Civit\'{a} tensor. 
However, this term is also written as 
a total divergence, and hence it does not contribute to the 
equations of motion. 

One interesting extension is to introduce a scalar field coupled 
to these higher order curvature terms. 
The two cases with a scalar field coupling to the Gauss-Bonnet term 
and the Chern-Simons term have been extensively studied. 
The Lagrangian is given by 
\begin{equation}
    L=\frac{1}{16\pi G_N}\left[R+\frac{\ell^2}{4}\phi 
    (R_{GB}~\mbox{or}~R\tilde R)-\frac{1}{2} 
    (\partial\phi)^2-V(\phi)\right]\,.
    \label{eq:eq22}
\end{equation}
where $\ell$ is the length scale that characterizes the coupling of 
the axion field to the Chern-Simons term. 
The deviation from GR in such theories can be easily hidden 
in the weak gravity regime, 
An interesting phenomena in this setup is that BHs 
can have a scalar hairs. The constraint coming from 
the GW observations is discussed in Sec.~\ref{sec:A01yamada}. 

\paragraph{Massive gravity}
If we simply add a mass to the graviton, the equations for the 
GW propagation in flat spacetime 
$\Box h_{\mu\nu}=0$ would be expected to 
be modified to $(\Box-m^2)h_{\mu\nu}=0$. 
However, this simple-minded extension is not theoretically 
very attractive. One consistent way that preserves the Lorentz 
symmetry on Minkowski background is the dRGT ghost-free massive gravity~\cite{deRham:2010kj}. 
However, it turned out to be difficult to construct even the homogeneous isotropic universe model
in this framework, and hence various generalizations were proposed. 
In any case, the dRGT ghost-free massive gravity model requires auxiliary metric or tetrad that 
do not transform as dynamical fields. As a result, 
the general covariance is not maintained.

In this class of models of massive gravity, linear perturbation predict 
a large excitation of a scalar graviton around a star, even if we send the 
graviton mass to a small value, which is known as the van Dam-Veltman-Zakharov discontinuity~\cite{vanDam:1970vg,Zakharov:1970cc}. 
However, in this limit, the linear analysis cannot be trusted any further
\blue{and the non-linear effects become large, strongly suppressing the excitation of the scalar graviton. This is the very same mechanism
as the Vainshtein mechanism introduced above~\cite{Vainshtein:1972sx}. }
As a result, the gravity around massive objects is thought to behave very similarly to GR. 

\paragraph{Bigravity}
A natural way to recover the general covariance in the theory with massive graviton 
is to promote the auxiliary metric field in the dRGT ghost-free massive gravity to a 
dynamical one. Such a model is know as dRGT bi-gravity theory, which is also 
shown to be ghost free by Hassan and Rosen~\cite{Hassan:2011zd}. 
However, the general covariance guarantees the presence of a massless graviton. 
In this model the field that becomes massive is the second graviton, which 
corresponds to the difference of two metrices, roughly speaking. 

The generation and propagation of GWs from binaries were studied in 
Ref.~\cite{DeFelice:2013nba}, in which the possibility of graviton oscillation 
between two graviton modes: one is massless while the other is massive, was pointed out. 
In the ordinary situation, only the metric directly coupled to the matter field 
is curved, while the other remains to be almost flat. 
In a particular choice of the model parameters, however, the massive graviton becomes 
very heavy around a star, and hence the difference between the two metrices 
is suppressed. In this case, the graviton oscillation becomes possible, without 
violating the test of gravity, say, tests in the solar system.
  
\paragraph{Einstein-aether theory}
The Einstein-aether theory, which for hypersurface-orthogonal configurations of the aether vector can be considered as the low energy limit of a candidate theory of quantum gravity called non-projectable Ho\v{r}ava-Lifshitz gravity and which also serves as a testing ground of GR. We thus review the theoretical and observational  constraints on the Einstein-aether theory
The basic variables of the gravity sector are~\cite{JM01} the four-dimensional metric $g_{\mu\nu}$, the aether vector $u^{\mu}$ and a Lagrange multiplier $\lambda$, where $\mu,\nu = 0, \cdots, 3$ and we adopt the signatures $(-, +,+,+)$ for the metric. The total action is~\cite{Jacobson} $S = S_{\textit{\ae}} + S_{m}$, where 
\begin{equation}
 S_{\textit{\ae}} = \frac{1}{16\pi G_{\textit{\ae}}}\int{\sqrt{- g} \; d^4x \Big[R(g_{\mu\nu}) + {\cal{L}}_{\textit{\ae}}\left(g_{\mu\nu}, u^{\lambda}\right)\Big]}\,,\ 
S_{m} = \int{\sqrt{- g} \; d^4x \Big[{\cal{L}}_{m}\left(g_{\mu\nu}, \psi\right)\Big]}\,.
\end{equation}
Here $\psi$ denotes matter fields, $R$ and $g$ are the Ricci scalar and the determinant of $g_{\mu\nu}$, and
\begin{equation}
 {\cal{L}}_{\textit{\ae}}  \equiv - M^{\alpha\beta}_{~~~~\mu\nu}\left(D_{\alpha}u^{\mu}\right) \left(D_{\beta}u^{\nu}\right) + \lambda \left(g_{\alpha\beta} u^{\alpha}u^{\beta} + 1\right)\,,
\end{equation}
where $D_{\mu}$ denotes the covariant derivative compatible with $g_{\mu\nu}$, and 
\begin{eqnarray}
M^{\alpha\beta}_{~~~~\mu\nu} = c_1 g^{\alpha\beta} g_{\mu\nu} + c_2 \delta^{\alpha}_{\mu}\delta^{\beta}_{\nu} +  c_3 \delta^{\alpha}_{\nu}\delta^{\beta}_{\mu} - c_4 u^{\alpha}u^{\beta} g_{\mu\nu}\,.
\end{eqnarray}
It is convenient to introduce linear combinations of $c_i$'s ($i=1,\dots,4$) as $c_{ij} \equiv c_i + c_j$ and $c_{ijk} = c_i + c_j + c_k$. The Newton constant $G_{N}$ is related to $G_{\textit{\ae}}$ as~\cite{CL04} $G_{N}= G_{\textit{\ae}}/(1 - \frac{1}{2}c_{14})$. 
 
The Minkowski spacetime with $u_{\mu} = \delta^{0}_{\mu}$ is a solution of the Einstein-aether theory. It is then straightforward to analyze linear perturbations around the Minkowski background and investigate properties of spin-$0$, -$1$ and -$2$ modes. The coefficients of the time kinetic term of each excitation must be positive, 
\begin{equation}
q_{S,V,T} > 0\,; \quad
 q_S = \frac{\left(1-c_{13}\right)\left(2+ c_{13} + 3c_2\right)}{c_{123}}\,,\quad q_V = c_{14}\,,\quad  q_T = 1-c_{13}\,.
\end{equation}
In addition to the ghost-free condition, we must also require the absence of gradient instabilities by demanding that the squared speeds of propagation be non-negative. Moreover, the almost simultaneous detection of GW170817~\cite{GW170817} and GRB 170817A~\cite{Monitor:2017mdv} sets a stringent constraint on the speed of the spin-2 mode as $- 3\times 10^{-15} < c_T -1 < 7\times 10^{-16}$, where $c_T^2=1/(1-c_{13})$. This implies 
\begin{equation}
\label{EAtheory:constraint-c13} 
\left |c_{13}\right| < 10^{-15}\,.
\end{equation}
As for spin-$0$ and -$1$ modes, the squared speeds must be greater than $1-\mathcal{O}(1)\times 10^{-15}$, in order to avoid vacuum gravi-\v{C}erenkov radiation~\cite{EMS05}. We thus impose
 \begin{equation}
c_{S,V}^2 \gtrsim 1\,; \quad
 c_S^2 = \frac{c_{123}(2-c_{14})}{c_{14}(1-c_{13}) (2+c_{13} + 3c_2)}\,,\quad
 c_V^2 = \frac{2c_1 -c_{13} (2c_1-c_{13})}{2c_{14}(1-c_{13})}\,, 
\end{equation}
in addition to (\ref{EAtheory:constraint-c13}).

Applying the theory to cosmology, one can find that the gravitational constant appearing in the effective Friedman equation is given by~\cite{CL04},
\begin{equation}
G_{{\mbox{cos}}} = \frac{G_{\textit{\ae}}}{1+\frac{1}{2}(c_{13} + 3c_2)} = \frac{1-\frac{1}{2}c_{14}}{1+\frac{1}{2}(c_{13} + 3c_2)}G_N\,.
\end{equation}
One then needs to impose the nucleosynthesis bound as 
\begin{equation}
\left|\frac{G_{{\mbox{cos}}}}{G_N} - 1\right| \lesssim \frac{1}{8}\,.
 \end{equation}

Among the 10 parameterized post-Newtonian (PPN) parameters~\cite{Will06},  the only two parameters that deviate from GR in this context are
\begin{equation}
\alpha_1 = -  \frac{8(c_3^2 + c_1c_4)}{2c_1 - c_1^2 +c_3^2 }\,, \quad 
\alpha_2 = \frac{1}{2}\alpha_1  - \frac{(c_1 +2c_3 - c_4)(2c_1 + 3c_2+c_3+c_4)}{c_{123}(2-c_{14})}\,. 
\end{equation}
In the weak-field regime, Solar System observations set the constraints~\cite{Will06}
\begin{equation}
\left| \alpha_1\right| \le 10^{-4}\,, \quad 
 \left|\alpha_2\right| \le 10^{-7}\,.
 \end{equation}

In the strong-field regime, the isolated millisecond pulsars PSR B1937 + 21~\cite{SPulsarA} and PSR J17441134~\cite{SPulsarB} set the constraint~\cite{Shao:2012eg,Shao:2013wga} as
\begin{equation}
\left|\hat\alpha_1\right| \le 10^{-5}\,, \quad 
 \left|\hat\alpha_2\right| \le 10^{-9} \quad  (95\% \mbox{ confidence})\,,
 \end{equation}
where ($\hat\alpha_1, \hat\alpha_2$) denotes the strong-field generalization of  ($\alpha_1, \alpha_2$)~\cite{DEF92} given by~\cite{Yagi14}, 
\begin{equation}
\hat\alpha_1 = \alpha_1 + \frac{c_-(8+\alpha_1)\sigma_{\textit{\ae}}}{2c_1}\,, \quad 
\hat\alpha_2 = \alpha_2 + \frac{\hat\alpha_1 - \alpha_1}{2}   
- \frac{(c_{14} -  2)(\alpha_1 -2\alpha_2)\sigma_{\textit{\ae}}}{2(c_{14} - 2c_{13})}\,.
\end{equation}
Here, $\sigma_{\textit{\ae}}$ is the sensitivity. Unfortunately, the sensitivities $\sigma_{\textit{\ae}}$ of a neutron star, which depend on $c_i$'s and the equation of state of nuclear matter~\cite{Yagi14}, are  not known for the new ranges of the parameters. We thus leave the analysis of these two constraints to a future work.

\begin{figure}
{
 \includegraphics[width=8.7cm]{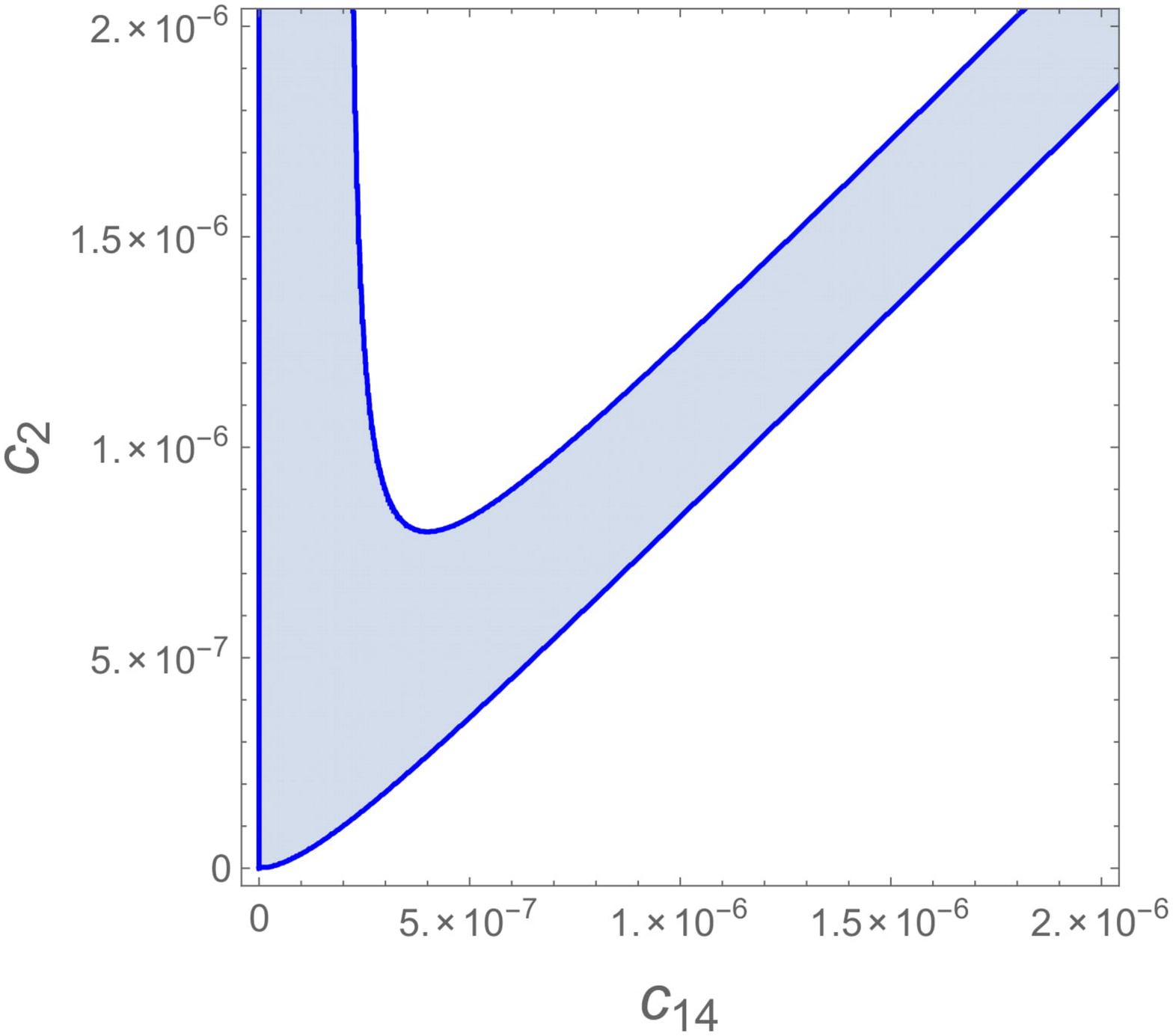}
 \includegraphics[width=7.5cm]{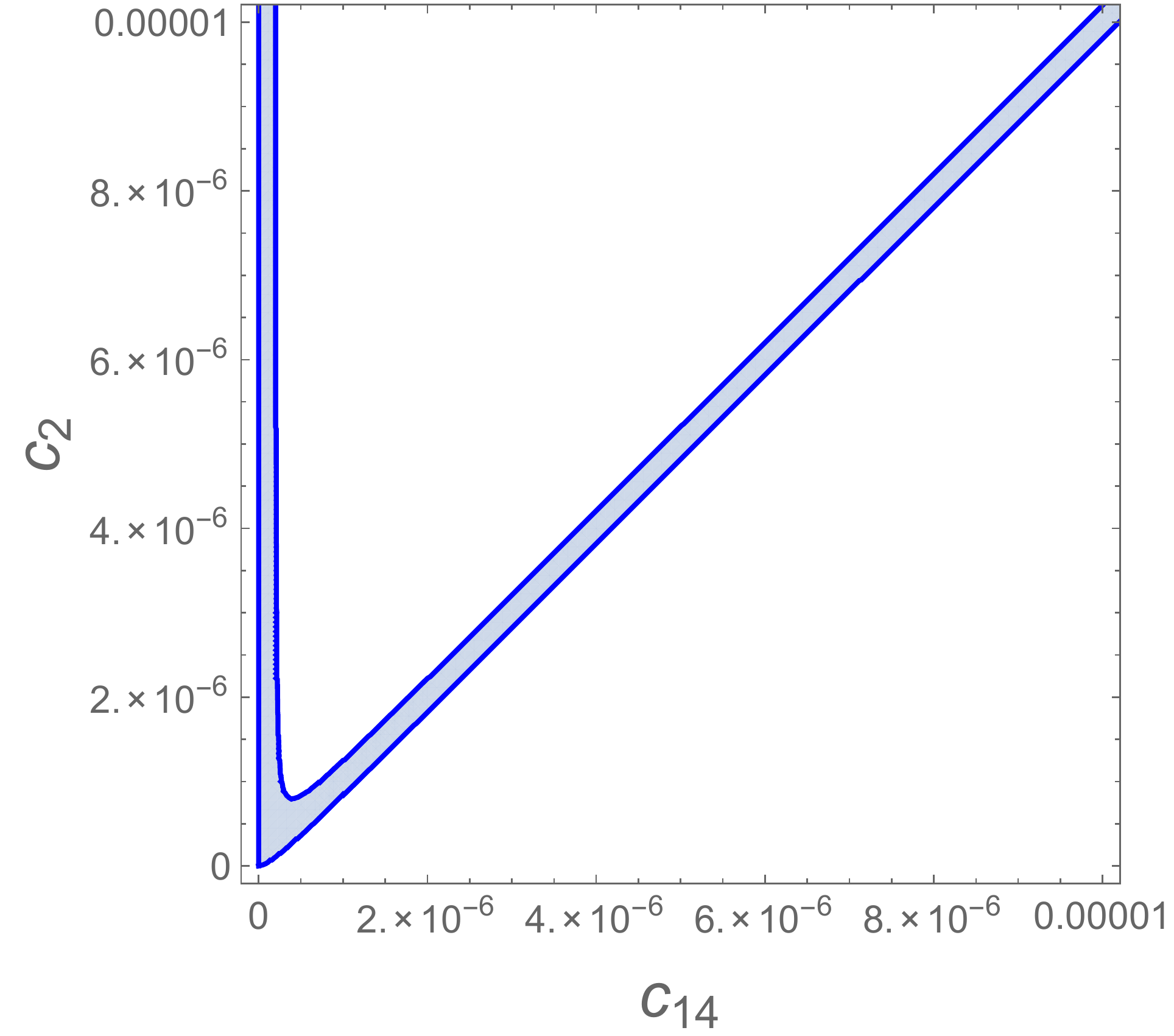}
}
\caption{This figure, adopted from~\cite{Oost:2018tcv}, shows the constraints in the ($c_2, c_{14}$)-plane summarized in (\ref{EAtheory:constraints-c2-c14-plane}) with two different scales.}
\label{fig1}
\end{figure}

Putting all theoretical and observational constraints together, we have (\ref{EAtheory:constraint-c13}),  as well as
\begin{eqnarray}
0 < c_{14} \le 2.5\times 10^{-5}, \quad 
c_{14} \lesssim c_1
\end{eqnarray}
in the ($c_1, c_{14}$)-plane and
\begin{equation}
\label{EAtheory:constraints-c2-c14-plane}
 0 < c_{14} \lesssim c_2 \lesssim  0.095\,, \quad
- 10^{-7} \le  \frac{c_{14}\left(c_{14} + 2c_2c_{14} - c_2\right)}{c_2\left(2-c_{14}\right)} \le 10^{-7}
\end{equation}
in the ($c_2, c_{14}$)-plane. In Fig.~\ref{fig1}, we show the constraints (\ref{EAtheory:constraints-c2-c14-plane}). 
All the constraints except (\ref{EAtheory:constraint-c13}) are divided into two classes, those in the ($c_1, c_{14}$)-plane and those in the ($c_2, c_{14}$)-plane. For further discussions about the constraints such as the comparison with those on the hypersurface-orthogonal theory~\cite{GSS18}, see~\cite{Oost:2018tcv}.

%


\subsubsection{Dedicated test for particular models}\label{sec:A01yamada}
To optimally constrain particular models, it would be 
better to employ dedicated test. 
One can imagine various extensions of the simple GW waveform 
adopted in the PPE framework.  

\paragraph{Massive dipolar radiation}
One simple example is to add a mass to the field that 
extracts energy from the system through the dipole radiation. 
Since the radiation is emitted only in the frequency range above the mass scale, 
the modifications in the orbital evolution and hence in 
the GW phase evolution show up only at high frequencies~\cite{Sampson:2013jpa, Sampson:2014qqa}. 
Therefore, even if a NS has a scalar hair, such scalar-tensor 
theories can evade the constraint from pulsar observations due to the mass of the scalar field. A simple model of  scalar-tensor theories with a massive scalar field has been constrained by several observations: (1) $m \lesssim 10^{-27}$eV for stability on cosmological scales~\cite{Ramazanoglu:2016kul}, (2) $m \lesssim 10^{-16}$eV from the observations of a pulsar-white dwarf binary~\cite{Antoniadis:2013pzd}, and (3) $10^{-13}$eV $\lesssim m \lesssim 10^{-11}$eV~\cite{Arvanitaki:2014wva, Brito:2017zvb}, which relies on the measurements of high spins of stellar mass BHs~\cite{Narayan:2013gca}.

As mentioned earlier, Einstein-dilaton-Gauss-Bonnet (EdGB) theory, 
is an interesting candidate of modification of gravity obtained by 
adding a quadratic term in curvature. We need to introduce a 
scalar field, which we call dilaton here, 
to make the Gauss-Bonnet term relevant. 
In this model a BH can possess a scalar charge and hence 
dipole radiation is expected. 
The constraint from observed GW data on the 
model in which the scalar field is massive was examined in 
Ref.~\cite{Yamada:2019zrb}. In the massless limit, on the other hand, the current constraint on EdGB theory is $\sqrt{\alpha_{\rm EdGB}} \lesssim 1.9\,$km, which is obtained by using low-mass X-ray binaries~\cite{Yagi:2012gp}.

Activation of dipole radiation due to massive fields and modifications to GW waveform has been analyzed for the GW events in the catalog GWTC-1~\cite{GWOSCweb} to constrain the magnitude of the modification nd~\cite{Yamada:2019zrb}. 
Since the ground based detectors, LIGO, Virgo, and KAGRA, are sensitive at frequencies around 10-1000 Hz, GW signals for those are useful to test dipole radiation of the massive field in the range $10^{-14}$eV $\lesssim m \lesssim 10^{-13}$eV. For $m \lesssim 10^{-14}$eV the modification of GWs effectively 
reduces to the massless limit.

Taking into account the dominant correction to the energy flux due to the dipole radiation, which appears at $-1\,$PN order, the GW waveform in the frequency domain, $\tilde{h} (f)$, can be expressed as~\cite{Stein:2013wza, Yunes:2016jcc}
\begin{align}
  \label{Eq:ModGWofMassiveField}
  \tilde{h} (f) = \tilde{h}_{\rm GR} ( 1 + \delta \mathcal{A} ) \, e^{i \delta \Psi} .
\end{align}
The corrections to the amplitude and the phase are, respectively, expressed as~\cite{Yamada:2019zrb}
\begin{align}
  \label{Eq:ModGWamp}
  \delta \mathcal{A} &= - \frac12 A \frac{( \hat{\omega}^2 - 1 )^{3/2}}{\hat{\omega}^3} \, u^{-2} \, \Theta (\hat{\omega}^2 - 1) , \\
  \label{Eq:ModGWphase}
  \delta \Psi &= \frac{5}{48} A \left( \frac{1}{m \, \mathcal{M}} \right)^{7/3}  F ( \hat{\omega} ) \, \Theta (\hat{\omega}^2 - 1) ,
\end{align}
with
\begin{align}
  \label{Eq:exactF}
  F (\hat{\omega}) &\equiv - \int_{\hat{\omega}}^{\infty} d\!\!\:\hat{\omega}' \,  \frac{\hat{\omega} - \hat{\omega}'}{{\hat{\omega}'}{}^{22/3}} \left[ \hat{\omega}'{}^2 - 1 \right]^{3/2} , 
\end{align}
where $\hat{\omega} = \pi f / m$, $m$ is the mass of the field, $\Theta$ is the Heaviside step function, 
$A$ is a parameter denoting the amplitude of the dipole radiation relative to the quadrupole one. 
The modified waveform, Eq.~\eqref{Eq:ModGWofMassiveField} with Eqs.~\eqref{Eq:ModGWamp} and \eqref{Eq:ModGWphase}, can model the modification due to the vector field as well as the scalar one. In this parametrization, there are two free parameters, i.e., the mass of the field $m$ and the relative amplitude of dipole radiation $A$. Note that $A$ is a function of the system parameters, such as the masses and spins, and the parameters of the modified gravity theory that one considers.

Using the above waveform as a template, 
90\% confidence level (CL) constraints on $A$ for each LVC GW event has been derived~\cite{Yamada:2019zrb}.
Since the modified waveform is valid only for the inspiral phase, the calculations were terminated when the mass scale reaches an 
appropriately chosen threshold frequency for each event. 
We should recall that 
in general a strong constraint on $A$ 
does not always imply a strong constraint on the modification to theory. 
This is because $A$ is proportional to the squared difference of the scalar charges of constituents of the binary as we will see in Eq.~\eqref{Eq:EdGBtypeCoupling}. 
Therefore, GW events with vanishing of this difference are not sensitive to the modification, 
even for a quite large coupling constant.

The constraints on $A$ can be translated to a constraint on the coupling strength, using the relation
~\cite{Yagi:2011xp, Berti:2018cxi, Prabhu:2018aun}
\begin{align}
  \label{Eq:EdGBtypeCoupling}
  A = \frac{5 \pi}{6} \frac{\alpha_{\rm EdGB}^2}{M^4} \frac{( m_1^2 s_2^{\rm EdGB} - m_2^2 s_1^{\rm EdGB} )^2}{M^4 \eta^{18/5}} ,
\end{align}
where
\begin{align}
  s_i^{\rm EdGB} = \frac{2 \sqrt{1 - \chi_i^2}}{ 1 + \sqrt{1 - \chi_i^2}} \,, ~~~ (i = 1,2)
\end{align}
describes the spin-dependence of the BH's scalar charges. 
After combining five BBH events possessing relatively high SNR, i.e., GW150914, GW170814, GW170608, GW170104, GW151226.
$\sqrt{\alpha_{\rm EdGB}}$ is constrained as $\sqrt{\alpha_{\rm EdGB}} \lesssim 2.47\,$km for all mass below $6 \times 10^{-14}$eV for the first time with 90\% CL including $\sqrt{\alpha_{\rm EdGB}} \lesssim 1.85\,$km in the massless limit. This constraint in the massless limit is much stronger than the results of in Ref.~\cite{Nair:2019iur, Tahura:2019dgr}, while it is accidentally almost the same with that by low-mass X-ray binaries~\cite{Yagi:2012gp}.

Since the number of events with a lighter chirp mass, such as BNSs, must increase in the near future, it is interesting to stack those events by assuming various theories in which charged NSs are allowed consistently. Furthermore, the NS-BH binary is also interesting~\cite{Carson:2019fxr}, because, in addition to their light chirp mass, 
$A$ can be large in a theory such that only either NS or BH has a charge, 
as in the case of EdGB theory. 
Moreover, future multiband observations will improve the current upper bounds of various modified theories~\cite{Gnocchi:2019jzp}.

\paragraph{Propagation of GWs in parity violating gravity}

The nearly simultaneous detection of GWs and the gamma-ray burst
from the merger of neutron stars, GW170817/GRB170817 A~\cite{GW170817,Monitor:2017mdv},
offered us an unprecedented opportunity to measure the speed of GWs,
$c_{\rm GW}$, at the level of one part in $10^{15}$.
This puts a tight constraint on modified gravity as an alternative
to dark energy. We now discuss the impact of this constraint
on parity-violating sector of gravity~\cite{Nishizawa:2018srh}

Let us consider parity-violating gravity whose action is of the form
\begin{align}
S=\frac{1}{16\pi G}\int d^4x\sqrt{-g}\left[
R+{\cal L}_{\rm PV}+{\cal L}_\phi
\right] \,,
\end{align}
where ${\cal L}_{\rm PV}$ is the parity-violating Lagrangian and
${\cal L}_\phi$ is the Lagrangian for a scalar field $\phi$ which
may be coupled nonminimally to gravity.
The most frequently studied example is Chern-Simons (CS) gravity,
for which ${\cal L}_{\rm PV}$ is given by
\begin{align}
{\cal L}_{\rm PV}={\cal L}_{\rm CS}:=f(\phi)R\tilde R,
\end{align}
Recently, it was found that one can construct other parity-violating
Lagrangians for the gravity sector such as~\cite{Crisostomi:2017ugk}
\begin{align}
{\cal L}_{\rm PV}=\sum_{I=1}^4a_I(\phi,\phi_\mu\phi^\mu)L_A \,,
\label{eq:new_L_parity}
\end{align}
with
$L_1:=\varepsilon^{\mu\nu\alpha\beta}R_{\alpha\beta\rho\sigma}%
R_{\mu\nu~\lambda}^{~~\;\rho}\phi^\sigma\phi^\lambda$,
$L_2:=\varepsilon^{\mu\nu\alpha\beta}R_{\alpha\beta\rho\sigma}%
R_{\mu\lambda}^{~~\;\rho\sigma}\phi_\nu\phi^\lambda$,
$L_3:=\varepsilon^{\mu\nu\alpha\beta}R_{\alpha\beta\rho\sigma}%
R^\sigma_{~\;\nu}\phi^\rho\phi_\mu$,
$L_4:=\phi_\lambda\phi^\lambda P$, and $\phi_\mu:=\nabla_\mu\phi$.
If $4a_1+2a_2+a_3+8a_4=0$ is satisfied, the dangerous Ostrogradsky ghosts are
removed in the unitary gauge.\footnote{The
ghost degrees of freedom might not be problematic
if the theory is treated as a low-energy truncation
of a fundamental theory. Although these ghost modes do
cause instabilities if the theory is regarded as a complete theory,
it is argued in~\cite{DeFelice:2018ewo}
that the would-be ghost mode
that is not manifest in the unitary gauge may be
an instantaneous mode which does not in fact propagate
and hence is not dangerous.}

We consider GWs $h_{ij}(t,\Vec{x})$ propagating on a cosmological background.
The above examples yield the quadratic Lagrangian of the form
\begin{align}
{\cal L}_{\rm PV}^{(2)}=\frac{1}{4}\left[
\frac{\alpha(t)}{a\Lambda}\epsilon^{ijk}\partial_t h_{il}
\partial_j \partial_t h_{kl}+\frac{\beta(t)}{a^3\Lambda}\epsilon^{ijk}\partial^2 h_{il}
\partial_j h_{kl}\right] \,,
\label{eq:L2_parity_gws}
\end{align}
where $\epsilon^{ijk}$ is the antisymmetric symbol, $\Lambda$ is some energy scale,
and $\alpha$ and $\beta$ are dimensionless functions of time
determined essentially by the cosmological evolution of the scalar field.
In the most familiar case of CS gravity, we identically have $\alpha=\beta$.
However, we may have $\alpha\neq\beta$ in general.
The Lagrangian~\eqref{eq:L2_parity_gws} can thus be used to describe
the GW propagation in parity-violating gravity in a unifying
way~\cite{Nishizawa:2018srh}  
(see also~\cite{Creminelli:2014wna} for an effective-field-theory approach
to arrive at the same Lagrangian).

In the presence of the parity-violating interactions~\eqref{eq:L2_parity_gws},
the $+$ and $\times$ polarization modes are coupled
in the equations of motion, but one can obtain two decoupled equations
by employing the circular polarization basis. In general parity-violating gravity,
the equations of motion for GWs in Fourier space
can be written as
\begin{align}
( h_{\Vec{k}}^A )^{\prime \prime} + (2+\nu_A) {\cal H} ( h_{\Vec{k}}^A )^{\prime}
+ (c_{\rm GW}^A)^2 k^2 h_{\Vec{k}}^A =0 \,,
\end{align}
where the prime denotes differentiation with respect to the conformal time,
${\cal H}:=a'/a$, and
\begin{align}
\nu_A =
\frac{ \lambda_A (\alpha-\alpha^{\prime} {\cal H}^{-1})}{1-\lambda_A \alpha k/a\Lambda}
\frac{k}{a\Lambda} \,,
 \quad
 (c_{\rm GW}^A )^2 = \frac{1-\lambda_A \beta k/a\Lambda}{1-\lambda_A\alpha k/a\Lambda } \,.
\end{align}
Here, $A$ labels left and right circular polarization modes
and $\lambda_A=\pm 1$.
It can be seen that CS gravity corresponds to the special case with
$(c_{\rm GW}^A)^2=1$. Therefore, from the propagation speed of gravity
no constraint can be imposed on CS gravity.
However, parity violation in the gravity sector in general leads to
$(c_{\rm GW}^A)^2\neq 1$ and thus is tightly constrained by GW170817.
For $k/a\sim 100\,$Hz, one has
\begin{align}
\Lambda^{-1}|\alpha-\beta|\lesssim 10^{-11}\,{\rm km} \,.
\end{align}
This is the new limit on parity violation in the gravity sector
derived in~\cite{Nishizawa:2018srh}.

By parametrizing the modification of GW waveforms during propagation in this theory, the data in the O1/O2 catalog has been reanalyzed by a grid search with the matched filtering method~\cite{Yamada:2020zvt}. The results imply that the above constraint improves by 7 digits since the matched filtering is sensitive to the phase modifications, while the constraints on the amplitude modification are similar to the previous result by Yagi and Yang~\cite{Yagi:2017zhb}.

For a face-on binary, the effects of parity violation of gravity cannot be constrained because only one of the circular polarization modes can be observed. Therefore, obtaining a more stringent constraint or find the signature of parity-violating gravity requires the detection of GW signals from a nearly edge-on binary.

\paragraph{Echoes from merger remnant black holes}
Exotic compact objects are proposed as alternative models to BHs.
They are assumed to be compact enough to possess the light ring, but without the event horizon.
In such models, the feature of GWs after the ringdown phase of compact binary coalescences is considered to be different from the BH case.
The infalling waves generated during merger-ringdown phase 
might be reflected at the surface of the object, 
which would be just absorbed if the horizon exists.
The reflected waves will be partly reflected back again 
at the angular momentum barrier but partly transmitted.
This process occurs repeatedly, which phenomenon is dubbed as GW
``echoes''~\cite{Cardoso:2016oxy,Cardoso:2016rao}
(see also Ref.~\cite{Hui:2019aox} for a simple explanation
with the causal structure of the Green's function
of the wave equation with two potentials
having disjoint bounded support).
Therefore, the detection of echoes implies the existence of 
an exotic compact objects.
The simplest model of the compact objects is that the horizon is replaced by a reflecting membrane at $\sim$ Planck proper length from the horizon radius in Kerr spacetime.
Abedi {\it et al.} have analyzed three O1 BBH events and reported that they found echo signals at 3$\sigma$ significance level (0.011 in p-value)~\cite{Abedi:2016hgu}.
In their study, they assume the GW waveform in the frequency domain as 
\begin{equation}
\tilde{h} (f) =   \tilde{h}_0(f) 
 \sum^{N}_{n=1} \gamma^{n-1} e^{- i \left [2\pi f \Delta t_{\rm echo}+ \phi(f) \right ] (n-1)}  \, , 
\label{template1}
\end{equation}
where $\gamma$ is the overall reflection rate, $\Delta t_{\rm echo}$ is the interval between neighboring echoes, $N$ is the number of echoes, and $\phi(f)$ is the phase shift.
The function $\tilde{h}_0(f) $ is a merger-ringdown waveform extracted by cutting off the 
inspiral part smoothly from the inspiral-merger-ringdown waveform.
The parameters $\gamma$ and $\Delta t_{\rm echo}$ are the most relevant parameters to characterize the echo waveforms on 
 which the significance of the signal depends most. 
Since Abedi {\it et al.} used only 32 seconds for the background estimation,
Westerweck {\it et al.}~\cite{Westerweck:2017hus} have improved the background estimation using 4096 seconds and shown lower significance, 0.032 in p-value, with the same template used in the paper by Abedi {\it et al.}~\cite{Abedi:2016hgu}.
The Bayesian analysis is also performed with the same template and small Bayes factors are
reported~\cite{Nielsen:2018lkf,Lo:2018sep}.
The spacetime considered in Abedi {\it et al.} is exactly Kerr spacetime outside the surface, while the assumption of reflection rate in their study is not physically reasonable.
Nakano {\it et al.} have derived the reflection rate at the angular momentum potential barrier numerically, 
and calculated a fitting formula for the reflection rate $R(f)$~\cite{Nakano:2017fvh}.
The reflection rate becomes frequency dependent, unlike in the assumption in Abedi {\it et al.}, and behaves as a high pass filter, i.e., $R(f) \to 0 \, \left [R(f) \to 1 \right ]$ for larger (smaller) $f$.
As a result, the lower frequency part of the waveform is strongly suppressed in the template given by Nakano {\it et al.} compared to that given by Abedi {\it et al.}.
The modified template becomes
\begin{equation}
\tilde{h} (f) =  \sqrt{1-R^2(f)}\, \tilde{h}_0(f) 
 \sum^{N}_{n=1} R(f)^{n-1} e^{- i \left [2\pi f \Delta t_{\rm echo}+ \phi(f) \right ] (n-1)}  \, . 
\label{template1}
\end{equation}
In this setup, Uchikata {\it et al.} have shown that the significance of echo 
signals becomes much lower, p-value is larger than 0.9 for eight BBH events in O1 and O2~\cite{Uchikata:2019frs}, 
although neither reduction nor increase of the significance was observed by increasing the number of samples 
when the original waveform proposed by Abedi {\it et al.} is employed. 

So far, we have focused on a model dependent search.
A morphology independent search was done
by Tsang {\it et al.}~\cite{Tsang:2018uie,Tsang:2019zra}, where they assumed that echo waveforms for any model can be effectively described as superpositions of sine-Gaussian waveforms. 
In their study, no significant echo signal is found in O1 and O2 events, including the BNS merger event.
The reflection rate at the surface can also vary due to the model of the object. 
In the analytical template proposed by Refs.~\cite{Testa:2018bzd,Maggio:2019zyv}, 
the reflection rate at the object's surface is also assumed as a parameter.
The reflection rate at the potential barrier is given from BH perturbations.
Further details of testing exotic compact objects includes the echo phenomenon are reviewed in Ref.~\cite{Cardoso:2019rvt}.

%% file: ptepGW_sectionA02.tex
Arguably, one of the most important questions in science is to elucidate the nature of dark matter.
Identifying the nature of dark matter definitely brings huge impact on cosmology,
particle physics, and astronomy.
Many dark matter candidates have been proposed in the literature.
Since different dark matter candidates yield different type of observational signals,
various experiments targeted for particular dark matter have been conducted.
In this subsection, we give a brief overview of how the GW experiments can shed light 
on some dark matter candidates. 

\subsubsection{Axion}

An axion is a strong candidate for the dark matter\cite{Marsh:2015xka}.
To probe the axion dark matter, it is important to study electromagnetic waves~\cite{Yoshida:2017ehj}
and GWs
propagating in the axion dark matter~\cite{Yoshida:2017cjl,Soda:2017sce}. 
In particular,  the gravitational Chern-Simons term allows us to probe the
parity violation in the gravity sector.
We consider the model specified by the Lagrangian~\eqref{eq:eq22}
%
with $V=m^2\phi^2/2$.
%

%
It is instructive to write down the quadratic action 
deduced from Eq.~\eqref{eq:L2_parity_gws} using the 
circular polarization modes as 
\begin{eqnarray}
S^{(2)} = \frac{M_p^2}{4}
 \int{\rm d}^3k\, dt \  \left(1 - \frac{ k\lambda_{A}}{M_p } \ell^2 \dot{\phi} \right) 
         \left[  \dot{h}^{A}_{\bf k} \dot{h}^{A}_{\bf -k} -  k^2 h^{A}_{\bf k}  h^A_{\bf -k} \right]  \,.
\end{eqnarray}
%
To avoid a ghost, due to the signature flip 
of the factor $\left(1 - k\lambda_{A} \ell^2 \dot{\phi}/M_p \right)$, 
we need the cutoff
scale below which we can use this effective action
\begin{eqnarray}
k < k_g = 6.7\times10^{9}{\rm Hz}\left(\frac{10^{8}{\rm km}}
{\ell}\right)^{2}\sqrt{\frac{0.3{\rm GeV/cm^{3}}}{\rho}} \,.
\end{eqnarray}
The axion is coherently oscillating as
$\phi=\phi_0 \cos(m t) $
with the amplitude determined by
\begin{eqnarray}
\phi_{0}\simeq2.1\times10^{7}{\rm eV}\,
\left(\frac{10^{-10}{\rm eV}}{m}\right)\sqrt{\frac{\rho}{0.3{\rm GeV/cm^{3}}}} \,.
\end{eqnarray}
Here, we used the observed local dark matter density $\rho$.
The coherent oscillation induces the parametric resonance of GWs with a frequency corresponding to the axion mass.
Using the current upperbound $\ell \leq 10^8 {\rm km}$~\cite{AliHaimoud:2011fw},
we  can estimate the length scale $R_{\times10}$ at which the GW
grows ten times larger as follows
\begin{eqnarray}
R_{\times10}
=5.2\times10^{-8}{\rm pc}\,\left(\frac{10^{-10}{\rm eV}}{m}
\right)^{2}\,\left(\frac{10^{8}{\rm km}}{\ell}\right)^{2}
\sqrt{\frac{0.3{\rm GeV}/{\rm cm}^{3}}{\rho}} \,.
\end{eqnarray}
Since the
Jeans length $r_{{\rm J}}$ of the axion dark matter, which is the coherence length of the axion oscillation 
\begin{eqnarray}
r_{{\rm J}} 
 =4.3\times10^{-3}{\rm
  pc}\,\left(\frac{m}{10^{-10}\,{\rm
    eV}}\right)^{-\frac{1}{2}}\left(\frac{\rho}{0.3\,{\rm
    GeV/cm^{3}}}\right)^{-\frac{1}{4}}
\end{eqnarray}
is much larger than the growth length,
there occurs a huge enhancement
of the amplitude of GWs, 
which gives rise to the strong constraint on the Chern-Simons coupling constant or the abundance of the axion dark matter.
We refer the reader to the original
paper~\cite{Yoshida:2017cjl} for the details.
\blue{For the detection strategy to probe this effect is under investigation.}

\subsubsection{Primordial black holes}
Primordial black holes (PBHs) are a dark matter candidate that is not an elementary particle.
The idea of PBHs dates back to 1971 when Hawking proposed a possibility that
BHs are produced directly by the gravitational collapse of an overdense region in the primordial
universe much prior to the recombination era. 
PBHs heavier than about $10^{15} {\rm g}$ do not lose their mass significantly due to the Hawking radiation
over the age of the Universe.
Since such PBHs are non-relativistic objects, interact with baryonic matter only gravitationally, and
do not emit light, they have been considered as a dark matter candidate.
Since the original proposal of PBHs, various observational searches for PBHs have been continuously conducted 
for a wide PBH mass range.
So far, there are no observational evidence of PBHs, and upper limits on the PBH abundance have been placed,
which, for some mass range, excluded the possibility that PBHs comprise all dark matter under the assumption that
all PBHs have the same mass.
Yet, there is still an open window where PBHs can provide all dark matter (see Ref.~\cite{Carr:2020gox} for the most
recent summary).

LIGO/Virgo detection of binary BHs provoked explosive research activities on PBHs (see Ref.~\cite{Sasaki:2018dmp} for a review
in this respect).
An exciting possibility is that these BHs (or a part of them) are PBHs.
Indeed, features of the LIGO BHs such as comparatively large masses and small spins do not contradict with PBHs.
It is known that a dominant formation channel of the PBH binaries is by a tidal force by 
the surrounding PBHs in the radiation dominated
era~\cite{Nakamura:1997sm}.
This channel can explain the merger rate of the BH binaries derived by the LIGO/Virgo observations if
the PBHs constitute about $10^{-3}$ fraction of dark matter.
Alternatively, a conservative assumption that LIGO BHs were produced by the standard astrophysical processes 
yields an upper limit on the abundance of stellar mass PBHs.
Irrespective of whether LIGO BHs are PBHs or not, that PBHs can comprise at most 
$\sim 10^{-3}$ of dark matter is the most severe constraint for PBHs in the stellar mass range.
This vividly demonstrates the powerfulness of the GW observations to test PBHs as dark matter.
It is worth to mention that LIGO/Virgo Collaboration extended searches of the PBH mergers to sub-solar mass
range~\cite{Authors:2019qbw}.

Although stellar mass PBHs are unlikely to explain all the dark matter, a question still remains that
such PBHs constitute a tiny fraction of dark matter.
Answering to that question definitely gives a profound consequence to cosmology.
To this end, it is essential to discriminate between PBHs and the astrophysical BHs from GW observations.
In what follows, we give a short review of the several ideas along this direction proposed in the literature. 

\paragraph{Mergers at high redshifts}
Contrary to astrophysical BH binaries which formed long after the Big Bang, 
PBH binaries exist since the Universe was still dominated by radiation.
Due to a broad distribution of orbital radius and eccentricity of the PBH binaries,
merger rate of the PBH binaries continuously extends to high redshift even beyond $z\sim 100$.
On the other hand, the merger rate of the astrophysical BHs sharply drops beyond $z\sim 10$.
Thus, searches for mergers beyond $z\sim 10$ provide a clean test of the existence of PBHs.
Reference~\cite{Nakamura:2016hna} has shown that the proposed space-interferometer called B-DECIGO is able to achieve this goal.

\paragraph{Mass distribution}
Mass distribution is another observable that can be used to test PBHs.
Let $m_1$ and $m_2$ be masses of the individual BHs in a binary and ${\cal R}(m_1,m_2,t)$ be the distribution of
the merger rate of the BH binaries at cosmic time $t$.
Notice that ${\cal R}$ includes all the mergers and the observational bias, which depends on the detectors,
should be taken into account in translating ${\cal R}$ to directly measurable merger distribution. 
Since this bias is possible to handle, we here regard ${\cal R}$ as an observable quantity.

Reference~\cite{Kocsis:2017yty} showed that PBH binaries formed in the radiation dominated era gives a characteristic mass 
dependence to ${\cal R}$ as
\begin{equation}
{\cal R}=C \psi (m_1) \psi (m_2) {(m_1+m_2)}^\alpha \,.
\end{equation}
Here $C$ is a constant independent of $m_1, m_2$, $\psi (m)$ is a function which depends on the PBH mass function, 
and $\alpha \approx 1$.
Quite interestingly, ${\cal R}$ almost linearly depends on the total mass $m_1+m_2$ irrespective of the PBH mass function
which differs for each inflation model and is unknown\footnote{
Recent study generalized the original work and found that $\alpha$ takes value different from unity if the
dark matter perturbation provides the dominant torque to produce the angular momentum of the PBH
binaries~\cite{Garriga:2019vqu}.}.
It is worth mentioning that other physical mechanisms leading to BH mergers predict different $\alpha$ values. 
For instance, the PBH binaries formed by two-body encounters in the low-redshift universe yield
$\alpha \approx 1.43$~\cite{Bird:2016dcv}.
Astrophysical BH binaries formed in dense star clusters give $\alpha\sim 4$~\cite{OLeary:2016ayz}. 
BH binaries formed in mass-segregated environments such as galactic nuclei give $\alpha$ that vary with the total binary mass $m_{\rm t}$~\cite{Gondanetal2017}.
Thus measurement of $\alpha$ provides a unique test of the PBH scenario which is independent of the
assumption for the PBH mass function.

Alternative approach is to directly compare the observed mass distribution with the theoretical
ones in the PBH hypothesis computed for several representative PBH mass functions.
This program has been already applied to the observational data, and some PBH functions are excluded~\cite{Gow:2019pok}.

\paragraph{Spin of PBHs}
Another potentially powerful GW observable to test the PBH scenario is spin distribution.
GW observations are primarily sensitive to a quantity defined by Eq.~\eqref{definition-chi-eff}.
PBHs form binaries dynamically and there is no correlation between the spin of individual PBHs and
the orbital angular momentum.
Thus, probability density of $\chi_{\rm eff}$ is an even function, which is a robust prediction of the PBH scenario.
Concrete shape of the probability density of $\chi_{\rm eff}$ is determined by the spin distribution of 
the individual PBHs $W(J)$.
Formally, the spin distribution of PBHs at cosmic time $t$ is given by
\begin{equation}
W(J,t) =\int dJ'~Q(J,J',t) \int_{\delta_{\rm th} (J')}~P(\delta_M, J') d \delta_M \,,
\end{equation}
where $P(\delta_M,J)$ is the probability distribution of the density contrast $\delta_M$
and the angular momentum $J$ of an overdense region that collapses to PBH if $\delta_M \ge \delta_{\rm th}$.
$Q(J,J',t)$ represents the evolution of the PBH spin from its initial value $J'$ to $J$ at time $t$.
$P(\delta_M,J)$ is determined once the statistical properties of the primordial density perturbations are specified.
All these quantities must be determined in order to derive $W(J,t)$.
In Refs.~\cite{Chiba:2017rvs, Mirbabayi:2019uph, DeLuca:2019buf}, analyses related to $P(\delta_M,J)$ and $Q(J,J',t)$ have been performed.
For the Gaussian primordial density perturbation, the angular momentum of PBHs coming from $P(\delta_M,J)$ 
is estimated to be at most a few percent~\cite{Mirbabayi:2019uph, DeLuca:2019buf}.
In Ref.~\cite{He:2019cdb}, $\delta_{\rm th}(J)$ for PBHs formed in the radiation dominated epoch was obtained as
\begin{equation}
\delta_{\rm th} \approx 0.62+ 0.015a_K^2 \,,
\end{equation}
where $a_K$ is the Kerr parameter.
This result shows that the threshold increases as the angular momentum is increased
and the formation of highly spinning PBHs are suppressed compared to slowly spinning PBHs. 
On the other hand, PBHs formed in the matter-dominated phase, which could happen prior to the
radiation dominated epoch, likely to have significant spin~\cite{Harada:2017fjm}.

\vspace{5mm}
So far, we have considered GWs from PBHs in the LIGO frequency band.
This band corresponds to the merger phase of the stellar mass PBH binaries. 
However, GWs in different frequency bands could be also generated. 
The PBH binaries in the inspiral phase, due to their enormous number and weak signals,
constitute the low-frequency stochastic GW background.
Furthermore, primordial density perturbations as a seed of the PBHs provide another stochastic 
GW background by the non-linear mode couplings.
These low-frequency GWs are not probed by the LIGO-like detectors, but could be detected
by future detectors such as LISA.

In what follows, we introduce possible constraints on the abundance
of PBHs expected in future experiments by evaluating energy-density
spectra of stochastic GW backgrounds in two independent ways. 
We show that the experiments
are sensitive to constrain the fraction to cold dark matter (CDM) for
$10^{-5} \lesssim f_{\rm PBH} \lesssim 1$
($10^{-13} \lesssim f_{\rm PBH} \lesssim 1$) in case of the GWs from
coalescing events (curvature perturbations)~\cite{Wang:2019kaf}.

By using the merger rate of {PBH} binaries~\cite{Sasaki:2016jop}, we
can calculate the energy-density spectrum of stochastic gravitational
wave background (SGWG), which is discussed in Ref.~\cite{Wang:2016ana} to
be
\begin{align}
\Omega_{\textrm{GW}}(\nu) = \frac{\nu}{\rho_\textrm{c}} \int_{0}^{\frac{\nu_{\mathrm{cut}}}{\nu}-1} \frac{R_\textrm{PBH}(z)}{(1+z)H(z)}  \frac{dE_\textrm{GW}}{d\nu_\textrm{s}}(\nu_\textrm{s}) dz \,,
\label{GW_spectrum1}
\end{align}
where $\frac{dE_\textrm{GW}}{d\nu_\textrm{s}}(\nu_\textrm{s})$ is the
energy spectrum of GW produced by a BBH
coalescence~\cite{Ajith:2007kx,Ajith:2009bn},
$\nu_\textrm{s}=(1+z)\nu$, and $\nu_{\mathrm{cut}}$ is the cutoff
frequency.  

On the other hand, according to Appendix D of
Ref.~\cite{Wang:2019kaf}, we obtain the dimensionless energy-density
spectrum of the induced {SGWB} to be
\begin{eqnarray}
  \label{GW_spectrum2}
\Omega_{\rm{IGW}} 
\left(\nu =
 \frac{k}{2 \pi}\right)
&=&  \Omega_{\rm{r}, 0}\left(\frac{g_{*}(T(k))}{g_{*}\left(T_{\rm{eq}}\right)}\right)\left(\frac{g_{*, s}(T(k))}{g_{*, s}\left(T_{\rm{eq}}\right)}\right)^{-4 / 3}  \\
&\times& \frac{3 A^{2}}{64}\left(\frac{4-\tilde{k}^{2}}{4}\right)^{2} \tilde{k}^{2}\left(3 \tilde{k}^{2}-2\right)^{2}  \nonumber \\
 &\times& \left[\pi^{2}\left(3 \tilde{k}^{2}-2\right)^{2} \Theta(2-\sqrt{3} \tilde{k})+\left(4+\left(3 \tilde{k}^{2}-2\right) \ln \left|1-\frac{4}{3 \tilde{k}^{2}}\right|\right)^{2}\right]
\cr
 &\times& \Theta(2-\tilde{k}) \,,
\nonumber
\end{eqnarray}
where $\nu=k/2\pi $ denotes the frequency of GW, and the dimensionless
wavenumber $\tilde{k}=k/k_{0}$ is introduced for simplicity. We used
notations of physical quantities written in Ref.~\cite{Wang:2019kaf}.

\begin{figure}[tbh]
 \centering
\includegraphics[width=0.80 \columnwidth]{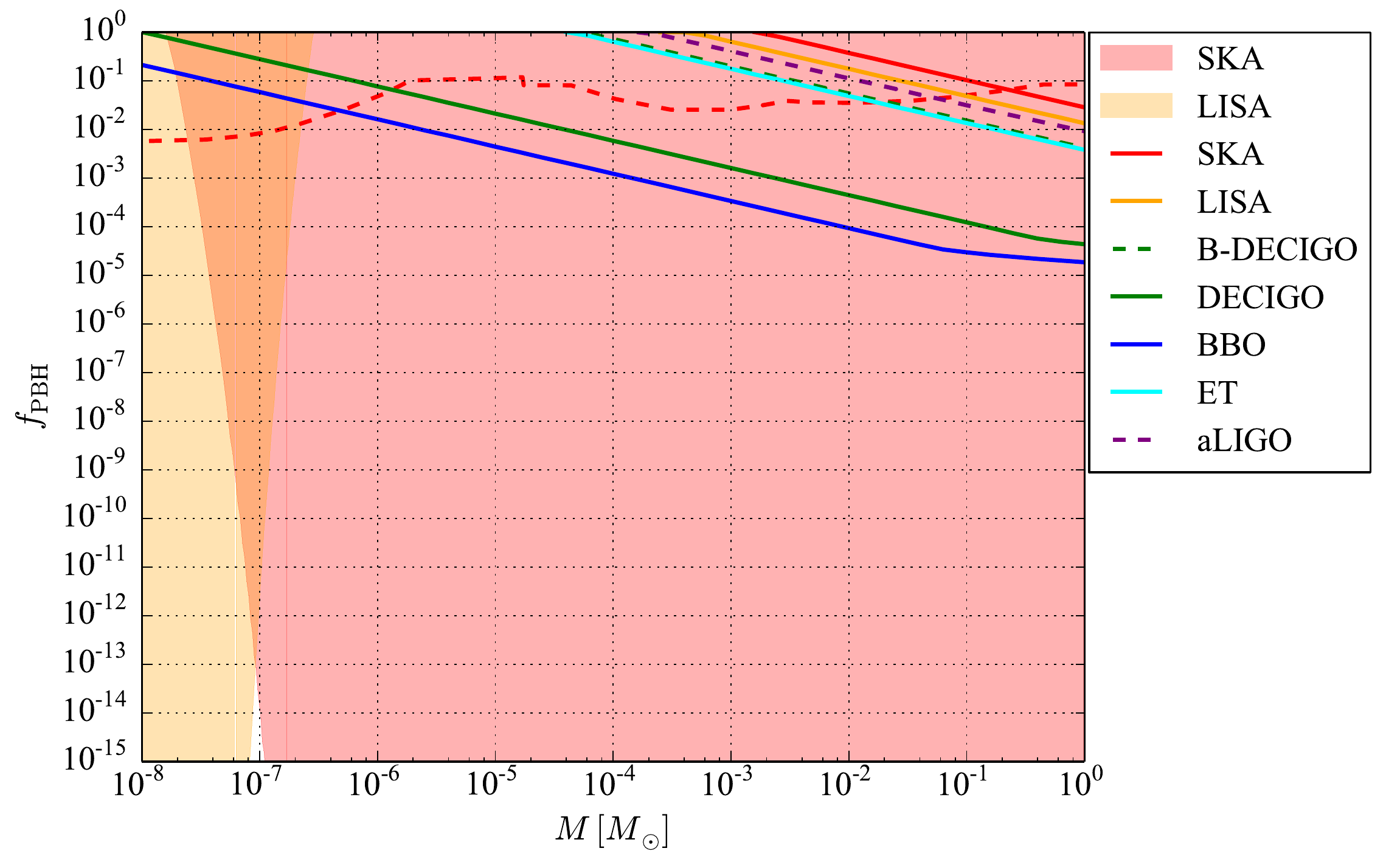}
\caption{Expected constraints on the {PBH} abundance versus the {PBH}
  mass from the null detection of the two kinds of {SGWB}. The
  existing observational constraint (red dashed) is plotted for
  comparison~\cite{Carr:2009jm}. This figure is taken by Fig.~7 of
  Ref.~\cite{Wang:2019kaf}.}
 \label{fig:expectedconstraints-2}
\end{figure}

Assuming a null detection of the GW signals from coalescing events in
Eq.~\eqref{GW_spectrum1} or curvature perturbations in
Eq.~\eqref{GW_spectrum2}, we can constrain the abundance of PBHs,
which is plotted in Fig.~\ref{fig:expectedconstraints-2}. The solid
lines give the expected upper limits on the abundance by the
non-detection of the GW signals from the coalescing events, and the
shaded regions mean the parameter space excluded by the non-detection
of the GW signals from curvature perturbations.

%% file: ptepGW_conclusion.tex
Gravitational wave physics and astronomy have finally become science that can be compared with 
the actual observational data after a long period of preparation.
In this paper, we introduced important findings such as the discovery of massive binary black 
holes and binary neutron stars, which bring us a new implication about physical properties of nuclear matter, heavy element synthesis in the Universe, and the origin of short gamma-ray 
bursts. 
It is expected that this trend will be further accelerated as the international network of gravitational wave detectors on the ground expands and space gravitational wave antennas extends.
Further improvements in sensitivity are planned for ground-based gravitational wave interferometers, \blue{including the construction of the third generation detectors\cite{CE,ET}}.
An increase in sensitivity by an order of magnitude increases the chance to detect an event having a very high signal-to-noise ratio, and at the same time increases the event rate by a factor of 1000 \blue{or more}, which enables highly statistically accurate analyses of gravitational waves.
It is certain that future gravitational wave observations will revolutionize the understanding of the physical universe that we have accumulated so far, powered by the improvement of the accuracy of gravitational wave source position and multi-messenger follow-up observations for gravitational wave events.
It is clear that we have not seen the great potential of this new probe of physics, gravitational waves, yet.

As the observation technology advances, it becomes more important to analyze and interpret the accumulated data correctly.
The real benefit from gravitational wave observations can be extracted only when we can construct an appropriate physical picture to explain the observed signal, by combining the data from other observational means to follow-up gravitational wave detections.
To that end, we need to further promote multi-messenger astronomy, and advance the theoretical research for understanding gravitational wave sources.
In addition, in order to detect gravitational wave signals that have not been discovered yet to open up a new paradigm, it is necessary to further develop gravitational wave data analysis methods,  making full use of the accumulated theoretical knowledge.

%% file: reference_PTEPGW.tex
\newcommand{\ApJL}[3]{{{Astrophys. J. Lett.} {\bf #1}, #2 (#3).}}
\newcommand{\ApJ}[3]{{{Astrophys. J.} {\bf #1}, #2 (#3).}}
\newcommand{\ApJS}[3]{{{Astrophys. J. Suppl.} {\bf #1}, #2 (#3).}}
\newcommand{\MNRAS}[3]{{{Mon. Not. R. Astron. Soc.} {\bf #1}, #2 (#3).}}
\newcommand{\MNRASL}[3]{{{Mon. Not. R. Astron. Soc. Lett.} {\bf #1}, #2 (#3).}}
\newcommand{\arxiv}[1]{{arXiv:#1.}}
\newcommand{\JCAP}[4]{{{J. Cosmol. Astropart. Phys.} {\bf #1}, #4 (#2).}}
\newcommand{\CQG}[3]{{{Class. Quant. Grav.} {\bf #1}, #2 (#3).}}
\newcommand{\Nat}[3]{{{Nature} {\bf #1}, #2 (#3).}}
\newcommand{\GRG}[3]{{{Gen. Rel. Grav.} {\bf #1} #2 (#3).}}
\newcommand{\AnA}[3]{{{Astronomy \& Astrophysics} {\bf #1}, #2 (#3).}}
\newcommand{\aap}[3]{{{Astronomy \& Astrophysics} {\bf #1}, #2 (#3).}}
\newcommand{\ARAA}[3]{{{Ann. Rev. Astron. Astrophys.} {\bf #1}, #2 (#3).}}
\newcommand{\PRep}[3]{{{Physics Reports} {\bf #1}, #2 (#3).}}
\newcommand{\Sci}[3]{{{Science} {\bf #1}, #2 (#3).}}
\newcommand{\LRR}[3]{{{Liv. Rev. Rel.} {\bf #1}, #2 (#3).}}
\newcommand{\PTEP}[3]{{{Prog. Theor. Exp. Phys.} {\bf #1}, #2 (#1).}}
\newcommand{\JPCS}[3]{{{J. Phys. Conf. Ser.} {\bf #1} #2 (#3).}}
\newcommand{\NatCom}[3]{{{\it Nature Communications} {\bf #1}, #2 (#3).}}
\newcommand{\NatAst}[3]{{{\it Nature Astron.} {\bf #1}:#2 (#3).}}
\newcommand{\PASJ}[3]{{{\it PASJ} {\bf #1}, #2 (#3).}}
\newcommand{\PASA}[3]{{{\it PASA} {\bf #1} #2 (#3).}}
\newcommand{\MPLA}[3]{{{\it Modern Physics Letters A} {\bf #1}, #2 (#3).}}
\newcommand{\NucPA}[3]{{{Nuc. Phys. A} {\bf #1}, #2 (#3).}}
\newcommand{\ARNPS}[3]{{{\it Annual Review of Nuclear and Particle Science} {\bf #1} #2 (#3).}}
\newcommand{\IJMPD}[3]{{{\it Int.\ J.\ Mod.\ Phys.\ D} {\bf #1}, #2 (#3).}}
\newcommand{\CRP}[3]{{{\it Comptes Rendus Physique} {\bf #1} #2 (#3).}}

\newcommand{\EPL}[4]{\href{https://doi.org/10.1209/epl/#4}{{\it Europhys. Lett.} {\bf #1}, #2 (#3).}}
\newcommand{\JOptB}[4]{\href{https://doi.org/10.1088/1464-4266/#4}{{\it J. Opt. B: Quantum Semiclass. Opt.}{\bf #1}, #2 (#3).}}
\newcommand{\LIGOdoc}[2]{\href{https://dcc.ligo.org/LIGO-#1}{{\it LIGO Document} {\bf #1} (#2).}}
\newcommand{\VIRdoc}[3]{\href{https://tds.virgo-gw.eu/?content=3&r=#3}{{\it Virgo document} {\bf VIR-#1} (#2).}}
\newcommand{\JGWdoc}[3]{\href{https://gwdoc.icrr.u-tokyo.ac.jp/cgi-bin/private/DocDB/ShowDocument?docid=#3}{{\it KAGRA document} {\bf JGW-#1} (#2)}}
\newcommand{\JETPL}[3]{{\it JETP Lett.} {\bf #1}, #2 (#3) [{\it Pisma Zh.\ Eksp. Teor. Fiz.}  {\bf #1}, #2 (#3).]}
\newcommand{\OME}[3]{\href{https://doi.org/10.1364/OME.#1.00#2}{{\it Opt. Mater. Express} {\bf #1}, #2 (#3)}}
\newcommand{\RPP}[4]{\href{https://doi.org/10.1088/1361-6633/#4}{{\it Rep. Prog. Phys.} {\bf #1}, #2 (#3).}}
\newcommand{\JOSAB}[4]{\href{https://doi.org/10.1364/JOSAB.#1.#4}{{\it J. Opt. Soc. Am. B} {\bf #1}, #2 (#3).}}
\newcommand{\AO}[4]{\href{https://doi.org/10.1364/AO.#1.#4}{{\it Appl. Opt.} {\bf #1}, #2 (#3).}}
\newcommand{\PSPIE}[4]{\href{https://doi.org/10.1117/#4}{{\it Proc. SPIE} {\bf #1}, #2 (#3).}}

\newcommand{\Authname}[2]{#2 #1} 
\newcommand{\LSC}{\Authname{Abbott}{B. P.} \etal (LIGO Scientific Collaboration) }
\newcommand{\LVC}{\Authname{Abbott}{B. P.} et al., (LIGO Scientific and Virgo Collaboration) }
\newcommand{\LVK}{\Authname{Abbott}{B. P.} \etal (LIGO Scientific, Virgo and KAGRA Collaboration) }